# Quantum Private Information Retrieval from Coded Storage Systems

**Matteo Allaix**

A doctoral thesis completed for the degree of Doctor of Science (Technology) to be defended, with the permission of the Aalto University School of Science, at a public examination held at the lecture hall H304 of the school on 29 November 2023 at 12:00.

**Aalto University**
**School of Science**
**Department of Mathematics and System Analysis**
**Algebra, Number Theory, and Applications**

**Supervising professor**
Professor Camilla Hollanti, Aalto University, Finland

**Thesis advisors**
Professor Camilla Hollanti, Aalto University, Finland
Doctor Tefjol Pllaha, University of Nebraska-Lincoln, United States of America

**Preliminary examiners**
Professor Gretchen Matthews, Virginia Tech, United States of America
Professor Alberto Ravagnani, Eindhoven University of Technology, Netherlands

**Opponent**
Professor Gretchen Matthews, Virginia Tech, United States of America

**Aalto University**
Aalto University, P.O. Box 11000, FI-00076 Aalto www.aalto.fi

# Abstract

**Author**
Matteo Allaix

**Abstract**
In the era of extensive data growth, robust and efficient mechanisms are needed to store and manage vast amounts of digital information, such as Data Storage Systems (DSSs). Concurrently, privacy concerns have arisen, leading to the development of techniques like Private Information Retrieval (PIR) to enable data access while preserving privacy. A PIR protocol allows users to retrieve information from a database without revealing the specifics of their query or the data they are accessing.

With the advent of quantum computing, researchers have explored the potential of using quantum systems to enhance privacy in information retrieval. In a Quantum Private Information Retrieval (QPIR) protocol, a user can retrieve information from a database by downloading quantum systems from multiple servers, while ensuring that the servers remain oblivious to the specific information being accessed. This scenario offers a unique advantage by leveraging the inherent properties of quantum systems to provide enhanced privacy guarantees and improved communication rates compared to classical PIR protocols.

In this thesis we consider the QPIR setting where the queries and the coded storage systems are classical, while the responses from the servers are quantum. This problem was treated by Song et al. for replicated storage and different collusion patterns. This thesis aims to develop QPIR protocols for coded storage by combining known classical PIR protocols with quantum communication algorithms, achieving enhanced privacy and communication costs. We consider different storage codes and robustness assumptions, and we prove that the achieved communication cost is always lower than the classical counterparts.

# Preface

I first want to express the biggest thanks to my supervisor and thesis advisor Camilla Hollanti. She has been the greatest supervisor I could have wished for, as even though she was always busy with her many projects, she could always find an hour or two in her schedule to support me on my path toward the end of my PhD. Our discussions were always helpful and would always give me some clarity when I was lost in my million thoughts. Furthermore, I am really grateful she's been very supportive and encouraging lately towards my choice of working full-time for a company, even if it means I don't have much time to work more on my research.

Next, I want to thank Tefjol Pllaha, who officially became my other thesis advisor only recently. When I first arrived in Finland for my Erasmus I knew nothing about quantum computation and communication, but thanks to his guidance and our countless discussions I learned everything I needed to fulfill the research I've conducted throughout the last 4 years.

Let me also thank my pre-examiners, Prof. Gretchen Matthews and Prof. Alberto Ravagnani, for agreeing to review my dissertation and approving it with kind words. I further thank Alberto for agreeing to be my opponent, even if it was on such a short notice and that means coming to the dark end of November in Finland.

A special thanks goes to Prof. Syed Ali Jafar who agreed to host me at University of California, Irvine to conduct research together for 3 months. Working with him and his research group was really inspiring, as it gave me a whole new perspective on how to do research. It was a once-in-a-lifetime experience to live in California for 3 months, and I'm really grateful to everyone I met there for the amazing moments spent together.

I want to express my gratitude towards my collaborators, Lukas Holzbaur, Seunghoan Song, Prof. Masahito Hayashi, Perttu Saarela, Ragnar Freij-Hollanti, Yuxiang Lu, Yuhang Yao for our many discussions that led to the works that support this dissertation. This dissertation wouldn't be even close to being as good as it is without their knowledge and support.

I'm thankful for being part of the great environment of the Department of Mathematics and System Analysis and of the ANTA group especially. It was always great to go to the coffee room, find somebody to discuss with during the coffee break, and share the current highs or lows of our research life.

I now want to thank all the amazing people I met throughout the last 4 years of my life in Finland for all the fun times we shared: Luca, David, Katherine, and the Bad Jokes Masters for “supporting” my bad jokes; Valentina, who first put me in contact with Camilla for an Erasmus in Finland; Francesco, for being a great flatmate and sharing endless conversations about startup ideas, and the other members of VMX; Franz, the cousin found in Finland, for making me meet so many other Allaix people; Pavel, for being an amazing friend even after moving away from Finland; Dawid and Álvaro, for being so inspiring with their constant motivation in life; Jake, Danilo, Daniel, and all the beach volleyball guys and coaches, for the great games and trainings played together throughout these years.

Voglio ora ringraziare tutti i miei amici in Italia che sono sempre stati pronti per vedersi ogni volta che sono sceso giù, nonostante scendo solo per un paio di settimane ogni 3/4 mesi. Voglio particolarmente ringraziare tutta la mia famiglia per esserci sempre nonostante la distanza, e per il costante incoraggiamento a continuare la mia vita qui in Finlandia, anche se vuol dire vedersi solo una volta ogni 3/4 mesi. Mamma, papà, non potrò mai smettere di ringraziarvi per tutti i sacrifici che avete fatto per permettermi di arrivare fin dove sono arrivato. Nonna, ti devo ringraziare anche solo per il fatto di esistere. Zii, Laura, Chri, Diana, Rosetta, grazie per essere la famiglia più piena di amore che potessi avere. Nonno, zio Dani, mi mancate, ma grazie per essere stati parte della mia vita. Mi mancate tutti.

A very special thanks goes to my Finnish family, Minna, Harppa, and Eme, just for being a real family in a country so far away from my hometown in Italy. Since I met you, you have been making me feel like a son to you.

The biggest biggest biggest thanks goes to Betsy, my partner, best friend, and so much more than that. Without her constant love and support I wouldn’t be the person I am right now. Bee, thank you so much for always being there, and for pushing me every day to be a better version of myself. I’m very thankful that we have Pepe in our life.

Everyone, thank you from the bottom of my heart.

Helsinki, December 10, 2023,

Matteo Allaix

# Contents

# List of Publications

This thesis consists of an overview and of the following publications which are referred to in the text by their Roman numerals.

**I** Matteo Allaix, Lukas Holzbaur, Tefjol Pllaha, and Camilla Hollanti. Quantum Private Information Retrieval from Coded and Colluding Servers. *IEEE Journal on Selected Areas in Information Theory* , 1, no. 2: 599–610, August 2020.

**II** Matteo Allaix, Seunghoan Song, Lukas Holzbaur, Tefjol Pllaha, Masahito Hayashi, and Camilla Hollanti. On the Capacity of Quantum Private Information Retrieval from MDS-Coded and Colluding Servers. *IEEE Journal on Selected Areas in Communications*, 40, no. 3: 885–898, January 2022.

**III** Matteo Allaix, Yuxiang Lu, Yuhang Yao, Tefjol Pllaha, Camilla Hollanti, and Syed Jafar. $N$-Sum Box: An Abstraction for Linear Computation over Many-to-one Quantum Networks. Submitted to *IEEE Transactions on Information Theory*, June 2023.

**IV** Perttu Saarela, Matteo Allaix, Ragnar Freij-Hollanti, and Camilla Hollanti. Private Information Retrieval from Colluding and Byzantine Servers with Binary Reed–Muller Codes. In *2022 IEEE International Symposium on Information Theory (ISIT)*, Espoo, Finland, pp. 2839–2844, June 2022.

# Author's Contribution

### Publication I: "Quantum Private Information Retrieval from Coded and Colluding Servers"

CH proposed the problem. TP contributed to broadening the team's understanding of quantum computation. MA derived the scheme construction for the Generalized Reed–Solomon (GRS) coded Quantum Private Information Retrieval (QPIR) scheme, and LH derived the scheme construction for the scheme with Locally Repairable Codes (LRC). All the co-authors contributed to the derivation of the technical results and writing the paper.

### Publication II: "On the Capacity of Quantum Private Information Retrieval from MDS-Coded and Colluding Servers"

CH and MH proposed the problem. TP contributed to broadening the team's understanding of stabilizer formalism for the scheme construction. MA derived the scheme construction for the Generalized Reed–Solomon (GRS) coded Quantum Private Information Retrieval (QPIR) scheme, and SS derived the converse results for the capacity of coded QPIR. All the co-authors contributed to the derivation of the technical results and writing the paper.

### Publication III: "$N$-Sum Box: An Abstraction for Linear Computation over Many-to-one Quantum Networks"

SJ proposed and formulated the problem. TP contributed to broadening the team's understanding of stabilizer formalism. YY contributed to establishing equivalence of different $N$-sum box forms, YL derived the maximal stabilizer-based $N$-sum box abstraction and used it to derive the construc-tion of the Quantum Private Information Retrieval (QPIR) scheme with a Cross-Subspace Alignment (CSA) coded storage, and MA derived the non-maximal stabilizer-based $(\kappa, N)$-sum box abstraction and used it to guarantee additional server privacy to the aforementioned scheme. All the co-authors contributed to the derivation of the technical results and writing of the paper.

### Publication IV: "Private Information Retrieval from Colluding and Byzantine Servers with Binary Reed–Muller Codes"

CH and RFH proposed the problem. PS derived the requirements for the Reed–Muller (RM) coded Private Information Retrieval (PIR) scheme. PS and MA derived the construction of the main example. All co-authors contributed to the derivation of the technical results and wrote the paper.

# List of Figures

# List of Tables

# Abbreviations

**DSS** Distributed Storage System

**PIR** Private Information Retrieval

**SPIR** Symmetric Private Information Retrieval

**QPIR** Quantum Private Information Retrieval

**SDMM** Secure Distributed Matrix Multiplication

**RSA** Rivest, Shamir, Adleman

**MDS** Maximum Distance Separable

**GRS** Generalized Reed–Solomon

**PRS** Primitive Reed–Solomon

**CSA** Cross Subspace Alignment

**BRM** Binary Reed–Muller

**LRC** Locally Repairable Code

**AG** Algebraic Geometry

**POVM** Positive-Operator Valued Measure

**PVM** Projective-Valued Measure

# 1. Introduction

In the era of big data, the exponential growth of digital information has required the development of robust and efficient mechanisms to store, manage, and access vast amounts of data. Traditional storage solutions, such as physical paper-based systems and early electronic databases, have proven inadequate in handling the sheer volume of information generated by modern organizations. On the other hand, regulatory requirements and compliance standards necessitate the secure and long-term storage of data. Additionally, disaster recovery and business continuity strategies rely on robust data storage systems to prevent data loss, minimize downtime, and protect against potential disruptions. Thus, reliable Data Storage Systems (DSSs) have become critical.

In parallel with the need for reliable DSSs, privacy has emerged as a critical concern in the digital age. Amongst the data collected for various purposes, a substantial amount of personal and sensitive information has been stored across multiple servers around the world. Hence, privacy breaches have the potential to inflict significant harm on individuals, ranging from financial fraud and identity theft to reputational damage and psychological distress.

To address the growing concerns surrounding data access and privacy, researchers have developed techniques that allow for the retrieval of information while preserving privacy, *e.g.*, Private Information Retrieval (PIR), which represents a promising solution to reconcile the conflicting objectives of data access and privacy. Introduced in the seminal paper by Chor *et al.* [7], PIR allows users to retrieve specific information from a database without revealing which specific item was being accessed, ensuring that the database server cannot determine the particular query or link it to the user making the request. In the PIR problem, two privacy settings have been considered: *information-theoretic* PIR aims to provide unconditional privacy, meaning that even an adversary with unlimited computational capacity cannot gain any information about the requested elements, while *computational* PIR aims to achieve privacy against adversaries with limited computational resources, relying on cryptographic techniques such as secure multiparty computation and homomorphic encryption.

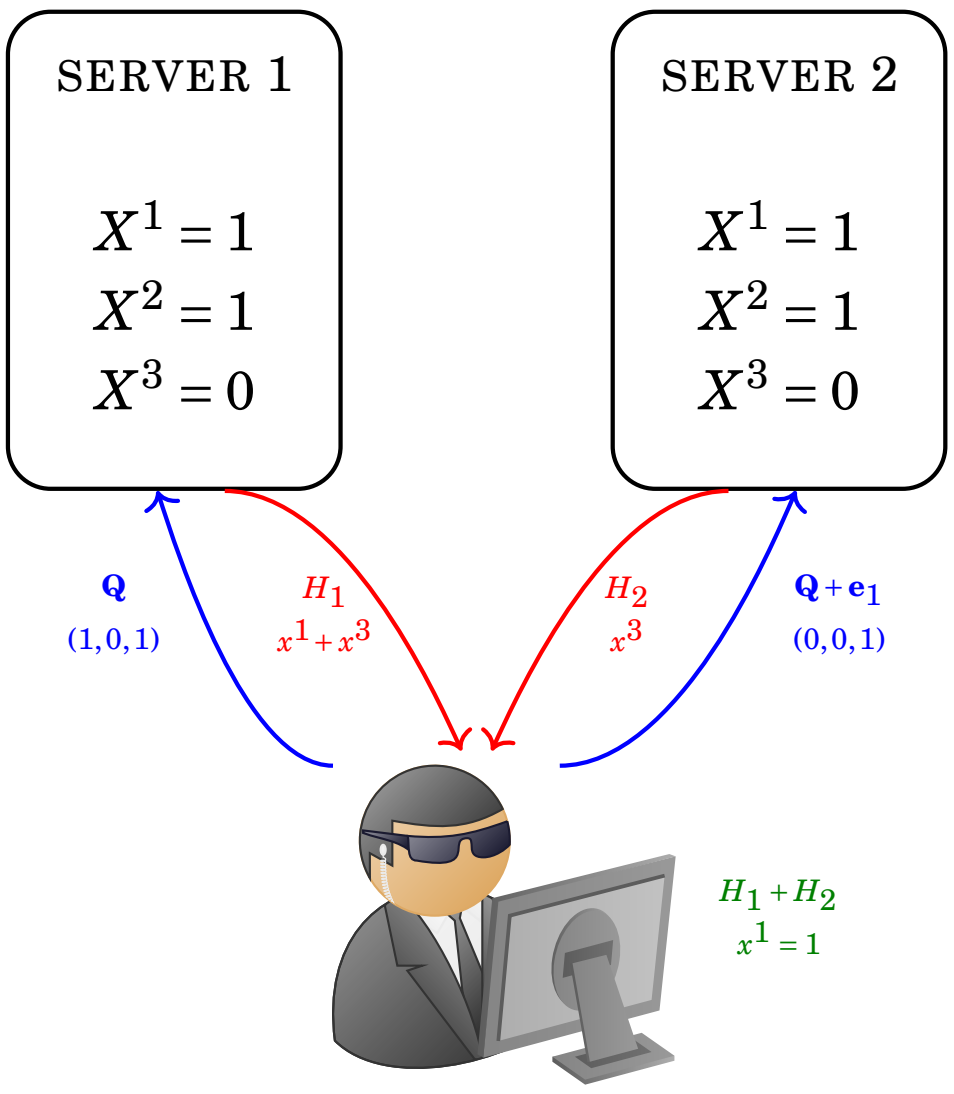

**Figure 1.1.** Trivial PIR example with replicated storage over $N = 2$ servers storing $M = 3$ files.

In the single-server case, the trivial solution of retrieving the whole database is optimal if we want to guarantee information-theoretic privacy [7], whereas there exist solutions that download only a part of the database if we want to guarantee only computational privacy [5, 8, 27]. The PIR problem becomes more interesting when we consider multiple servers, *e.g.*, in a DSS, as shown in the following example.

**Example 1.** *Consider the setting of a DSS with replicated storage over* $N = 2$ *servers storing* $M = 3$ *files that can be represented as a column vector* $\mathbf{X} \in \mathbb{F}_2^M$*, as shown in Figure 1.1. The user wants to retrieve the file* $X^1$ *with index 1. To achieve this, the user first generates a random query* $\mathbf{Q} = (1,0,1)$ *of* $M = 3$ *bits. The user sends* $\mathbf{Q}$ *to the first server and* $\mathbf{Q} + \mathbf{e}_1 = (0,0,1)$ *to the second server, as* $\mathbf{e}_1$ *is the query vector he would use to retrieve the first file. The servers compute the inner product between* $\mathbf{X}$ *and their received query, and they send their result* $H_i \in \mathbb{F}_2$ *to the user. After summing the responses, the user obtains* $H_1 + H_2 = X^1 = 1$ *as desired.*

Subsequent to the initial developments, notable progress has been made in constructing PIR schemes that exhibit low communication complexity in replicated systems with multiple servers, without relying on assumptions about their computational capabilities [4, 10, 21, 47]. More recent research has focused on PIR settings wherein a reliable DSS holds a collection of $M$ files $\{\mathbf{X}^1, \ldots, \mathbf{X}^M\}$, possibly encoded and distributed over multiple servers, where each file's size significantly exceeds the total number of files. In such scenarios, the size of a query is typically independent of the file size, thereby allowing estimation of the number of communicated bits solely based on the downloaded bits. The PIR *rate*, consequently, is defined as the ratio of the desired file's size over the downloaded data's total size. In recent years the PIR rate has been improved accounting for scenarios that include different assumptions (cf. Section 4.1.1) [11, 22, 24, 40], such as *T-privacy*, *symmetry*, *storage coding*, *X-security*, and *robustness against byzantine and unresponsive servers*. The PIR *capacity* represents the supremum of all the achievable rates in a given setting. The capacity has not been found yet for the most general PIR setting (cf. Section 4.1.3), but many results are available considering a subset of all the possible assumptions [3, 19, 20, 37–39, 43–45].

While classical PIR schemes have been progressively improved to achieve capacity results, the advent of quantum computing has created opportunities for exploring the potential of Quantum Private Information Retrieval (QPIR) protocols [14,26,28]. In the context of this thesis, the QPIR problem addresses the problem of retrieving a classical file in a DSS by enabling quantum communication from the servers to the user. Each server contains a set of codewords encoding the matrix of files, while ensuring that the index of the retrieved file remains undisclosed to any individual server or to any $T$-subset of servers. It is assumed that there is no transmission of quantum systems from the user to the servers. Previously, Song *et al.* have studied this QPIR setting where each server holds the entire classical file matrix [33–35]. This thesis aims to develop QPIR protocols for the case of coded storage with some possibly additional assumptions. Furthermore, we propose a result for the QPIR capacity when the QPIR assumptions are inherited by a classical setting. The QPIR capacity, as one might expect, is defined as the supremum of the ratios of the file size over the total dimension of the downloaded quantum systems, representing the maximal information that can be possibly carried by the quantum systems according to the well known Holevo bound [18].

This thesis is structured as follows.

- In Chapter 2 we provide an introduction to coding-theoretic concepts essential for describing the QPIR protocols. Specific emphasis is placed on the properties of Generalized Reed–Solomon (GRS) codes, Cross Subspace Alignment (CSA) codes, binary Reed–Muller (BRM) codes, and Locally Repairable Codes (LRCs).

- In Chapter 3 we offer an overview of quantum systems and explain some key properties in quantum computation, including fundamental quantum protocols. Remarkably, quantum computation can be modeled and formulated using postulates based on linear algebra.

- In Chapter 4 we discuss previous results in classical PIR. Furthermore, we propose the results of the publications in this thesis that improve the rates and capacities of classical PIR protocols using quantum computation.

- In Chapter 5 we summarize the results shown in this thesis and we provide some open problems and future research directions.

# 2. Coding Theory

In recent years, the rapid growth of digital data has brought data storage to the forefront of technological concerns. To address this challenge, DSSs have emerged as a powerful solution. The main idea behind DSSs is to implement data storage in a server network, introducing redundancy to ensure the reliable recovery of lost data. One way to achieve this redundancy is through data replication, where the same data is stored on multiple servers. This enables the repair of a malfunctioning server or the recovery of a file by interacting with any other server in the network that holds a copy of the data. The replication is the simplest type of encoding and it is achieved by the so-called *repetition code*.

However, while replication is effective, it comes at the cost of storage overhead. To mitigate this, erasure codes provide a more efficient solution by splitting files into $k$ information symbols and further encoding those into $n$ information symbols. These codes can be constructed such that any subset of $k$ encoded information symbols is enough to reconstruct the original file, even if some of the symbols are lost or damaged. This reduces the storage overhead required for redundancy.

As a result, erasure codes have become a popular choice for DSSs that deal with large amounts of data. By providing high reliability with lower storage requirements, erasure codes offer a scalable and cost-effective solution for distributed data storage.

In this chapter, we give some basic properties of codes and introduce some linear codes that we use in the next chapters. We mainly follow the results shown in [29, 41].

## 2.1 Finite fields

Let $q = p^{\mu}$, where $p$ is a prime number and $\mu$ is a positive integer. We denote the field of $q$ elements with $\mathbb{F}_q$ and the $n$-dimensional vector space over $\mathbb{F}_q$ by $\mathbb{F}_q^n$. For $x \in \mathbb{F}_q$, the map $y \in \mathbb{F}_q \mapsto xy \in \mathbb{F}_q$ is linear. Let $\Gamma = \{\gamma_1, ...., \gamma_\mu\}$ be a basis of $\mathbb{F}_{q^\mu}$ over $\mathbb{F}_q$. For $\alpha \in \mathbb{F}_{q^\mu}$ we denote by $\varphi$ the bijective, $\mathbb{F}_q$-linear

map

$$\varphi : \mathbb{F}_{q^\mu} \to \mathbb{F}_q^\mu$$
$$\alpha = \sum_{i=1}^{\mu} \alpha_i \gamma_i \mapsto (\alpha_1, \dots, \alpha_\mu). \tag{2.1}$$

For a vector $\mathbf{v} \in \mathbb{F}_{q^\mu}^n$ we write $\varphi(\mathbf{v}) = (\varphi(v_1), \varphi(v_2), \dots, \varphi(v_n))$ for the component-wise application of the mapping, by slight abuse of notation. Since $\mathbb{F}_q$ is isomorphic to the vector space $\mathbb{F}_p^\mu$ through $\varphi$, the map $\varphi^{-1}(y) \in \mathbb{F}_p^\mu \mapsto \varphi^{-1}(xy) \in \mathbb{F}_p^\mu$ for some $x \in \mathbb{F}_q$ is also linear and we define its matrix representation as $\mathbf{T}_x$. Then we can define the trace operator $\mathrm{tr}\, x := \mathrm{Tr}\, \mathbf{T}_x \in \mathbb{F}_p$ for $x \in \mathbb{F}_q$.

## 2.2 Linear codes

*Codewords* are used to ensure the reliable storage of messages. An *encoding* is an injective map $\varepsilon : \mathbb{F}_q^k \to \mathbb{F}_q^n$, where $n \geq k$, and a codeword is formally defined as $\varepsilon(\mathbf{m})$ for some $\mathbf{m} \in \mathbb{F}_q^k$. The collection of codewords is called a *code*, denoted by $\mathscr{C}$. A $q$-ary code of length $n$ is a non-empty subset of $\mathbb{F}_q^n$, *i.e.*, $\mathscr{C} := \left\{\varepsilon(\mathbf{m}) : \mathbf{m} \in \mathbb{F}_q^k, k \neq 0\right\}$. *Linear codes* are a specific type of codes such that any linear combination of two codewords in the code $\mathscr{C}$ is also a codeword in $\mathscr{C}$, making $\mathscr{C}$ a subspace of $\mathbb{F}_q^n$. This work only considers linear codes, so the term "code" is used to refer to linear codes. For non-linear codes, we refer the reader to [48].

We now present several standard definitions and results from coding theory needed for the following sections. We denote the set $\{1, 2, \dots, n\}$ by $[n]$ and the vector of length $n$ containing $a$ in every entry by $\mathbf{a}^n$. We denote matrices and vectors by bold characters, and their sub-indices indicate their dimensions if they are unclear from context. Vectors are initialized as row vectors unless stated otherwise in the context.

**Definition 1.** *The* support of a vector $\mathbf{v} \in \mathbb{F}_q^n$ *is defined as the set of positions of non-zero entries of* $\mathbf{v}$, i.e., $\mathrm{supp}(\mathbf{v}) := \{i : v_i \neq 0\}$. *The* support of a code $\mathscr{C} \subseteq \mathbb{F}_q^n$ *is the set of positions where the non-zero symbols of all codewords in* $\mathscr{C}$ *are located,* i.e., $\mathrm{supp}(\mathscr{C}) := \bigcup_{\mathbf{c} \in \mathscr{C}} \mathrm{supp}(\mathbf{c})$.

*The* Hamming distance *between two codewords* $\mathbf{c}, \mathbf{d} \in \mathscr{C}$ *is the number of positions where they differ,* i.e., $d_H(\mathbf{c}, \mathbf{d}) := |\{i : c_i \neq d_i\}| = |\mathrm{supp}(\mathbf{c} - \mathbf{d})|$. *The Hamming distance defines a metric on* $\mathscr{C}$.

*The* Hamming weight *of a codeword* $\mathbf{c} \in \mathscr{C}$ *is defined as the number of non-zero symbols in the codeword,* i.e., $\mathrm{wt}(\mathbf{c}) := |\mathrm{supp}(\mathbf{c})| = d_H(\mathbf{c}, \mathbf{0})$. *Notice that* $\mathrm{wt}(\mathbf{c} - \mathbf{d}) = d_H(\mathbf{c}, \mathbf{d})$.

*Given a codeword* $\mathbf{c} \in \mathscr{C}$*, the* Hamming ball *of radius r around* $\mathbf{c}$ *is the set* $\mathscr{B}(\mathbf{c}, r) := \{\mathbf{d} \in \mathscr{C} : d_H(\mathbf{c}, \mathbf{d}) \leq r\}$.

Three parameters characterize a code $\mathscr{C}$:

- the *dimension* $\dim(\mathscr{C}) = k$ of the code corresponds to the number of information symbols in a message that can be stored using $\mathscr{C}$, which is equivalent to the dimension of the message space $\mathbb{F}_q^k$;

- the *length* $n$ is the length of the codewords in the codebook, which is a subspace of $\mathbb{F}_q^n$;

- the *minimum distance* $d$ of the code is defined as the smallest Hamming distance between any two distinct codewords in the codebook:

$$d := \min_{\substack{\mathbf{c},\mathbf{d}\in\mathscr{C} \\ \mathbf{c}\neq\mathbf{d}}} d_H(\mathbf{c},\mathbf{d}). \tag{2.2}$$

**Definition 2.** *A code with the three parameters described above is called an $[n,k,d]_q$ code. The* redundancy *of such a code is defined to be $n-k$.*

For a linear code $\mathscr{C}$, its minimum distance equals the minimum Hamming weight of non-zero codewords of $\mathscr{C}$, *i.e.*, $d = \min_{\mathbf{c}\in\mathscr{C}\setminus\{0\}} \mathrm{wt}(\mathbf{c})$. Let us give some simple yet important examples of binary linear codes:

- the $[n,n-1,2]_2$ code, also called the *binary parity-check code*, contains all the vectors in $\mathbb{F}_2^n$ with even Hamming weight;

- the $[n,1,n]_2$ code, also called the *binary repetition code*, contains only the two vectors $\mathbf{0}^n$ and $\mathbf{1}^n$;

- the $[n,n,1]_2$ code contains all the vectors in $\mathbb{F}_2^n$.

These codes are also called the *trivial binary codes*. One example of a trivial $q$-ary code can be given by the $q$-ary repetition code $\mathscr{R}_q(n)$, which is an $[n,1,n]_q$ code consisting of the $q$ vectors $\mathbf{0}^n,\cdots,(\mathbf{q-1})^n$.

There exists a fundamental relationship between a code's minimum distance $d$ and its redundancy $n-k$, known as the *Singleton bound* [32].

**Theorem 1** (Singleton bound)**.** *For any code $\mathscr{C}$, the redundancy is always greater than or equal to the minimum distance minus one, which can be written as*

$$n-k \geq d-1. \tag{2.3}$$

### 2.2.1 Error detection and error correction

The redundancy of code enables error detection and correction, which are techniques used to protect digital information from errors that can occur during transmission or storage. The goal is to ensure that the receiver can recover the original message even if some bits of the message have been corrupted by noise or other factors. In particular, linear codes are equipped with error-detection and error-correction capabilities.

**Definition 3.** *A code $\mathscr{C}$ of length $n$ is said to* correct *$t$ errors if for every $\mathbf{x} \in \mathbb{F}_q^n$ there is at most one $\mathbf{c} \in \mathscr{C}$ such that $d_H(\mathbf{c},\mathbf{x}) \le t$ or, equivalently, if $\mathscr{B}(\mathbf{c},t) \cap \mathscr{B}(\mathbf{d},t) = \emptyset$ for each pair of distinct $\mathbf{c},\mathbf{d} \in \mathscr{C}$.*

*A code $\mathscr{C}$ of length $n$ is said to* detect *$t$ errors (or correct $t$ erasures) if for every $\mathbf{x} \in \mathbb{F}_q^n, \mathbf{c} \in \mathscr{C}$ such that $d_H(\mathbf{c},\mathbf{x}) \le t$ we have that $\mathbf{x} \notin \mathscr{C}$ or, equivalently, if $\mathbf{d} \notin \mathscr{B}(\mathbf{c},t)$ for each pair of distinct $\mathbf{c},\mathbf{d} \in \mathscr{C}$.*

These notions are related to the minimum distance as follows.

**Theorem 2.** *An $[n,k,d]_q$ code can detect up to $d-1$ errors and can correct up to $\lfloor \frac{d-1}{2} \rfloor$ errors.*

### 2.2.2 The generator matrix and the parity-check matrix

In this subsection, we describe how to define a linear code through a linear mapping.

**Definition 4.** *We define a* generator matrix *$\mathbf{G}_{\mathscr{C}} \in \mathbb{F}_q^{k \times n}$ for a linear code $\mathscr{C}$ of length $n$, dimension $k$, and minimum distance $d$ as a matrix with rank $k$ whose rows span $\mathscr{C}$,* i.e., *form a vector space basis. The encoding of a message $\mathbf{m} \in \mathbb{F}_q^k$ into a codeword $\mathbf{c} \in \mathscr{C}$ is performed by the linear transformation $\mathbf{m} \mapsto \mathbf{m}\mathbf{G}_{\mathscr{C}}$. Thus, for a linear code $\mathscr{C}$, there exists an* encoding map *$\varepsilon : \mathbb{F}_q^k \to \mathbb{F}_q^n$ that maps each message $\mathbf{m}$ to its corresponding codeword $\mathbf{c} = \mathbf{m}\mathbf{G}_{\mathscr{C}}$.*

The generator matrix $\mathbf{G}_{\mathscr{C}}$ allows us to efficiently encode a message $\mathbf{m} \in \mathbb{F}_q^k$ into a codeword $\mathbf{c} \in \mathscr{C}$ using the relation $\mathbf{c} = \mathbf{m}\mathbf{G}_{\mathscr{C}}$. It also provides a way to generate all the codewords of the code by taking linear combinations of the rows of $\mathbf{G}_{\mathscr{C}}$.

**Theorem 3.** *Suppose $\mathscr{C}$ is an $[n,k,d]_q$ code. Then, $\mathscr{C}$ has a generator matrix $\mathbf{G}_{\mathscr{C}}$ that can be expressed in the* systematic form *$(\mathbf{I}_k \,|\, \mathbf{G}')$, where $\mathbf{I}_k$ is the $k \times k$ identity matrix and $\mathbf{G}'$ is a $k \times (n-k)$ matrix, after permuting the coordinates if necessary.*

The systematic form ensures that the first $k$ columns of $\mathbf{G}_{\mathscr{C}}$ correspond to the $k$ message symbols and the remaining columns correspond to the parity-check symbols. This form is particularly useful since it allows us to easily recover the original message from a received codeword by performing elementary row operations on the matrix. For example, the systematic form of a generator matrix of the binary parity-check $[n,n-1,2]_2$ code is $\mathbf{G} = (\mathbf{I}_{n-1} \,|\, \mathbf{1}_{n-1}^\top)$.

**Definition 5.** *In the context of an $[n,k,d]_q$ code $\mathscr{C}$, an* information set *is any $\mathscr{I}_k \subseteq [n]$ of size $k$ such that the columns of the generator matrix $\mathbf{G}_{\mathscr{C}}$ indexed by $\mathscr{I}_k$ are linearly independent.*

In other words, the set of positions indexed by $\mathscr{I}_k$ corresponds to the coordinates of the received codeword that can be used to recover the original message, without knowing the values of the remaining coordinates.

**Definition 6.** *The* parity-check matrix *of an* $[n,k,d]_q$ *code* $\mathscr{C}$ *is a full-rank matrix* $\mathbf{H}_{\mathscr{C}} \in \mathbb{F}_q^{n\times(n-k)}$ *that satisfies the equation* $\mathbf{c}\mathbf{H}_{\mathscr{C}} = \mathbf{0}$ *for all* $\mathbf{c}$ *in* $\mathscr{C}$.

The parity-check matrix $\mathbf{H}_{\mathscr{C}}$ allows us to efficiently check if a given vector $\mathbf{c} \in \mathbb{F}_q^n$ is a codeword in $\mathscr{C}$ by verifying if $\mathbf{c}\mathbf{H}_{\mathscr{C}} = \mathbf{0}$. Also, notice that $\mathscr{C}$ is the nullspace of $\mathbf{H}_{\mathscr{C}}$, since we can write $\mathscr{C} = \{\mathbf{c} \in \mathbb{F}_q^n : \mathbf{c}\mathbf{H}_{\mathscr{C}} = \mathbf{0}\}$. Thus, a linear code can be compactly represented by either its generator matrix or its parity-check matrix. The minimum distance of the code $\mathscr{C}$ is related to the properties of the parity-check matrix $\mathbf{H}_{\mathscr{C}}$: it is equal to the minimum number of linearly dependent rows of $\mathbf{H}_{\mathscr{C}}$.

### 2.2.3 Dual codes

A linear code has a natural complement, which we call its dual code, and is defined as follows.

**Definition 7.** *The* dual code $\mathscr{C}^\perp := \left\{\mathbf{x} \in \mathbb{F}_q^n : \mathbf{x}\cdot\mathbf{c} = 0 \ \forall \mathbf{c} \in \mathscr{C}\right\} \subseteq \mathbb{F}_q^n$ *is the dual vector space of* $\mathscr{C}$*,* i.e.*, is the set of all the vectors in* $\mathbb{F}_q^n$ *that are orthogonal to every codeword in* $\mathscr{C}$.

The dual code has many properties that depend on the properties of the original code.

**Theorem 4.** *If* $\mathscr{C}$ *is an* $[n,k,d]_q$ *code, then* $\mathscr{C}^\perp$ *is an* $[n,n-k,-]_q$ *code and* $\left(\mathscr{C}^\perp\right)^\perp = \mathscr{C}$. *If* $\mathbf{H}_{\mathscr{C}}$ *is a parity-check matrix for* $\mathscr{C}$*, then* $\mathbf{H}_{\mathscr{C}}^\top$ *is a generator matrix for* $\mathscr{C}^\perp$*; equivalently, if* $\mathbf{G}_{\mathscr{C}}$ *is a generator matrix for* $\mathscr{C}$*, then* $\mathbf{G}_{\mathscr{C}}^\top$ *is a parity-check matrix for* $\mathscr{C}^\perp$.

**Corollary 1.** *Let* $\mathscr{C}$ *be an* $[n,k,d]_q$ *code. If* $\mathbf{G}_{\mathscr{C}}$ *is a generator matrix for* $\mathscr{C}$ *and* $\mathbf{G}_{\mathscr{C}^\perp}$ *is a generator matrix for* $\mathscr{C}^\perp$*, then* $\mathbf{G}_{\mathscr{C}}\mathbf{G}_{\mathscr{C}^\perp}^\top = \mathbf{0}$.

While in vector spaces of characteristic 0 (such as $\mathbb{R}$) the dual (or orthogonal complement) $\mathscr{W}^\perp$ of a subspace $\mathscr{W} \subseteq \mathbb{R}^n$ satisfies $\mathscr{W} \cap \mathscr{W}^\perp = \{0\}$ and $\mathscr{W} + \mathscr{W}^\perp = \mathbb{R}^n$, subspaces of $\mathbb{F}_q^n$ can behave differently. Specifically, the dual code $\mathscr{C}^\perp$ of a linear code $\mathscr{C}$ can intersect non-trivially, meaning $\mathscr{C} \cap \mathscr{C}^\perp \neq \{0\}$. It is even possible for $\mathscr{C}$ to be equal to its dual code $\mathscr{C}^\perp$. In this case, we refer to it as a *self-dual* code.

### 2.2.4 MDS codes

*Maximum Distance Separable* (MDS) codes are a type of erasure code that has a minimum distance of $d = n-k+1$, *i.e.*, they achieve the Singleton bound with equality (cf. Equation (2.3)). MDS codes are able to correct up to $n-k$ erasures and $\lfloor \frac{n-k}{2} \rfloor$ errors. We use the notation $[n,k]_q$ code to denote an MDS code as the minimum distance is implied by its definition. Additionally, the dual of an MDS code is also an MDS code of length $n$ and dimension $n-k$, *i.e.*, the class of MDS codes is closed under the dual operation.

A very important property of MDS codes is given in the following theorem.

**Theorem 5.** *Let $\mathscr{C}$ be an $[n,k]_q$ MDS code. Any $k$-subset of the $n$ columns of the generator matrix $\mathbf{G}_{\mathscr{C}}$ forms an information set. In other words, any $k \times k$ submatrix of its generator matrix $\mathbf{G}_{\mathscr{C}}$ is invertible.*

One example of an $[n,k]_q$ MDS code is the parity-check code, where $k = n-1$ and its generator matrix is given by $\mathbf{G}_{\mathscr{C}} = (\mathbf{I}_{n-1} \,|\, \mathbf{1}_{n-1}^{\top})$. This code stores the uncoded information symbols in the first $k$ coordinates and the sum of all symbols in the last coordinate. It is straightforward to observe that any $k \times k$ submatrix of $\mathbf{G}_{\mathscr{C}}$ is invertible. Since this code has a minimum distance $d = 2$, it can correct only one erasure and it cannot correct any errors.

More generally, all the trivial binary codes are MDS, and they are the only binary codes that have such a property [42, Proposition 9.2].

### 2.2.5 The star product of codes

We now define the *star product* of two subvector spaces, also referred to as the *wedge product*, *Hadamard product* or *Schur product*, and we denote it by $\star$.

**Definition 8.** *For any two vector spaces $\mathscr{V}, \mathscr{W} \subseteq \mathbb{F}_q^n$, we can define $\mathscr{V} \star \mathscr{W}$ to be the subspace of $\mathbb{F}_q^n$ generated by the Hadamard products $\mathbf{v} \star \mathbf{w} := (v_1w_1, \ldots, v_nw_n)$,* i.e.*,*

$$\mathscr{V} \star \mathscr{W} := \operatorname{span}\{\mathbf{v} \star \mathbf{w} : \mathbf{v} \in \mathscr{V}, \mathbf{w} \in \mathscr{W}\}. \tag{2.4}$$

The star product has interesting properties:

- for any linear code $\mathscr{C} \subseteq \mathbb{F}_q^n$, we have that $\mathscr{C} \star \mathscr{R}_q(n) = \mathscr{C}$;
- if $\mathscr{C}$ and $\mathscr{C}'$ are any linear codes in $\mathbb{F}_q^n$ with $\operatorname{supp}(\mathscr{C}) = \operatorname{supp}(\mathscr{C}') = [n]$, then $d_{(\mathscr{C} \star \mathscr{C}')^\perp} \geq d_{\mathscr{C}^\perp} + d_{\mathscr{C}'^\perp} - 2$;
- for any MDS code $\mathscr{C} \subseteq \mathbb{F}_q^n$ we have that $\left(\mathscr{C} \star \mathscr{C}^\perp\right)^\perp = \mathscr{R}_q(n)$;
- if $\mathscr{C}$ and $\mathscr{C}'$ are linear codes in $\mathbb{F}_q^n$, it holds that

$$(\mathscr{C} \times \mathscr{C}) \star (\mathscr{C}' \times \mathscr{C}') = (\mathscr{C} \star \mathscr{C}') \times (\mathscr{C} \star \mathscr{C}'),$$

where $\times$ represents the cartesian product of two codes.

### 2.2.6 Codes for storage

Let us define an encoded DSS with $n$ servers as follows. A *storage code* is a type of code that first divides files into $\beta$ stripes of $k$ information symbols, then encodes each stripe into a vector of $n$ information symbols, and finally stores each symbol of the encoded vector in the corresponding server. We can represent the $i$-th file, denoted as $\mathbf{x}^i$, as a $\beta \times k$ matrix as follows

$$\mathbf{x}^i := \begin{pmatrix} x^i_{1,1} & x^i_{1,2} & \cdots & x^i_{1,k} \\ x^i_{2,1} & x^i_{2,2} & \cdots & x^i_{2,k} \\ \vdots & \vdots & \ddots & \vdots \\ x^i_{\beta,1} & x^i_{\beta,2} & \cdots & x^i_{\beta,k} \end{pmatrix}.$$

In a storage code $\mathscr{C}$, the $k$ information symbols in the $b$-th stripe of the file $\mathbf{x}^i$ are mapped to the $n$ encoded symbols $\mathbf{y}^i_b = \mathbf{x}^i_b \mathbf{G}_{\mathscr{C}}$. Thus, the encoded symbols of file $\mathbf{x}^i$ are stored as a $\beta \times n$ matrix $\mathbf{y}^i = \mathbf{x}^i \mathbf{G}_{\mathscr{C}}$, where the $s$-th column of $\mathbf{y}^i$ is stored on server $s$.

In order to decode an encoded file, we need to choose an information set $\mathscr{I}$ of the generator matrix $\mathbf{G}_{\mathscr{C}}$ of the code. For simplicity, consider $\beta = 1$. Let $\mathbf{y}^i_{\mathscr{I}}$ be the vector obtained by selecting the elements of $\mathbf{y}^i$ in the positions specified by $\mathscr{I}$. Then we can decode to the original file by multiplying $\mathbf{y}^i_{\mathscr{I}}$ by the inverse of the $k \times k$ submatrix obtained by the $k$ columns of $\mathbf{G}_{\mathscr{C}}$ corresponding to $\mathscr{I}$.

For an MDS code, any $k \times k$ submatrix of its generator matrix $\mathbf{G}_{\mathscr{C}}$ is invertible. Thus, if we want to decode the file $\mathbf{x}^i$, we just need to use any $k$ out of $n$ columns of the matrix $\mathbf{y}^i$.

## 2.3 Families of linear codes

In this section we introduce some families of linear codes used for PIR purposes: Generalized Reed–Solomon (GRS) codes, Cross-Subspace Alignment (CSA) codes, Binary Reed–Muller (BRM) codes, and Locally Repairable codes (LRC). Notice that GRS and BRM codes are examples of evaluation codes, of which another important subclass are Algebraic Geometric (AG) codes [36].

### 2.3.1 Generalized Reed–Solomon codes

*Generalized Reed–Solomon* (GRS) codes constitute a popular and useful family of MDS codes. To construct a GRS code we start with a field $\mathbb{F}_q$ of size $q \geq n$, a set $\mathscr{A} = \{\alpha_1, \ldots, \alpha_n\}$ of $n$ distinct elements from $\mathbb{F}_q$, and a vector $\mathbf{v} = (v_1, \ldots, v_n)$ of $n$ non-zero elements from $\mathbb{F}_q$.

**Definition 9.** *A GRS code* $\mathrm{GRS}^q_k(\mathscr{A}, \mathbf{v})$ *is the set of all the evaluations of*

*polynomials* $p \in \mathbb{F}_q[z]^{<k}$ *at the elements of* $\mathscr{A}$ *weighted by the vector* $\mathbf{v}$, i.e.,

$$\mathrm{GRS}_k^q(\mathscr{A},\mathbf{v}) := \left\{\mathbf{v} \star \mathrm{eval}_{\mathscr{A}}(p) : p \in \mathbb{F}_q[z]^{<k}\right\}, \tag{2.5}$$

In the definition, $\mathbb{F}_q[z]^{<k}$ is the set of single-variable polynomials of degree less than $k$ over $\mathbb{F}_q$. The evaluation map $\mathrm{eval}_{\mathscr{A}}$ takes a polynomial $p$ and maps it to the $n$-tuple $(p(\alpha_1),\dots,p(\alpha_n)) \in \mathbb{F}_q^n$ given $\mathscr{A} = \{\alpha_1,\dots,\alpha_n\}$, *i.e.*,

$$\mathrm{eval}_{\mathscr{A}} : \mathbb{F}_q[z]^{<k} \to \mathbb{F}_q^n, \quad \mathrm{eval}_{\mathscr{A}}(p) = (p(\alpha_1),\dots,p(\alpha_n)) \in \mathbb{F}_q^n. \tag{2.6}$$

We can interpret the $\mathrm{GRS}_k^q(\mathscr{A},\mathbf{v})$ code through its encoding map. To encode a message $\mathbf{m} = (m_0,\dots,m_{k-1})$, we can consider it as a polynomial $p(z) = m_0 + m_1 z + \cdots + m_{k-1}z^{k-1} \in \mathbb{F}_q[z]^{<k}$. Next, we evaluate the polynomial $p$ at the points $\alpha_1,\dots,\alpha_n$ and multiply each evaluation $p(\alpha_i)$ by $v_i$ to obtain the codeword corresponding to $\mathbf{m}$. We can easily evaluate the polynomial $p$ at these points by left-multiplying the message vector $\mathbf{m}$ by a Vandermonde matrix, which is defined as follows.

**Definition 10.** *Given a set* $\mathscr{A} = \{\alpha_1,\dots,\alpha_n\} \subset \mathbb{F}_q$ *of distinct elements of* $\mathbb{F}_q$*, a* Vandermonde matrix $\mathbf{V}(\mathscr{A}) \in \mathbb{F}_q^{k\times n}$ *is a matrix with entries* $V_{ij} = \alpha_i^{j-1}$*, which can be explicitly written as*

$$\mathbf{V}(\mathscr{A}) := \begin{pmatrix} 1 & 1 & \cdots & 1 \\ \alpha_1 & \alpha_2 & \cdots & \alpha_n \\ \alpha_1^2 & \alpha_2^2 & \cdots & \alpha_n^2 \\ \vdots & \vdots & \ddots & \vdots \\ \alpha_1^{k-1} & \alpha_2^{k-1} & \cdots & \alpha_n^{k-1} \end{pmatrix}. \tag{2.7}$$

Denoting by $\mathrm{diag}(\mathbf{v})$ the diagonal matrix with the values $v_i$ on the diagonal, $\mathbf{V}(\mathscr{A}) \cdot \mathrm{diag}(\mathbf{v})$ is clearly a generator matrix for a $\mathrm{GRS}_k^q(\mathscr{A},\mathbf{v})$ code. Let the generator matrix for $\mathscr{C} = \mathrm{GRS}_k^q(\mathscr{A},\mathbf{v})$ be $\mathbf{G}_{\mathscr{C}} = \mathbf{V}(\mathscr{A}) \cdot \mathrm{diag}(\mathbf{v})$, then

$$\mathrm{eval}_{\mathscr{A}}(p) \star \mathbf{v} = \mathbf{m}\mathbf{G}_{\mathscr{C}}. \tag{2.8}$$

GRS codes have a useful property where the star product of two GRS codes defined on the same set $\mathscr{A}$ is again a GRS code defined on $\mathscr{A}$. Suppose we have two GRS codes $\mathscr{C} = \mathrm{GRS}_k^q(\mathscr{A},\mathbf{v})$ and $\mathscr{C}' = \mathrm{GRS}_{k'}^q(\mathscr{A},\mathbf{w})$, then their star product is given by a GRS code $\mathscr{C} \star \mathscr{C}'$ defined on $\mathscr{A}$ as follows:

$$\mathscr{C} \star \mathscr{C}' = \mathrm{span}\left\{\mathbf{v} \star \mathbf{w} \star \left(\mathrm{eval}_{\mathscr{A}}(pp') : p \in \mathbb{F}_q[z]^{<k}, p' \in \mathbb{F}_q[z]^{<k'}\right)\right\}. \tag{2.9}$$

The resulting code is a $\mathrm{GRS}^q_{\min\{k+k'-1,n\}}(\mathscr{A},\mathbf{v}\star\mathbf{w})$ code. This is due to the fact that the degree of the polynomial $p'' = pp'$ is at most $\deg(p'') = \deg(p) + \deg(p') = k + k' - 2$ [12].

**Remark.** *A GRS code with vector* $\mathbf{v} = \mathbf{1}^n$ *is simply called a* Reed–Solomon *code. In this case, since* $\mathscr{C} = \mathrm{GRS}_k^q(\mathscr{A},\mathbf{1}^n)$*, the generator matrix is simply the Vandermonde matrix,* i.e., $\mathbf{G}_{\mathscr{C}} = \mathbf{V}$.

GRS codes also have an explicit encoding matrix that is systematic, defined as

$$\widetilde{\mathbf{G}}_{\mathscr{C}} := \begin{pmatrix} g_1(\alpha_1) & \cdots & g_1(\alpha_n) \\ \vdots & \ddots & \vdots \\ g_k(\alpha_1) & \cdots & g_k(\alpha_n) \end{pmatrix}, \tag{2.10}$$

where

$$g_i(z) := v_i^{-1} \prod_{j \in [k]\setminus\{i\}} \frac{z - \alpha_j}{\alpha_i - \alpha_j}$$

for $i \in [k]$. To see that it is in the systematic form, notice that $g_i(\alpha_i) = v_i^{-1}$ and $g_i(\alpha_j) = 0$ for $j \in [k]\setminus\{i\}$.

**Corollary 2.** *The dual of a* $\mathrm{GRS}_k^q(\mathscr{A}, \mathbf{v})$ *code is given by a* $\mathrm{GRS}_{n-k}^q(\mathscr{A}, \mathbf{u})$ *code [12], where*

$$u_i = \left( v_i \prod_{j \neq i} (\alpha_i - \alpha_j) \right)^{-1}. \tag{2.11}$$

**Remark.** *The dual of a Reed–Solomon code is not a Reed–Solomon code. This is why we introduced the family of GRS codes.*

### 2.3.2 Cross-Subspace Alignment codes

*Cross-Subspace Alignment* (CSA) codes are a family of codes that have been introduced in the context of PIR [24] and further used to improve some Secure Distributed Matrix Multiplication (SDMM) results [6, 22, 23]. These codes are characterized by a Cauchy–Vandermonde matrix structure that facilitates interference alignment along Vandermonde terms, while the desired computations remain resolvable along the Cauchy terms.

**Definition 11.** *Given two sets* $\mathscr{F} = \{f_1, \ldots, f_l\} \subseteq \mathbb{F}_q^l$ *and* $\mathscr{A} = \{\alpha_1, \ldots, \alpha_n\} \subseteq \mathbb{F}_q^n$ *of distinct elements and such that* $\mathscr{F} \cap \mathscr{A} = \emptyset$, *a* Cauchy–Vandermonde matrix *has the following structure:*

$$\mathbf{V}(\mathscr{A}, \mathscr{F}) := \begin{pmatrix} \frac{1}{f_1 - \alpha_1} & \frac{1}{f_1 - \alpha_2} & \cdots & \frac{1}{f_1 - \alpha_1} \\ \vdots & \vdots & \ddots & \vdots \\ \frac{1}{f_l - \alpha_1} & \frac{1}{f_l - \alpha_2} & \cdots & \frac{1}{f_l - \alpha_1} \\ 1 & 1 & \cdots & 1 \\ \alpha_1 & \alpha_2 & \cdots & \alpha_n \\ \alpha_1^2 & \alpha_2^2 & \cdots & \alpha_n^2 \\ \vdots & \vdots & \ddots & \vdots \\ \alpha_1^{k-1} & \alpha_2^{k-1} & \cdots & \alpha_n^{k-1} \end{pmatrix} \in \mathbb{F}_q^{(k+l) \times n}. \tag{2.12}$$

*A CSA code* $\mathrm{CSA}_k^q(\mathscr{A}, \mathscr{F})$ *is the set of all the codewords generated by a Cauchy–Vandermonde matrix,* i.e.,

$$\mathrm{CSA}_k^q(\mathscr{A}, \mathscr{F}) := \left\{ \mathbf{m}\mathbf{V}(\mathscr{A}, \mathscr{F}) : \mathbf{m} \in \mathbb{F}_q^{k+l} \right\} \subseteq \mathbb{F}_q^n. \tag{2.13}$$

Clearly, CSA codes generalize GRS codes as a Cauchy–Vandermonde matrix generalizes the Vandermonde matrix that is used as their generator matrix. If we choose $k+l=n$, the Cauchy–Vandermonde matrix is invertible [22], which implies that any square submatrix is invertible. This implies that CSA codes are $[n,k+l]_q$ MDS codes.

### 2.3.3 Binary Reed–Muller codes

*Binary Reed–Muller* (BRM) codes are a well-known family of non-trivial binary codes [29]. Being non-trivial binary codes, BRM codes are not MDS by the result shown in Section 2.2.4. A BRM code is specified by two parameters denoted by serif letters $\mathsf{r},\mathsf{m}$ as follows.

**Definition 12.** *Let $\mathsf{r},\mathsf{m}$ be two integers such that $0 \le \mathsf{r} \le \mathsf{m}$ and let $\mathscr{P} = \{P_1,\dots,P_{2^{\mathsf{m}}}\}$ be the set of all the points of $\mathbb{F}_2^{\mathsf{m}}$. The $\mathsf{r}$-th order Reed–Muller code of length $2^{\mathsf{m}}$ is defined as*

$$\mathrm{BRM}(\mathsf{r},\mathsf{m}) := \left\{\mathrm{eval}_{\mathscr{P}}(p) : p \in \mathbb{F}_2[z_1,\dots,z_{\mathsf{m}}]^{\le \mathsf{r}}\right\}. \tag{2.14}$$

A $\mathrm{BRM}(\mathsf{r},\mathsf{m})$ code has the following properties [29]:

- it is a linear code of dimension $k = \sum_{i=0}^{\mathsf{r}} \binom{\mathsf{m}}{i}$;
- it has length $n = 2^{\mathsf{m}}$ and minimum distance $d = 2^{\mathsf{m}-\mathsf{r}}$;
- for $\mathsf{r}' \le \mathsf{r}$, $\mathrm{BRM}(\mathsf{r}',\mathsf{m}) \subseteq \mathrm{BRM}(\mathsf{r},\mathsf{m})$;
- it has a dual code $\mathrm{BRM}(\mathsf{m}-\mathsf{r}-1,\mathsf{m})$.

In a similar way as for GRS codes, a $\mathrm{BRM}(\mathsf{r},\mathsf{m})$ code can be interpreted via its encoding map. Let $\mathbf{m} = (m_0,\dots,m_{k-1})$ be a message. Then we can see it as a polynomial $p(z_1,\dots,z_{\mathsf{m}}) = m_0 + m_1 z_1 + \cdots + m_{\mathsf{m}} z_{\mathsf{m}} + m_{\mathsf{m}+1} z_1 z_2 + \cdots + m_{k-1} z_{\mathsf{m}-r+1} \cdots z_{\mathsf{m}} \in \mathbb{F}_2[z_1,\dots,z_{\mathsf{m}}]^{\le \mathsf{r}}$. We can then evaluate $p$ at the points $P_1,\dots,P_n$ to obtain the $n$ symbols of the encoded message. It follows that a generator matrix for $\mathscr{C} = \mathrm{BRM}(\mathsf{r},\mathsf{m})$ is given by

$$\mathbf{G}_{\mathscr{C}} = \begin{pmatrix} \mathrm{eval}_{\mathscr{P}}(1) \\ \mathrm{eval}_{\mathscr{P}}(z_1) \\ \vdots \\ \mathrm{eval}_{\mathscr{P}}(z_{\mathsf{m}}) \\ \mathrm{eval}_{\mathscr{P}}(z_1 z_2) \\ \vdots \\ \mathrm{eval}_{\mathscr{P}}(z_{\mathsf{m}-\mathsf{r}+1} \cdots z_{\mathsf{m}}) \end{pmatrix}. \tag{2.15}$$

An interesting property of BRM codes is that the star product of two BRM codes is still a BRM code, as the following theorem states.

**Theorem 6** (Lemma 7, [11]). *Given two BRM codes* $\mathrm{BRM}(\mathsf{r},\mathsf{m})$ *and* $\mathrm{BRM}(\mathsf{r}',\mathsf{m})$, *we have that* $\mathrm{BRM}(\mathsf{r},\mathsf{m}) \star \mathrm{BRM}(\mathsf{r}',\mathsf{m}) = \mathrm{BRM}(\mathsf{r}+\mathsf{r}',\mathsf{m})$. *Such a code has minimum distance* $d'' = 2^{\mathsf{m}-\mathsf{r}-\mathsf{r}'}$ *and dimension* $k'' = \sum_{i=0}^{\mathsf{r}+\mathsf{r}'} \binom{\mathsf{m}}{i}$.

### 2.3.4 Locally repairable codes

*Locally repairable codes* (LRCs) have been introduced to allow erased or lost servers to be recovered locally by only downloading data from a small subset of other servers that are in the same *repair set* [17, Chapters 14–16]. This does not mean that a message encoded into servers via an LRC can be recovered only by downloading data from servers in the same repair set: given a non-trivial LRC, any of its information sets consists of positions from multiple repair sets. Local recovery of data results in a reduction of the minimum distance of the code and, as a consequence, in the loss of the MDS property. This trade-off between locality and maximum number of server failures establishes a limit on the number of failures that can occur while ensuring that no data loss happens. For a set of integers $\mathscr{A} \subset [n]$ we denote the restriction of an $[n,k]_q$ code $\mathscr{C}$ to the coordinates indexed by $\mathscr{A}$ by $\mathscr{C}|_{\mathscr{A}}$.

**Definition 13.** *An* $[n,k]$*-LRC* $\mathscr{C} = \mathrm{LRC}_{n,k}^{q}(r,\rho)$ *is said to have* $(r,\rho)$*-locality if there exists a partition* $\mathscr{P} = \{\mathscr{P}_1,\ldots,\mathscr{P}_l\}$ *of* $[n]$ *into sets* $\mathscr{P}_p$ *such that* $|\mathscr{P}_p| \leq r+\rho-1$ *and* $d_p \geq \rho$ *for every* $p \in [l]$, *where* $d_p$ *is the distance of the code* $\mathscr{C}|_{\mathscr{P}_p}$. *The sets* $\mathscr{P}_1,\ldots,\mathscr{P}_l$ *are called* repair sets *and the code* $\mathscr{C}|_{\mathscr{P}_p}$ *is the* $p$*-th* local code. *A code* $\mathrm{LRC}_{n,k}^{q}(r,\rho)$ *is said to be* optimal *if its minimum distance* $d$ *attains with equality the Singleton-like bound [25]*

$$d \leq n-k+1-\left(\left\lceil \frac{k}{r} \right\rceil -1\right)(\rho-1). \tag{2.16}$$

Many optimal code families exist and, as one might expect, they can be built from local codes with the MDS property. More precisely, for an optimal LRC the local codes are $[r+\rho-1,r]_q$ MDS codes. Assuming for simplicity that $k \mid r$, the $r$ positions from each local code form a local information set, and the union of these information sets from $\frac{k}{r}$ local codes form an information set for the LRC.

# 3. Quantum Computation

The field of quantum computation is advancing rapidly and shows great potential, as it combines the principles of quantum mechanics with information theory to establish a new computing paradigm. Quantum mechanics, which was developed to explain the behavior of particles at the atomic and subatomic scale, has already led to many technological advancements such as transistors, lasers, and magnetic resonance imaging [13]. However, merging quantum mechanics and information theory did not gain much traction until the 1970s.

In 1994, Peter Shor derived a quantum algorithm that can efficiently factorize large composite numbers into their prime factors [31], something that classical algorithms might never be able to achieve within a reasonable amount of time. In recent years the security of many online transactions has been relying on the Rivest, Shamir, Adleman (RSA) cryptosystem which is based on the difficulty of the factoring problem by classical algorithms. This breakthrough sparked a surge of interest in quantum computing and opened up the possibility of finding solutions to problems that are currently exceedingly difficult for classical computers to handle.

The advancement of quantum computation has been driven significantly by the goal of developing a sophisticated quantum computer capable of executing Shor's algorithm for large numbers. It is important to highlight that quantum computers are expected to achieve significant speed-ups only for specific problems. Researchers are actively investigating which problems are suitable for quantum speed-ups and are currently racing to achieve *quantum supremacy*, which aims to prove that a programmable quantum computer can solve a problem that is beyond the reach of any classical computer, regardless of whether the problem has practical applications or usefulness. Nevertheless, it is believed that quantum computers will help with a wide variety of problems.

In this chapter, we give an overview of quantum systems and focus on quantum computation, describing some of the fundamental quantum protocols. To help us understand quantum coding theory from a classical perspective, we also introduce the stabilizer formalism. We mainly follow the results shown in [30,46].

## 3.1 Quantum systems

The concept of a *bit* forms the foundation of classical computation and information theory. It is essentially one of the two symbols in the binary system and serves as the smallest unit of information required to differentiate between two equiprobable events.

Its counterpart in quantum computation and quantum information theory is called *qubit*, which is short for a quantum bit. Qubits are realized as physical systems, but we can treat them as mathematical objects just like we do for bits. It is well-known that we can represent a bit in a state, either 0 or 1.

In quantum information theory, the standard notation used to denote states is called the Dirac notation (or *bra-ket* notation), in which a *ket* is the column vector denoted by $|\cdot\rangle$ and a *bra* is the row vector denoted by $\langle\cdot|$. For example, the states in a qubit corresponding to the two classical states can be represented by the following vectors:

$$|0\rangle := \begin{pmatrix} 1 \\ 0 \end{pmatrix}, \quad |1\rangle := \begin{pmatrix} 0 \\ 1 \end{pmatrix}. \tag{3.1}$$

On the other hand, qubits can also be in a linear combination of those two states, *i.e.*,

$$|\psi\rangle = \alpha\,|0\rangle + \beta\,|1\rangle\,, \tag{3.2}$$

where the *amplitudes* $\alpha, \beta \in \mathbb{C}$ are such that $|\alpha|^2 + |\beta|^2 = 1$. This state is called *superposition*. This implies that the state of a qubit is represented by a vector in a two-dimensional complex vector space, for which the two classical states $|0\rangle$ and $|1\rangle$ form an orthonormal basis called the *computational basis*.

The reason why we use vectors to describe quantum states is given by the first of the four postulates of quantum mechanics, established through a lengthy and mostly experimental process that involved a significant amount of trial and error by the founders of the theory.

**Postulate 1.** *Associated to any isolated physical system is a complex vector space with an inner product (*i.e.*, a Hilbert space) known as the* state space *of the system. The system is completely described by its* state vector*, which is a unit vector in the system's state space.*

This first postulate is now formulated on unit vectors for simplicity, assuming that *we know exactly what the state is on a qubit*. Often, our knowledge of the state is probabilistic, meaning that we do not know the exact state of the qubit. In those cases, we say that the state is *mixed* and we use *density operators* to describe them. We will treat this more in detail in Section 3.1.2.

The dimension of the Hilbert space indicates the number of distinguishable states of the quantum system. For example, a qubit is a quantum system with dimension two with $|0\rangle$ and $|1\rangle$ representing its distinguishable states. We call *qudit* a quantum system with dimension $d$ that has $d$ distinguishable states $|0\rangle, \ldots, |d-1\rangle$. In this thesis, we consider only qudits with $q$ distinguishable states corresponding to the $q$ elements of a finite field $\mathbb{F}_q$, so the Hilbert spaces considered have always finite dimension. Thus, the general superposition can be written as

$$|\psi\rangle = \sum_{i \in \mathbb{F}_q} \alpha_i |i\rangle, \tag{3.3}$$

where the amplitudes $\alpha_i \in \mathbb{C}$ satisfy the unitary condition $\sum_{i\in\mathbb{F}_q} |\alpha_i|^2 = 1$. For the sake of clarity, most of the examples shown in this chapter are in a two-dimensional Hilbert space with distinguishable states corresponding to the bits 0 and 1.

### 3.1.1 Density matrices

The state of a system can be described by a hermitian matrix called the density matrix. Here we follow the standard notation with Greek letters such as $\rho, \sigma$ for this kind of matrices.

**Definition 14.** *A* density matrix $\rho$ *is a positive semi-definite hermitian matrix with trace equal to one,* i.e.*, it is such that* $\rho \geq 0$ *and* $\mathrm{Tr}(\rho) = 1$*. Furthermore, given a finite-dimensional Hilbert space* $\mathcal{H}$*, we denote the set of density matrices acting on* $\mathcal{H}$ *by* $\mathcal{S}(\mathcal{H})$*,* i.e.*,*

$$\mathcal{S}(\mathcal{H}) := \left\{\rho \in \mathbb{C}^{d\times d} : \rho = \rho^\dagger,\ \rho \geq 0,\ \mathrm{Tr}(\rho) = 1,\ \dim(\mathcal{H}) = d\right\}. \tag{3.4}$$

This means that the eigenvalues of a density matrix must be non-negative and add up to one. For example, the density matrices associated with the two classical states when $\dim(\mathcal{H}) = 2$ are

$$\rho_0 := |0\rangle\langle 0| = \begin{pmatrix} 1 & 0 \\ 0 & 0 \end{pmatrix}, \quad \rho_1 := |1\rangle\langle 1| = \begin{pmatrix} 0 & 0 \\ 0 & 1 \end{pmatrix},$$

It is easy to see that $\rho_0\rho_1 = \mathbf{0}$, which means that the two states are orthogonal to each other and, in terms of physics, that they are perfectly distinguishable.

Another example is given by any probabilistic mixture of two quantum states: for $\sigma_0, \sigma_1 \in \mathcal{S}(\mathcal{H})$ and $p \in [0,1]$ we have that $p\sigma_0 + (1-p)\sigma_1 \in \mathcal{S}(\mathcal{H})$. This implies that the set of density matrices is convex, but it is worth noticing that it is not linear, as $\frac{\mathbf{I}}{2} \in \mathcal{S}(\mathcal{H})$ but $\mathbf{I} \notin \mathcal{S}(\mathcal{H})$.

More generally, we can say that a density matrix subsumes a probability distribution, which describes the state of a classical system. All probability distributions can be embedded into a quantum state by placing the entries along the diagonal of the density matrix as in

$$p|0\rangle\langle 0| + (1-p)|1\rangle\langle 1| = \begin{pmatrix} p & 0 \\ 0 & 1-p \end{pmatrix}.$$

### 3.1.2 Pure and mixed states

A density matrix $\rho \in \mathscr{S}(\mathscr{H})$ can be obtained from a unit vector $|\psi\rangle \in \mathscr{H}$ with $\rho = |\psi\rangle\langle\psi|$. The converse is not necessarily true: given a density matrix, it is not always possible to decompose it as the product of a ket and a bra. The necessary condition to decompose a density matrix into the product of two unit vectors is that it has rank one or, equivalently, that its eigenvalues are either 0 or 1. For example, any non-trivial probability distribution embedded in a quantum system has some diagonal elements (and so, eigenvalues) that differ from 0 or 1.

**Definition 15.** *We call a quantum state* pure *if its density matrix can be decomposed as* $\rho = |\psi\rangle\langle\psi|$, *otherwise we call it* mixed.

From the discussion above, it is clear that a quantum state is pure if and only if it is represented by a rank-one density matrix. Rank-one density matrices are *projections*, which are matrices with the property $\mathbf{M}^2 = \mathbf{M}$. It follows that a quantum state is pure if and only if $\mathrm{Tr}(\rho^2) = 1$.

**Example 2.** *Consider the superposition state* $|\psi\rangle$ *defined in Equation* (3.2) *and define* $\langle\psi| = |\psi\rangle^\dagger$ *to obtain a complex inner scalar product of* $\langle\psi|\psi\rangle = |\alpha|^2 + |\beta|^2 = 1$. *The matrix*

$$|\psi\rangle\langle\psi| = \begin{pmatrix} |\alpha|^2 & \alpha\overline{\beta} \\ \beta\overline{\alpha} & |\beta|^2 \end{pmatrix}$$

*represents a pure state, which differs from the other quantum states we have considered thus far since it contains off-diagonal elements referred to as* quantum coherences. *These elements provide information about the environment,* i.e., *the quantum system. On the other hand, the state*

$$\sigma = \begin{pmatrix} |\alpha|^2 & 0 \\ 0 & |\beta|^2 \end{pmatrix}$$

*is a mixed state that is physically distinct from the previous one. We can write* $\sigma = |\alpha|^2|0\rangle\langle 0| + |\beta|^2|1\rangle\langle 1|$, *which indicates that* $\sigma$ *has a probability of* $|\alpha|^2$ *of being in the pure state* $|0\rangle\langle 0|$ *and a probability of* $|\beta|^2$ *of being in the pure state* $|1\rangle\langle 1|$. *We discuss this difference more carefully in Section 3.2.2.*

Thus, density matrices provide a generalization of state vectors. As stated in Section 3.1, we need this generalization to take into account the situations in which we do not know exactly the state on a qubit. For example, during a computation on a quantum computer, we manipulate the state initially prepared in the state $|0\rangle$, but the manipulation mechanism has finite precision, thus ending up in a probability distribution of final states. Another example is when the isolation of the qubits and their environment is not perfect: the qubits become entangled with the environment (cf. Section 3.2.1) and since we are not observing the environment, our description of the qubits is imprecise.

We thus reformulate Postulate 1 by saying that the system is completely described by a density matrix acting on the state space of the system. If two quantum states $|\psi\rangle$ and $|\psi'\rangle$ are related by a global phase $\omega$ such that $|\omega| = 1$, *i.e.*, $|\psi'\rangle = \omega|\psi\rangle$, then we can see that their associated density matrices are the same, *i.e.*, $\rho' = \rho$. Thus, if two pure quantum states differ by a global phase, then physically they represent the same object.

## 3.2 Composite quantum systems

In quantum information theory, the study of multiple quantum systems' interactions is a significant aspect that enhances the value of quantum information. In classical information theory we represent the state of multiple bits using the Cartesian product, *e.g.*, we use $\mathbb{F}_2 \times \mathbb{F}_2$ to represent pairs of bits. This is insufficient to capture quantum states: the following postulate describes how the state space of a composite system is built up from the state spaces of the component systems.

**Postulate 2.** *The state space of a composite physical system is the tensor product of the state spaces of the component physical systems. Moreover, if the joint system consists of $n$ parts each prepared to state $\rho_i$, then their joint system is in the state $\rho = \rho_1 \otimes \cdots \otimes \rho_n$.*

Clearly, if the states are pure, then we can reformulate the postulate with the joint system being in the state $|\psi\rangle = |\psi_1\rangle \otimes \cdots \otimes |\psi_n\rangle$. For simplicity, we denote $|i_1,\ldots,i_n\rangle := |i_1\rangle \otimes \cdots \otimes |i_n\rangle$. When the values $i_j$ are fixed, we usually avoid the usage of the comma. Sometimes it is useful to write system labels, which indicate which qubit belongs to a system $\mathscr{H}_i$:

$$|i_1,\ldots,i_n\rangle_{1,\ldots,n} := |i_1\rangle_1 \otimes \cdots \otimes |i_n\rangle_n .$$

For example, the superposition of two-qubit vectors is

$$|\psi\rangle_{1,2} = \alpha\,|00\rangle_{1,2} + \beta\,|01\rangle_{1,2} + \gamma\,|10\rangle_{1,2} + \delta\,|11\rangle_{1,2} ,$$

where $|\alpha|^2 + |\beta|^2 + |\gamma|^2 + |\delta|^2 = 1$.

### 3.2.1 Separable and entangled states

The idea behind entanglement is that, given a system of multiple quantum systems, we are not able to describe anymore each individual quantum system: we need to look at the whole system to get all the information about each part of it. This happens, for instance, when we bring two quantum systems close to each other and we let them interact with each other: in this scenario, they can get into a state where neither of these two quantum systems can be described by itself.

Suppose we are given a two-qubits system described by the state $|\psi\rangle$. If the state is

$$|\psi\rangle = a_0b_0|00\rangle + a_0b_1|01\rangle + a_1b_0|10\rangle + a_1b_1|11\rangle,$$

then it is easy to see that it can be decomposed into $|\psi\rangle = |\alpha\rangle \otimes |\beta\rangle$, where $|\alpha\rangle = a_0|0\rangle + a_1|1\rangle$ and $|\beta\rangle = b_0|0\rangle + b_1|1\rangle$. In general, given a state $|\psi\rangle = a_{00}|00\rangle + a_{01}|01\rangle + a_{10}|10\rangle + a_{11}|11\rangle$, it is not always possible to decompose it as $|\psi\rangle = |\alpha\rangle \otimes |\beta\rangle$.

**Definition 16.** *We call a state* separable *if it can be decomposed in the form* $|\psi\rangle = |\alpha\rangle \otimes |\beta\rangle$*. Otherwise, the state is said to be* entangled.

Let us consider a simple yet important example: the *Bell states*. We follow this example throughout this chapter to better understand the theory.

**Example 3.** *The Bell states are the following two-qubits states:*

$$\begin{aligned} |\beta_{00}\rangle &:= \frac{1}{\sqrt{2}}(|00\rangle + |11\rangle), \quad |\beta_{01}\rangle := \frac{1}{\sqrt{2}}(|00\rangle - |11\rangle), \\ |\beta_{10}\rangle &:= \frac{1}{\sqrt{2}}(|01\rangle + |10\rangle), \quad |\beta_{11}\rangle := \frac{1}{\sqrt{2}}(|01\rangle - |10\rangle), \end{aligned} \tag{3.5}$$

*which can be remembered via* $|\beta_{ij}\rangle = \frac{1}{2}|0,i\rangle + (-1)^j|1,\bar{i}\rangle$ *for* $i,j \in \mathbb{F}_2$ *and* $\bar{i} = i \oplus 1$. *It is easy to prove that each Bell state is an entangled state using the definition.*

### 3.2.2 Purification of quantum states

The concept of *purification* is a powerful tool in quantum information theory, which does not exist in classical information theory. This tool can be useful when we have a system in a mixed state: by introducing another system we can define a pure state for the joint system. Notice that the purification of a quantum system is a mathematical procedure that allows us to link pure states to mixed states. For this reason, the system introduced is called a *reference* system: it is a mathematical object that is physically inaccessible. A physical interpretation of purification states that the mixedness in a quantum state is due to entanglement with an inaccessible reference system.

In order to define purification mathematically, we need the partial trace.

**Definition 17.** *The* partial trace *of a density matrix* $\mathbf{X}$ *acting on* $\mathscr{H}_A \otimes \mathscr{H}_B$ *can be realized as*

$$\begin{aligned}\mathrm{Tr}_A : \mathscr{S}(\mathscr{H}_A \otimes \mathscr{H}_B) &\to \mathscr{S}(\mathscr{H}_B) \\ \mathbf{X} &\mapsto \mathrm{Tr}_A(\mathbf{X}) := \sum_i (\langle i|_A \otimes \mathbf{I}_B)\mathbf{X}(|i\rangle_A \otimes \mathbf{I}_B),\end{aligned} \tag{3.6}$$

*where* $\{|i\rangle_A\}$ *is an orthonormal basis for* $\mathscr{H}_A$ *and* $\mathbf{I}_B$ *is the identity matrix acting on* $\mathscr{H}_B$.

The partial trace is a useful tool to observe the state of a portion of a quantum system. For example, suppose that we have two quantum systems $A, B$ in the separable state $\rho_{AB} = \rho_A \otimes \rho_B$. We can compute the density matrix for system $A$ using the partial trace: $\rho_A = \mathrm{Tr}_B(\rho_{AB})$. When we apply the partial trace over system $B$, we say that we are *tracing out* system $B$, as the result describes the remaining system $A$. Then we can use $\rho_A$ to predict the outcome of any experiment performed on system $A$ alone. The partial trace is mostly useful for entangled states, as shown in the following example.

**Example 4.** *Let* $|\beta_{ij}\rangle_{AB}$ *be a Bell state. If we trace out system A (or, similarly, system B) we obtain* $\mathrm{Tr}_A\left(|\beta_{ij}\rangle\langle\beta_{ij}|_{AB}\right) = \frac{\mathbf{I}}{2}$, *which is called the* maximally mixed state. *This state is the embedding of the uniform distribution with two outcomes into a quantum system: this means that observing each system by itself tells us nothing about it, as it is equivalent to a fair-coin flip.*

The partial trace thus gives us information about a part of the system when we know the state of the whole system. On the other hand, if we know that a system $\mathscr{H}_A$ is in the state $\rho_A$, then there exists a pure state $|\psi\rangle\langle\psi|_{AR}$ on the system $\mathscr{H}_A \otimes \mathscr{H}_R$ such that $\rho_A = \mathrm{Tr}_R\left(|\psi\rangle\langle\psi|_{AR}\right)$. Such a state is called a purification of $\rho_A$ and $\mathscr{H}_R$ is the reference system. Notice that a purification is not unique, as we can see in Example 4: any of the four Bell states can be a purification to the maximally mixed state.

## 3.3 Quantum circuits

The language of *quantum computation* can be used to describe the changes that occur to a quantum state. Similar to a classical computer built from electrical circuits with wires and logic gates, a quantum computer is built from a *quantum circuit* comprising wires and elementary *quantum gates* to manipulate and transport quantum information. The classical NOT gate is the only non-trivial single-bit logic gate that sends a bit to its logical counterpart ($0 \mapsto 1,\ 1 \mapsto 0$). In quantum computation, there are several non-trivial single-qubit logic gates. For example, a "quantum NOT" gate transforms the state $\alpha|0\rangle + \beta|1\rangle$ into the corresponding state where $|0\rangle$ and $|1\rangle$ have been interchanged, resulting in $\alpha|1\rangle + \beta|0\rangle$.

Quantum gates operate on quantum states as inputs and outputs. Therefore, the matrices that describe them must satisfy the condition of unitarity to preserve the normalization constraint, which brings us to the next postulate of quantum mechanics.

**Postulate 3.** *The evolution of a* closed *quantum system is described by a* unitary transformation. *That is, the state $\rho_{t_1}$ of the system at time $t_1$ is related to the state $\rho_{t_2}$ of the system at time $t_2$ by a unitary operator* $\mathsf{U}$ *which depends only on the times $t_1$ and $t_2$,* i.e.,

$$\rho_{t_2} = \mathsf{U}\rho_{t_1}\mathsf{U}^{\dagger}. \tag{3.7}$$

### 3.3.1 Examples of quantum gates

The unitary transformations described in Postulate 3 are called *quantum gates* in the language of quantum computation. In quantum circuits, gates are usually represented as in Figure 3.1, where the black line denotes a quantum wire.

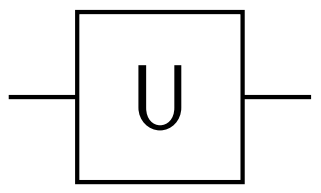

**Figure 3.1.** Gate in a quantum circuit.

Let $\omega_p = e^{\frac{2\pi\imath}{p}}$, $a, b \in \mathbb{F}_q$, where $q$ is a power of some prime $p$, and let $\oplus$ be the sum operation over $\mathbb{F}_q$. Throughout this thesis, we consider only the following gates over $q$-dimensional qudits:

- $\mathsf{I}_q := \sum_{i\in\mathbb{F}_q} |i\rangle\langle i| = \mathbf{I}$ is the *identity gate* that does not change the state of a qudit. It is useful, for example, when we want to describe the action of a single-qudit gate over a composite system, as shown in Example 5.

- $\mathsf{X}_q(a) := \sum_{i\in\mathbb{F}_q} |i \oplus a\rangle\langle i|$ generalizes the two-dimensional NOT gate for $a \in \mathbb{F}_q$ and maps the state $|i\rangle$ to $|i \oplus a\rangle$.

- $\mathsf{Z}_q(b) := \sum_{i\in\mathbb{F}_q} \omega_p^{\mathrm{tr}\, bi} |i\rangle\langle i|$ is called the *phase gate* for $b \in \mathbb{F}_q$ and maps the state $|i\rangle$ to $\omega_p^{\mathrm{tr}\, b}|i\rangle$, where tr is the trace function over finite fields defined in Section 2.1.

- $\mathsf{W}_q(a,b) = \mathsf{X}_q(a)\mathsf{Z}_q(b)$ is called the *Weyl gate*.

- $\mathsf{QFT}_q := \frac{1}{\sqrt{q}} \sum_{i,j\in\mathbb{F}_q} \omega_p^{\mathrm{tr}\, ij} |i\rangle\langle j|$ is the gate representation of the Quantum Fourier Transform and is the $q$-dimensional generalization of the Hadamard gate $\mathsf{H}$, as $\mathsf{H} = \mathsf{QFT}_2$. The Hadamard gate on a qubit maps bijectively $|0\rangle$ to $|+\rangle = \frac{1}{\sqrt{2}}(|0\rangle + |1\rangle)$ and $|1\rangle$ to $|-\rangle = \frac{1}{\sqrt{2}}(|0\rangle - |1\rangle)$.

- $\mathsf{CX}_q := \sum_{i,j\in\mathbb{F}_q} |i, i\oplus j\rangle\langle i,j|$ is the controlled-X gate and generalizes the two-dimensional CNOT gate over two qubits, which maps $|i,j\rangle$ to $|i,i\oplus j\rangle$ for $i,j\in\mathbb{F}_2$. The CNOT gate adds the value stored in the *control* qubit to the value stored in the *target* qubit.

In quantum circuits, the initial state is always the zero state $|\mathbf{0}\rangle$ and then it evolves accordingly to the gates applied to the qudits.

**Example 5.** *In order to generate the Bell state $|\beta_{00}\rangle$ over two qubits $\mathscr{H}_A, \mathscr{H}_B$ in the state $|00\rangle$ we need to apply first a Hadamard gate on $\mathscr{H}_A$ and then a* CNOT *gate with control $\mathscr{H}_A$ and target $\mathscr{H}_B$, as shown in Figure 3.2. It is also easy to see that the Bell states are related through the Weyl gate on $\mathscr{H}_A$ as follows:*

$$|\beta_{ij}\rangle = (\mathsf{W}(i,j)\otimes \mathsf{I})\,|\beta_{00}\rangle\,. \tag{3.8}$$

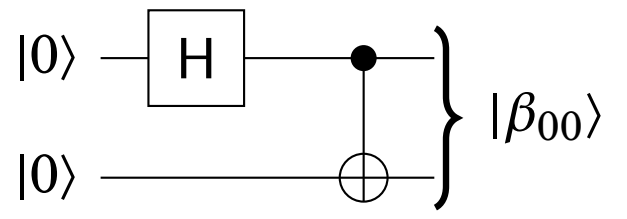

**Figure 3.2.** Quantum circuit to generate the Bell state $|\beta_{00}\rangle$.

### 3.3.2 Measurements

In a quantum circuit with zero error, we can mathematically predict the state of a system by applying all the gates applied to the system to the initial zero state. In reality, every gate introduces an error into the computation and we cannot know the state of the system until we *observe* it. This is due to the *uncertainty principle* of quantum mechanics.

A measurement is an operation that takes quantum systems as input and produces classical systems as output. Measurement in quantum mechanics is probabilistic in nature. When a measurement is made on a quantum system, the system can only be observed to be in one of a number of possible states, each with a certain probability. The probabilities are determined by the state of the system before the measurement and the measurement apparatus used to make the measurement. However, once the system is measured or observed, the state is determined and subsequent measurements would yield the same result (unless the state decays over time back to the zero state). This is why a measurement is said to *consume* a quantum system and *collapse* it to a definite classical state. Measurements are formalized by the last postulate of quantum mechanics:

**Postulate 4.** *Quantum measurements are described by a collection $\{\mathbf{M}_m\}$ of* measurement operators. *These are operators acting on the state space of the system being measured. The index $m$ refers to the measurement outcomes that may occur in the experiment. If the state of the quantum system is $\rho$* *immediately before the measurement, then the probability that result m occurs is given by*

$$\Pr(m) = \mathrm{Tr}\left(\mathbf{M}_m^\dagger \mathbf{M}_m \rho\right) \tag{3.9}$$

*and, if outcome m occurs (*i.e., $\Pr(m) \neq 0$*), the state of the system after the measurement is* $\frac{\mathbf{M}_m \rho \mathbf{M}_m^\dagger}{\Pr(m)}$. *This process is called* normalization *and it is needed in order to satisfy the unitarity of the resultant state. The measurement operators satisfy the* completeness equation*,* i.e.*,*

$$\sum_m \mathbf{M}_m^\dagger \mathbf{M}_m = \mathbf{I}. \tag{3.10}$$

Clearly, if the state $|\psi\rangle$ of the system is pure immediately before the measurement, the probability that result $m$ occurs can be easily computed as $\Pr(m) = \left\langle \psi \middle| \mathbf{M}_m^\dagger \mathbf{M}_m \middle| \psi \right\rangle$, and, if outcome $m$ occurs, the state of the system after the measurement is $\frac{\mathbf{M}_m |\psi\rangle}{\sqrt{\Pr(m)}}$.

The measurement operators can be obtained from an orthonormal basis to the Hilbert space $\mathscr{H}$ representing the state space. Suppose $\mathscr{H}$ has dimension $q$ and let $|i\rangle$ be the column vector with a 1 in position $i \in \mathbb{F}_q$ and zero otherwise. The set $\{|i\rangle : i \in \mathbb{F}_q\}$ is clearly an orthonormal basis and is called the *computational basis*, which was already mentioned in Section 3.1. This basis induces a collection of measurement operators $\mathscr{M}_0 = \left\{|i\rangle\langle i| : i \in \mathbb{F}_q\right\}$ that describes the measurement gate in quantum circuits, commonly represented as in Figure 3.3. Notice that on the right of the measurement box, the double line denotes a classical output.

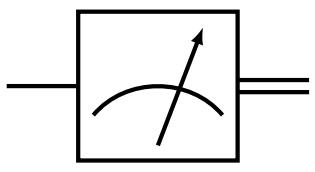

**Figure 3.3.** Measurement in a quantum circuit.

Let $|\psi\rangle = \sum_{i \in \mathbb{F}_q} \alpha_i |i\rangle$ be a generic superposition state as in Equation 3.3. Then the probability of measuring the outcome $i \in \mathbb{F}_q$ is given by the modulo squared of its respective amplitude, *i.e.*, $\Pr(i) = |\alpha_i|^2$. We now return to our example on the Bell state to show the surprising outcome of measuring an entangled state.

**Example 6.** *The Bell state* $|\beta_{00}\rangle = \frac{1}{\sqrt{2}}(|00\rangle + |11\rangle)$ *has probability* $\left(\frac{1}{\sqrt{2}}\right)^2 = \frac{1}{2}$ *of being in the state* $|00\rangle$ *and probability* $\left(\frac{1}{\sqrt{2}}\right)^2 = \frac{1}{2}$ *of being in the state* $|11\rangle$. *If we want to measure just the first qubit on the computational basis (which can be mathematically done using the partial trace defined in Section 3.2.2), the probability of observing 0 is* $\frac{1}{2}$*, and the new state would be* $|00\rangle$*, while the probability of observing 1 is* $\frac{1}{2}$*, and the new state would be* $|11\rangle$*. By symmetry, if we want to measure just the second qubit, we would have the same results. If we want to measure the second qubit after the measurement on the first one with outcome 0, we would observe 0 with probability 1, since the state is* $|00\rangle$ *after the first measurement. This is true no matter how far the two qubits of the system are until they are entangled. Thus, if two players share an entangled state and the first player measures their side of the system, then the second player's side becomes deterministic.*

### 3.3.3 Positive operator-valued measures

The computational basis is the most trivial orthonormal basis and is the standard choice in quantum circuits. However, one might notice that we can change an orthonormal basis into another just by multiplying each element by the same unitary $\mathsf{U}$. This means that measuring on a different basis is the same as first applying the unitary $\mathsf{U}^\dagger$ to the system to bring it in the computational basis and then applying the default measurement. Thus, when we say that we are measuring on a different basis, we really mean that first we apply the hermitian conjugate of the unitary that takes the computational basis to the new one, and then we measure on the computational basis.

**Example 7.** *The* Bell basis *is the set* $\{|\beta_{ij}\rangle : i,j \in \mathbb{F}_2\}$ *of the Bell states and it forms a basis for* $\mathscr{H}_A \otimes \mathscr{H}_B \simeq \mathbb{C}^4$. *As we saw in Example 5, the computational basis and the Bell basis are related via* $\mathsf{U} = \mathsf{CNOT}(\mathsf{H} \otimes \mathsf{I})$. *Thus, applying the* Bell measurement *is the same as first applying* $\mathsf{U}^\dagger = (\mathsf{H} \otimes \mathsf{I})\mathsf{CNOT}$ *and then measuring both qubits on the computational basis (cf. Figure 3.4).*

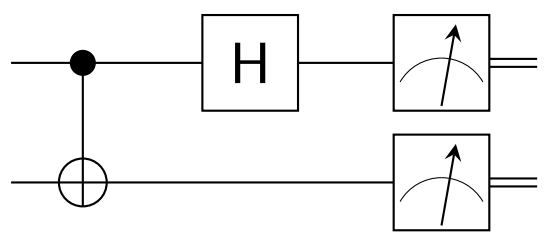

**Figure 3.4.** Quantum circuit for a Bell Measurement.

More generally, as stated in Postulate 4, we need a collection of measurement operators $\{\mathbf{M}_m\}$ satisfying Equation 3.10, *i.e.*, the completeness equation. In this thesis, we need only the following class of measurements.

**Definition 18.** *A* Positive-Operator Valued Measure *(POVM) is a set of Hermitian positive semi-definite operators* $\{\mathbf{E}_m\}$ *on a Hilbert space* $\mathscr{H}$ *that sum to the identity operator,* i.e.,

$$\sum_m \mathbf{E}_m = \mathbf{I}; \quad \mathbf{E}_m \geq 0, \quad \mathbf{E}_m^\dagger = \mathbf{E}_m \quad \forall m. \tag{3.11}$$

If we define $\mathbf{E}_m = \mathbf{M}_m^\dagger \mathbf{M}$, it is easy to see that each $\mathbf{E}_m$ is a positive semi-definite operator. Hence, a POVM is sufficient to determine the probabilities of the different measurement outcomes, and the operators $\mathbf{E}_m$ are known as the *POVM elements* associated with the measurement.

An important special class among POVMs is given by the projective-valued measures (PVM), which are defined as follows.

**Definition 19.** *A* Projective-Valued Measure *(PVM) is described by an* observable $\mathbf{O}$, i.e.*, a Hermitian operator on the state space of the system being observed. The observable has a spectral decomposition,*

$$\mathbf{O} = \sum_m m\mathbf{P}_m, \tag{3.12}$$

*where* $\mathbf{P}_m$ *is the projector onto the eigenspace of* $\mathbf{O}$ *with eigenvalue* $m$. *Thus, PVM can also refer to the set* $\{\mathbf{P}_m\}$.

Notice that any measurement induced by a basis over the Hilbert space $\mathscr{H}$ describing the quantum system is a PVM. For example, the computational basis measurement and the Bell measurement introduced in Example 7 are PVMs.

## 3.4 Stabilizer formalism

The *stabilizer formalism* [15] is a compact framework for quantum computation that establishes a helpful connection with classical computation. For this reason, it has recently been utilized to enhance various classical protocols. The basic idea of the stabilizer formalism is that many quantum states can be more easily described by working with the operators that stabilize them than by working explicitly with the state itself. This formalism involves a group of operators, known as *stabilizer operators*, that commute with each other and with the logical operators that encode the quantum information.

**Definition 20.** *Let* $\tilde{\mathbf{W}}(\mathbf{s}) := \mathsf{W}(s_1, s_{n+1}) \otimes \cdots \otimes \mathsf{W}(s_n, s_{2n})$ *be the Weyl operator acting on n qudits, where* $\mathbf{s} = (s_1, \ldots, s_{2n}) \in \mathbb{F}_q^{2n}$*. The* Heisenberg-Weyl group *is defined as* $\mathrm{HW}_q^n := \left\{c\tilde{\mathbf{W}}(\mathbf{s}) : \mathbf{s} \in \mathbb{F}_q^{2n},\ c \in \mathbb{C} \setminus \{0\}\right\}$*. A commutative subgroup of* $\mathrm{HW}_q^n$ *not containing* $c\mathbf{I}_{q^n}$ *for any* $c \neq 1$ *is called a* stabilizer group*. If such a group is generated by n independent elements, we say that the stabilizer is* maximal.

The reason why we say that a stabilizer with $n$ independent elements is maximal is due to commutativity, as a commutative group can have at most $n$ generators. Let us show yet another example involving Bell states.

**Example 8.** *Consider the Bell state* $|\beta_{00}\rangle = \frac{1}{\sqrt{2}}(|00\rangle + |11\rangle)$*: we say that this state is* stabilized *by the operators* $\mathsf{G}_1 = \mathsf{X} \otimes \mathsf{X}$ *and* $\mathsf{G}_2 = \mathsf{Z} \otimes \mathsf{Z}$ *because they satisfy the conditions* $\mathsf{G}_1 |\beta_{00}\rangle = |\beta_{00}\rangle$ *and* $\mathsf{G}_2 |\beta_{00}\rangle = |\beta_{00}\rangle$*. Furthermore, the state* $|\beta_{00}\rangle$ *is the unique quantum state (up to a global phase, which is irrelevant as we discussed in Section 3.1.2) which is stabilized by* $\mathsf{G}_1$ *and* $\mathsf{G}_2$*. The set generated by* $\mathscr{S} = \mathrm{span}\{\mathsf{G}_1, \mathsf{G}_2\}$ *is a multiplicative commutative group, so* $\mathscr{S}$ *is a stabilizer group. Furthermore, it is maximal, since the group is generated by two elements and the Bell state is defined over 2 qubits.*

### 3.4.1 Self-orthogonal subspaces

The classical object that corresponds to stabilizer groups is given by self-orthogonal subspaces of $\mathbb{F}_q^{2n}$, *i.e.*, subspaces that are self-orthogonal with respect to the symplectic inner product.

**Definition 21.** *For* $\mathbf{x}, \mathbf{y} \in \mathbb{F}_q^{2n}$*, the* symplectic inner product *is defined as*

$\langle \mathbf{x},\mathbf{y}\rangle_{\mathbb{S}} := \mathrm{tr}\left(\mathbf{x}\mathbf{J}^\top \mathbf{y}^\top\right)$, *where* $\mathbf{J}$ *is the* $2n \times 2n$ *matrix*

$$\mathbf{J} := \begin{pmatrix} \mathbf{0} & -\mathbf{I} \\ \mathbf{I} & \mathbf{0} \end{pmatrix}. \tag{3.13}$$

*The subspace* $\mathcal{V}^{\perp_{\mathbb{S}}} := \left\{\mathbf{s} \in \mathbb{F}_q^{2n} : \langle \mathbf{v},\mathbf{s}\rangle_{\mathbb{S}} = 0\ \forall \mathbf{v} \in \mathcal{V}\right\}$ *is the dual of a subspace* $\mathcal{V}$ *of* $\mathbb{F}_q^{2n}$ *with respect to this form. A subspace* $\mathcal{V} \subset \mathbb{F}_q^{2n}$ *is said to be* self-orthogonal *if* $\mathcal{V} \subseteq \mathcal{V}^{\perp_{\mathbb{S}}}$*, and* strongly self-orthogonal *if the equality holds,* i.e.*,* $\mathcal{V} = \mathcal{V}^{\perp_{\mathbb{S}}}$.

This means that for any self-orthogonal $\mathcal{V} \subset \mathbb{F}_q^{2n}$ we have that $\dim(\mathcal{V}) \le n$, which means that there are $k = \dim(\mathcal{V})$ independent vectors $\mathbf{s}_1,\dots,\mathbf{s}_k \in \mathbb{F}_q^{2n}$ such that $\mathcal{V} = \mathrm{span}(\mathbf{s}_1,\dots,\mathbf{s}_k)$.

**Definition 22.** *A matrix* $\mathbf{G} \in \mathbb{F}_q^{k\times 2n}$ *with independent vectors* $\mathbf{s}_1,\dots,\mathbf{s}_k \in \mathbb{F}_q^{2n}$ *as rows is said to be a* generator matrix *for* $\mathcal{V}$*. Since this matrix has the property that* $\mathbf{G}\mathbf{J}\mathbf{G}^\top = \mathbf{0}$*, we say that* $\mathbf{G}$ *is* self-orthogonal *if* $\mathrm{rank}(\mathbf{G}) = k < n$ *and* strongly self-orthogonal *if* $k = n$.

Considering the Heisenberg-Weyl group, there is a surjective homomorphism $c\tilde{\mathbf{W}}(\mathbf{s}) \in \mathrm{HW}_q^n \mapsto \mathbf{s} \in \mathbb{F}_q^{2n}$ with kernel $\left\{c\mathbf{I}_{q^n} : c \in \mathbb{C}\setminus\{0\}\right\}$. If we restrict the Heisenberg-Weyl group to the collection of stabilizer groups, the homomorphism is actually an isomorphism. Thus, a stabilizer group defines a self-orthogonal subspace and, conversely, given a self-orthogonal subspace $\mathcal{V}$, there exist complex numbers $c_{\mathbf{v}}$ so that

$$\mathcal{S}(\mathcal{V}) := \left\{\mathbf{W}(\mathbf{v}) := c_{\mathbf{v}}\tilde{\mathbf{W}}(\mathbf{v}) : \mathbf{v} \in \mathcal{V}\right\} \subseteq \mathrm{HW}_q^n \tag{3.14}$$

forms a stabilizer group. Furthermore, commutativity in $\mathrm{HW}_q^n$ is equivalent to symplectic orthogonality in $\mathbb{F}_q^{2n}$, as $\mathbf{W}(\mathbf{x})$ and $\mathbf{W}(\mathbf{y})$ commute if and only if $\langle \mathbf{x},\mathbf{y}\rangle_{\mathbb{S}} = 0$. It follows that a generator matrix $\mathbf{G}$ of $\mathcal{V}$ describes completely also the stabilizer group, which makes it a key concept that facilitates the understanding of quantum computation using classical computation techniques. Clearly, if $\mathbf{G}$ is strongly self-orthogonal, then the induced stabilizer is maximal.

Continuing Example 8, we show how to create such a generator matrix.

**Example 9.** *Let* $\mathbf{s}_1 = (1,1,0,0)$ *and* $\mathbf{s}_2 = (0,0,1,1)$*. Then we have that* $\mathsf{G}_1 = \mathbf{W}(\mathbf{s}_1)$ *and* $\mathsf{G}_2 = \mathbf{W}(\mathbf{s}_2)$*. This means that*

$$\mathbf{G} = \left(\begin{array}{cc|cc} 1 & 1 & 0 & 0 \\ 0 & 0 & 1 & 1 \end{array}\right) \in \mathbb{F}_2^{2\times 4}$$

*is a generator matrix for* $\mathcal{S}$ *and, since* $\mathrm{rank}(\mathbf{G}) = 2$*, it is strongly self-orthogonal. Notice that the first 2 columns correspond to the* $\mathsf{X}$ *gates of the generators* $\mathsf{G}_1,\mathsf{G}_2$*, while the last 2 columns correspond to their* $\mathsf{Z}$ *gates.*

### 3.4.2 Symplectic matrices

Symplectic matrices play an important role in the study of the *Clifford group* [16], which is a group of quantum operations that can be efficiently implemented on a quantum computer. Furthermore, symplectic matrices are the only ones that preserve the symplectic inner product and their columns form a symplectic basis for $\mathbb{F}_q^{2n}$.

**Definition 23.** *A matrix* $\mathbf{F} \in \mathbb{F}_q^{2n\times 2n}$ *is called* symplectic *if* $\mathbf{F}^\top \mathbf{J}\mathbf{F} = \mathbf{J}$.

If we write $\mathbf{F} = \left(\begin{smallmatrix}\mathbf{A} & \mathbf{C}\\ \mathbf{B} & \mathbf{D}\end{smallmatrix}\right)$, then $\mathbf{F}$ is symplectic if and only if $\mathbf{B}^\top\mathbf{A}$, $\mathbf{D}^\top\mathbf{C}$ are symmetric and $\mathbf{A}^\top\mathbf{D} - \mathbf{B}^\top\mathbf{C} = \mathbf{I}$. Thus,

$$\mathbf{F}^{-1} = \mathbf{J}^\top\mathbf{F}^\top\mathbf{J} = \begin{pmatrix}\mathbf{D}^\top & -\mathbf{C}^\top\\ -\mathbf{B}^\top & \mathbf{A}^\top\end{pmatrix}. \tag{3.15}$$

It is worth noticing that the transpose of the submatrix obtained by the first $k \le n$ columns of a symplectic matrix is a self-orthogonal matrix. Conversely, a self-orthogonal matrix can be completed to a symplectic matrix (*e.g.*, using the Gram–Schmidt Algorithm).

**Example 10.** *The matrix* $\mathbf{G}$ *of Example 9 is a self-orthogonal matrix obtained by the symplectic matrix*

$$\mathbf{F} = \begin{pmatrix}1 & 0 & 0 & 1\\ 1 & 0 & 0 & 0\\ 0 & 1 & 0 & 0\\ 0 & 1 & 1 & 0\end{pmatrix}.$$

## 3.5 Fundamental protocols

In this section, we discuss fundamental quantum protocols that have significant implications for quantum communication: *superdense coding* and *quantum teleportation*. These protocols have been widely studied and serve as building blocks for more complex quantum communication tasks, as we discuss in the next chapter. In the following, we consider players A, B and C as users of the protocols shown. An interesting fact is that they can be seen as dual protocols in the sense that they accomplish opposite tasks using the same resources: in superdense coding, A sends two classical bits of information using one qubit, while in quantum teleportation, A sends one qubit using two classical bits of information. For simplicity, both protocols are presented with 2-dimensional quantum systems, *i.e.*, qubits. We also present the *two-sum transmission* protocol, which is built upon superdense coding and is used to transmit two sums from two players, A and B, to a receiver.

### 3.5.1 Superdense coding

Superdense coding is a quantum communication protocol that allows two parties, A and B, to communicate two classical bits of information by sending only one qubit between them. The protocol makes use of quantum entanglement and a shared set of pre-established Bell states.

Let us assume here that A and B possess qubits $\mathscr{H}_A$ and $\mathscr{H}_B$, respectively, prepared in the Bell state $|\beta_{00}\rangle$. Furthermore, A wants to send bits $a_1, a_2 \in \mathbb{F}_2$ to B. The protocol, as depicted in Figure 3.5, works as follows:

1. A applies the Weyl operator $\mathsf{W}(a_1, a_2)$ on their qubit.

2. A sends their qubit $\mathscr{H}_A$ to B.

3. B performs a Bell measurement, as defined in Example 7, on the qubits $\mathscr{H}_A \otimes \mathscr{H}_B$ and obtains $a_1$ and $a_2$ as output.

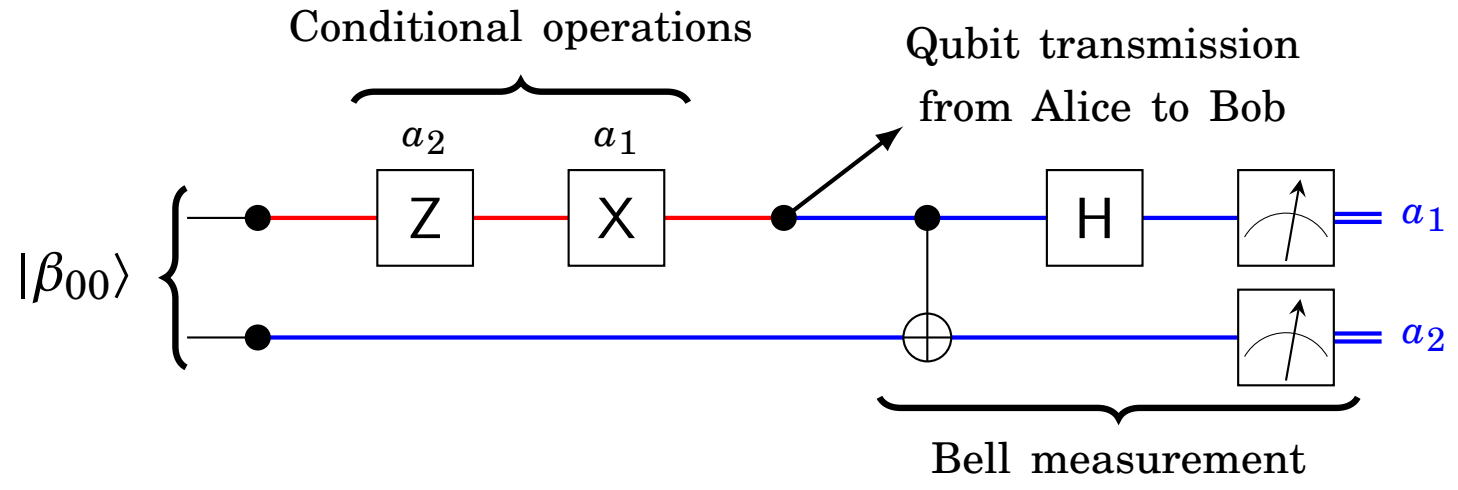

**Figure 3.5.** Quantum circuit for superdense coding.

The Holevo bound [18] in this case states that one cannot send more than 1 bit of information per qubit, so it might seem that superdense coding contradicts it. In reality, superdense coding does not contradict the Holevo bound because the protocol requires two qubits to be transmitted in order to send two classical bits of information. The sender has to send one qubit to the receiver first, and then the receiver can obtain two classical bits of information through the joint measurement on both qubits. Therefore, the amount of information transmitted is limited by the number of qubits that are possessed by the receiver at the moment of the measurement, not just by the number of qubits sent by the transmitter. In other words, the Holevo bound places a limit on the amount of classical information that can be transmitted per qubit, not per pair of entangled qubits used in a protocol like superdense coding.

### 3.5.2 Quantum teleportation

Quantum teleportation is a fundamental protocol in quantum communication that allows the transfer of an unknown quantum state from one location to another, without physically moving any quantum system, by communicating two classical bits of information. It is an essential building block for the realization of a quantum internet, where quantum information can be exchanged and processed between distant nodes with enhanced security and speed.

Let us assume here that A possesses two qubits $\mathscr{H}_{A_1} \otimes \mathscr{H}_{A_2}$, B possesses one qubit $\mathscr{H}_B$, and the qubits $\mathscr{H}_{A_2} \otimes \mathscr{H}_B$ are prepared in the Bell state $|\beta_{00}\rangle$. Furthermore, A wants to *teleport* the state $\rho$ on their qubit $\mathscr{H}_{A_1}$ to B's qubit by consuming the entanglement on their qubits and communicating two classical bits of information. Since $\rho$ might be a mixed state, we know from Section 3.2.2 that there exists a reference system $\mathscr{H}_R$ such that $\rho = \mathrm{Tr}_R\left(|\psi\rangle\langle\psi|\right)$ for some pure state $|\psi\rangle$ over $\mathscr{H}_{A_1} \otimes \mathscr{H}_R$. The protocol, as depicted in Figure 3.6, works as follows:

1. A performs a Bell measurement on the qubits $\mathscr{H}_{A_1} \otimes \mathscr{H}_{A_2}$, and obtains $a_1, a_2 \in \mathbb{F}_2$ as outcomes.

2. A sends the two bits $a_1$ and $a_2$ to B.

3. B applies the Weyl operator $\mathsf{W}(a_1, a_2)$ to their qubit $\mathscr{H}_B$. B's qubit is now in the state $\rho$.

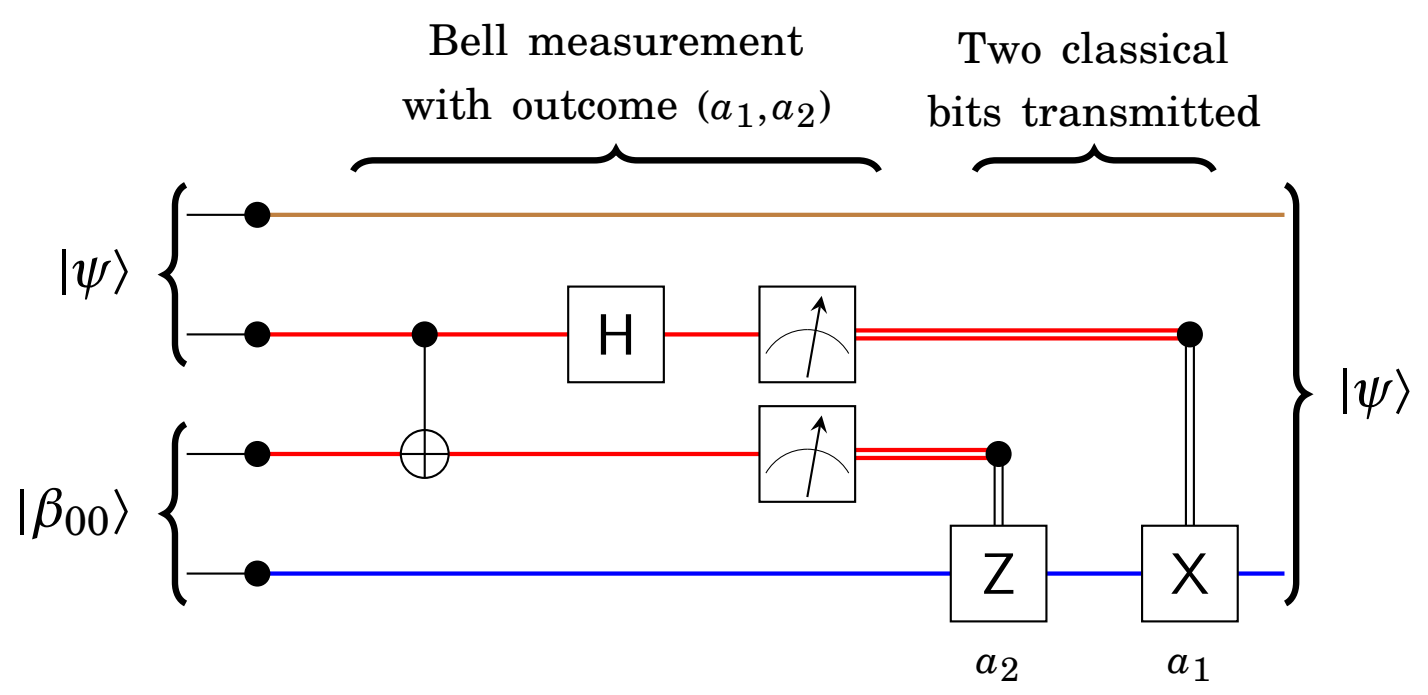

**Figure 3.6.** Quantum circuit for quantum teleportation.

One can also modify the protocol by allowing B to apply a Weyl operator $\mathsf{W}(b_1, b_2)$ for bits $b_1, b_2 \in \mathbb{F}_2$ either before or after applying $\mathsf{W}(a_1, a_2)$. In both cases, B's qubit ends up in the state $\mathsf{W}(b_1, b_2)\rho\mathsf{W}(b_1, b_2)^\dagger$. Such modification is referred to as quantum teleportation with an operation.

### 3.5.3 Two-sum transmission

The two-sum protocol allows sending the sum of two pairs of classical bits by sending two qubits. This protocol is a direct generalization of superdense coding, as one can see by comparing the quantum circuits for the protocols in Figure 3.5 and Figure 3.7.

Let us assume here that A and B possess qubits $\mathscr{H}_A$ and $\mathscr{H}_B$, respectively, prepared in the Bell state $|\beta_{00}\rangle$. Furthermore, A and B have bits $a_1, a_2 \in \mathbb{F}_2$, $b_1, b_2 \in \mathbb{F}_2$, respectively, and they want to send the two sums $a_1 \oplus b_1$ and $a_2 \oplus b_2$ to a receiver. The protocol, as depicted in Figure 3.7, works as follows:

1. A and B apply the Weyl operators $\mathsf{W}(a_1, a_2)$ and $\mathsf{W}(b_1, b_2)$ on their respective qubits.

2. A and B send their qubits to the receiver.

3. The receiver C performs a Bell measurement on the received qubits $\mathscr{H}_A \otimes \mathscr{H}_B$ and obtains $a_1 \oplus b_1$ and $a_2 \oplus b_2$ as output.

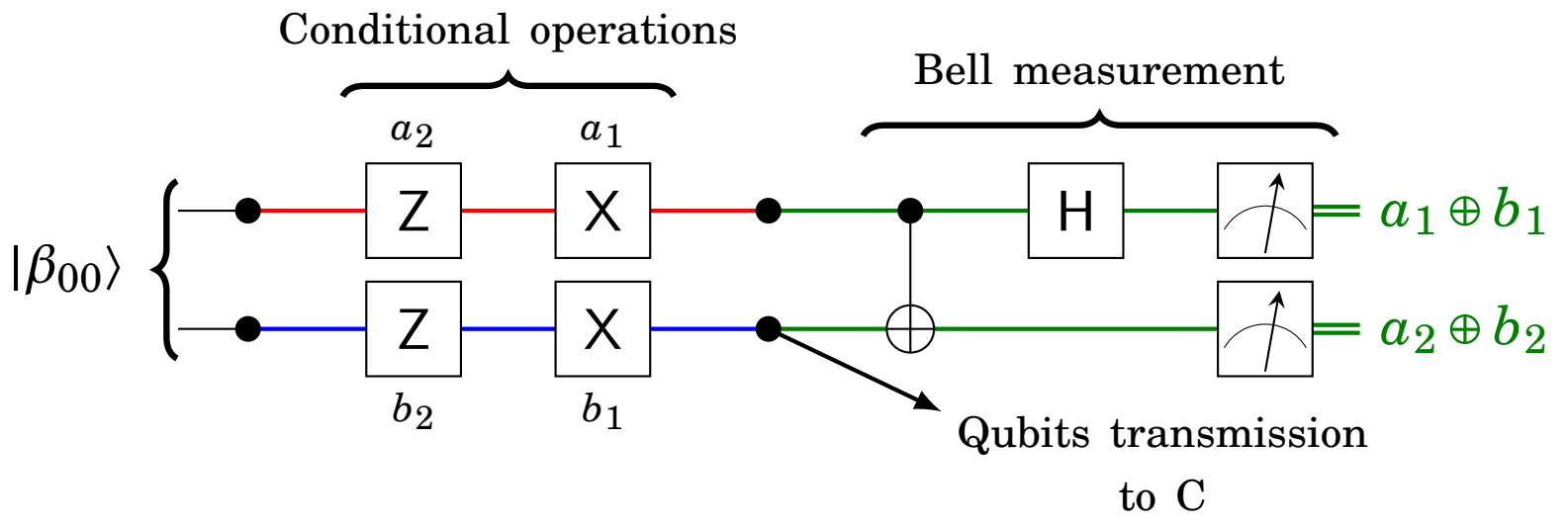

**Figure 3.7.** Quantum circuit for the two-sum transmission protocol.

In this protocol, it is worth noting that if A sets their data to all zeros ($x_1 = x_3 = 0$), the protocol allows B to transmit both of their classical input bits $(x_2, x_4)$ to C, even though only one qubit is sent, under the condition that A also sends their qubit to the receiver. This situation demonstrates the connection between this protocol and superdense coding. This protocol is of fundamental importance for the scope of this thesis, as it represents the base case of the core functionality of all the protocols presented in the next chapter.

# 4. New Private Information Retrieval schemes with Quantum Communication

*Private Information Retrieval* (PIR) is a research area in the field of secure information retrieval that addresses the challenge of enabling users to retrieve desired information from a database while preserving their privacy. PIR protocols aim to provide confidentiality by allowing users to retrieve specific data items without revealing their access patterns or the items of interest to the database server. By safeguarding the privacy of users, PIR protocols offer a means to perform efficient and secure information retrieval in scenarios where the confidentiality of queries and retrieved data is highly valued.

The concept of Private Information Retrieval (PIR) was originally introduced by Chor *et al.* [7,9]. In their classical PIR model [7], the database is represented as an $M$-bit binary vector, and the objective is to retrieve a single bit from the database without disclosing the index of the desired bit. In this work, we follow an extended model [3,12,20,22,37,44] to include a more realistic scenario where the database is structured as a matrix $\mathbf{X} = \left(\mathbf{X}^1, \dots, \mathbf{X}^M\right)$, composed of individual file vectors $\mathbf{X}^i$ containing $q$-ary digits from a finite field. Our goal is to enable the user to download a file $\mathbf{X}^\theta$ without revealing the corresponding index.

In this thesis, we focus on the problem of achieving PIR in the *information-theoretic* setting, where there are no constraints on the computational power of the servers. In this context, a fundamental result states that in the case of a single server, the only solution to achieve PIR is the *naïve* approach of downloading all the files in the system [7]. However, this approach incurs significant communication complexity.

On the other hand, considering multiple servers in the context of PIR allows for the possibility of designing more efficient and sophisticated solutions. In the classical PIR model with multiple servers, each server holds a replica or a subset of the database. By leveraging the collective information across the servers, it becomes possible to retrieve the desired data without having to download the entire database or revealing the specific file of interest. The key idea is to exploit the redundancy and overlapping information among the servers. Various coding schemes and information-theoretic techniques can be employed to distribute and retrieve the data across the servers in a way that minimizes the communication complexity while still preserving privacy. These schemes typically involve encoding the data and distributing encoded pieces among the servers in such a way that a user can retrieve the desired information by querying the servers without revealing which file is being requested.

*Quantum private information retrieval* (QPIR) builds upon classical PIR by leveraging the principles of quantum computation. For the purposes of this thesis, we focus solely on the setting where storage is classical, the user sends classical queries to the servers and the servers send a quantum response to the user. In earlier literature, Song *et al.* explored QPIR under the assumption of replicated storage [33–35]. The exploration of quantum techniques in PIR holds the potential for novel approaches and enhanced performance. By employing the unique properties of quantum systems, such as superposition and entanglement, QPIR offers advantages in terms of efficiency and security compared to classical PIR schemes.

In this chapter we first introduce the parameters and the definition of a PIR scheme that extends naturally to the definition of a QPIR scheme. After showing some known capacity results and the notation used throughout this chapter, we present the following protocols which constitute of the main contributions of this thesis. More details are given in the subsequent sections.

- an $[N,K]_{4^L}$ MDS-coded QPIR protocol that protects against $(N-K)$-collusion, assuming the servers possess only qubits [PI];

- an LRC-coded QPIR protocol that protects against $(\delta-1)$-collusion, where $\delta$ represents the locality of the Locally Repairable Code (LRC), assuming the servers possess only qubits [PI];

- a capacity-achieving GRS-coded QPIR protocol that protects against $T$-collusion, assuming the servers possess qudits [PII];

- an $X$-secure CSA-coded (*i.e.*, with a Cross Subspace Alignment storage code) QPIR protocol that protects against $T$-collusion, assuming the servers possess qudits [PIII];

- an example of a $(T,U,B)$-robust BRM-coded (*i.e.*, with a Binary Reed–Muller storage code) PIR protocol and an example of a BRM-coded QPIR protocol that protects against $T$-collusion [PIV].

## 4.1 PIR schemes

In this section, we formally define the PIR parameters and different classes of schemes that we consider in this work.

### 4.1.1 Parameters of a PIR scheme

We begin by introducing the concepts of user privacy and server privacy.

**Definition 24.** *A PIR scheme ensures* user privacy *if the servers are unable to determine the identity of the requested file from the user's queries. On the other hand, it guarantees* server privacy *when the user does not gain any information about files other than the one being retrieved.*
*A PIR scheme that guarantees server privacy is called* symmetric.

While user privacy is a fundamental requirement for any PIR scheme, server privacy represents an additional criterion specifically embedded in the definition of symmetric PIR.
Another fundamental requirement for any PIR scheme is its correctness, *i.e.*, its property to output the requested file.

**Definition 25.** *The* correctness *of a PIR scheme with M files $\Phi_M$ is evaluated by the error probability*

$$\Pr_{\mathrm{err}}(\Phi_M) := \max_{i\in[M]} \Pr\left(\mathbf{X}^i \neq \hat{\mathbf{X}}^i\right), \tag{4.1}$$

i.e.*, the maximum probability over any file that the output file $\hat{\mathbf{X}}^\theta$ differs from the requested file $\mathbf{X}^\theta$.*
*A PIR scheme $\Phi_M$ is said to be correct if* $\Pr_{\mathrm{err}}(\Phi_M) = 0$.

Multiple servers in PIR enable coded storage by distributing the files across the servers using a storage code, as described in Section 2.2.6. This coding scheme ensures that the files are redundantly encoded and dispersed among the servers so that the content of each file can be retrieved by accessing multiple servers simultaneously. By leveraging coding techniques, PIR systems with multiple servers achieve increased fault tolerance and improved retrieval efficiency. Furthermore, we might want to ensure the security and confidentiality of the stored files even if any subset of $X$ servers are compromised or behave maliciously.

**Definition 26.** *A PIR scheme is said to be* coded *if the files are distributed across servers based on a storage code (cf. Section 2.2.6). Furthermore, a PIR scheme is $X$-secure if no subset of $X$ servers can gain any meaningful information about the content of the files.*

**Remark.** *In the context of this thesis, privacy refers to hiding a file's identity, whereas security refers to hiding a file's content.*

As we consider multiple servers in a general PIR scheme, we allow collusion among subsets of servers, where they can share information about their interactions with the user. This scenario aligns with practical distributed storage systems, where server communication is necessary to recover data in the event of node failures. Furthermore, we might want to consider the cases of erroneous, malicious or unresponsive servers.

- *Adversarial* servers refer to servers that may exhibit malicious behavior or deviate from the intended protocol. Unlike honest servers that faithfully follow the protocol, adversarial servers can potentially behave arbitrarily, such as providing incorrect answers, modifying data, or colluding to compromise user privacy.

- *Unresponsive* servers, on the other hand, are servers that do not respond. They may be offline, experiencing network issues, or intentionally refusing to cooperate. Unresponsive servers can disrupt the retrieval process and hinder the successful execution of a PIR scheme.

**Definition 27.** *A PIR scheme is considered to protect against $T$-*collusion *if no subset of $T$ servers can deduce the index of the requested file from the user's queries. A PIR scheme is said to be $(T,U,B)$-*robust *if it protects against $T$-collusion, $U$ unresponsive servers (corresponding to erasures), and $B$ adversarial servers (corresponding to errors).*

Next, we define the measure of the efficiency of a PIR scheme that we consider in this work.

**Definition 28.** *The* rate $R = R(\Phi_M)$ *of a PIR scheme $\Phi_M$ with $M$ files is a measure of its efficiency, calculated as the ratio of the size $F$ of the requested file (the* file size*) to the amount $D$ of downloaded data in bits (the* download cost*), i.e., $R(\Phi_M) := \frac{F}{D}$. A rate $R$ is called $\epsilon$-*error achievable *if there exists a sequence of PIR schemes with $M$ files $\{\Phi_M^{(\ell)}\}_\ell$ such that the PIR rate $R\left(\Phi_M^{(\ell)}\right)$ approaches $R$ and the error probability satisfies $\lim_{\ell\to\infty} \Pr_{\mathrm{err}}\left(\Phi_M^{(\ell)}\right) \le \epsilon$.*

*The* capacity $C_M$ *represents the maximum achievable rate for a given PIR setting, such as linear coded PIR with $M$ files. It is defined as $C_M := \sup_{\Phi_M} R$, where $\Phi_M$ denotes any PIR scheme. The $\epsilon$-*error capacity $C_{R,\epsilon}$ *is the supremum of all the $\epsilon$-error achievable PIR rates with $M$ files. The* asymptotic capacity $C$ *is defined as $C := \lim_{M\to\infty} C_M$ and the* asymptotic $\epsilon$-error capacity $C_\epsilon$ *is defined as $C_\epsilon := \lim_{M\to\infty} C_{M,\epsilon}$.*

*The* upload cost *refers to the communication complexity of the upload phase, which is assumed to be negligible compared to the size of the files stored on the servers,* i.e.*, the download cost can be thought of dominating the upload cost.*

*The size of a file is defined as the length of its binary representation or the number of bits needed to represent the file.*

**Remark.** *We make the assumption that the files have the same size to avoid introducing additional privacy concerns. If the files have different lengths, zero-padding can be applied to achieve equal sizes.*

### 4.1.2 Definition of a PIR scheme

Now we are ready to define a PIR scheme and its natural extension to QPIR. We provide a summary of symbols in Table 4.1.

**Definition 29.** *A* PIR scheme *is characterized by the following components:*

- *The $M$ files are stored on a DSS using an $[N, K+X]$ storage code to possibly provide $X$-security to the encoded files.*
- *The user aims to retrieve a specific file with index $\theta$ while ensuring user privacy.*

| Symbol | Description |
|---|---|
| $N$ | Number of servers / Length of a code |
| $K$ | Dimension of files before encoding |
| $X$ | File security parameter |
| $T$ | Number of colluding servers / Dimension of query code |
| $U, B$ | Numbers of unresponsive and adversarial servers |
| $(i), M$ | (Index running over) Number of files |
| $(b), \beta$ | (Index running over) Number of stripes in a file |
| $(r), \rho$ | (Index running over) Number of rounds |
| $p,\ n$ | Indices for pair and server, respectively |
| $F$ | File size (in QPIR, $F = 2K\beta\log_2(q)$) |
| $\mathscr{C}, \mathscr{D}$ | Storage and query codes |
| $(\sigma), \mathscr{H}$ | (State of) Quantum system |
| $\mathcal{V}, \mathcal{V}^{\perp_{\mathbb{S}}}$ | Self-orthogonal subspace of $\mathbb{F}_q^{2N}$ and its symplectic dual |
| $\mathsf{X}, \mathsf{Z}$ | Pauli operators |
| $\mathsf{W}(a, b)$ | Weyl gate with inputs $a, b \in \mathbb{F}_q$ |
| $\mathbf{X}, \mathbf{Y}$ | Matrices of information and encoded symbols of a file |
| $\mathbf{Q}, \mathbf{A}$ | Matrices of queries and responses |
| $R, R^{\mathsf{Q}}$ | Rate of classical, quantum PIR scheme |
| $C_M, C_M^{\mathsf{Q}}$ | Capacity of classical, quantum PIR scheme with $M$ files |
| $C_{M,\epsilon}, C_{M,\epsilon}^{\mathsf{Q}}$ | $\epsilon$-error capacities |

**Table 4.1.** Summary of symbols.

- *The user submits a certain number of queries* $\mathbf{Q}$ *to the servers. These queries are designed to protect against* $T$*-collusion attacks trying to compromise privacy,* $U$ *erasures, and* $B$ *errors.*

- *Each server responds with an answer vector, which is a function of the queries and the stored files.*

- *The user combines the received answer vectors to reconstruct the requested file.*

- *The upload cost, referring to the communication complexity during the upload phase, is assumed to be negligible compared to the size of the files stored on the DSS.*

*A* QPIR scheme *is a PIR scheme that utilizes quantum principles and quantum channels to enable private and possibly secure and efficient retrieval of information from a database, offering potential advantages over its classical counterpart. The rate and capacity of a QPIR scheme are denoted by a* $\mathsf{Q}$ *in the superscript. In the context of a QPIR scheme the download cost is given by the binary logarithm of the dimension of the downloaded composite quantum system.*

This thesis focuses on strongly linear PIR schemes [20], which are PIR schemes where the queries, the answers from the servers, and the reconstruction of the targeted file are linear. It has been shown that the star-product PIR scheme [12] is optimal among strongly linear schemes.

**Definition 30.** *A PIR scheme is called* linear *if*

- *the overall query is represented by a matrix* $\mathbf{Q} \in \mathbb{F}_q^{M\beta \times \gamma N}$*, and*

- *the classical answer of server* $n$ *is represented by*

$$\mathbf{A}_n = \mathbf{Y}_n^\top \mathbf{Q}_n \in \mathbb{F}_q^{1\times\gamma},$$

  *where* $\mathbf{Y}_n \in \mathbb{F}_q^{M\beta \times 1}$ *is the column vector stored on server* $n$*.*

*A linear PIR scheme is called* strongly linear *if there exist linear maps* $\{f_{i,j} \mid (i,j) \in [\beta] \times [K]\}$ *such that*

$$\mathbf{X}_j^i = f_{i,j}\Big((\mathbf{A}_{(n-1)\gamma + t_{i,j}} \mid n \in [N])\Big) \quad \textit{for some } t_{i,j} \in [\gamma].$$

### 4.1.3 Known capacity results

In the following, we present several known capacity results in the classical PIR model. Some of those results can be compared in Table 4.2. It is worth noting that the capacity of the symmetric PIR schemes tends to not depend on the number $M$ of files stored on the servers.

| CAPACITIES | PIR | ref. | SPIR | ref. | QPIR | ref. |
|---|---|---|---|---|---|---|
| Replicated storage, no collusion | $1-\frac{1}{N}$ | [37] | $1-\frac{1}{N}$ | [39] | 1 | [34] |
| Replicated storage, $T$-collusion | $1-\frac{T}{N}$ | [38] | $1-\frac{T}{N}$ | [43] | $\min\left\{1, \frac{2(N-T)}{N}\right\}$ | [33] |
| $[N,K]$-MDS coded storage, no collusion | $1-\frac{K}{N}$ | [3] | $1-\frac{K}{N}$ | [44] | $\min\left\{1, \frac{2(N-K)}{N}\right\}$ | [PII] |
| $[N,K]$-MDS coded storage, $T$-collusion | $1-\frac{K+T-1}{N}$ | [12] | $1-\frac{K+T-1}{N}$ | [44] | $\min\left\{1, \frac{2(N-K-T+1)}{N}\right\}$ | [PII] |

**Table 4.2.** Known capacity results with $N$ servers. The reported capacity results for PIR are asymptotic ($M \to \infty$). The result in red is a conjecture in its full generality [12], but shown to hold for strongly linear [19] (in this case the capacity is not asymptotic as it does not depend on $M$) and full support-rank [20] PIR. A scheme achieving that rate was proposed in [12]. The results in green are proved in Publication II for strongly linear PIR.

- The capacity of a PIR scheme for a replicated storage system protecting against no collusion was proven to be $\frac{1-1/N}{1-(1/N)^M}$ [37].

- The capacity of a symmetric PIR scheme from non-colluding replicated servers is $1-1/N$ if the amount of shared common randomness among servers exceeds $\frac{1}{N-1}$ bits per file bit, otherwise it is 0 [39].

- The capacity of a PIR scheme for a replicated storage system protecting against any collusion of $T$ servers is $\frac{1-T/N}{1-(T/N)^M}$ [38].

- The capacity of a PIR scheme for a replicated storage system protecting against any collusion of $T$ servers and correcting errors from any $B$ adversarial servers is $\frac{N-2B}{N} \cdot \frac{1-T/(N-2B)}{1-(T/(N-2B))^M}$ in [2].

- The capacity of a symmetric PIR scheme with replicated servers protecting against $T$-collusion and correcting errors from any $B$ adversarial servers is $1-(T+2B)/N$ if the amount of shared common randomness among servers exceeds $\frac{T+2B}{N-T-2B}$ bits per file bit, otherwise it is 0 [43].

- The capacity of a coded PIR scheme using an $[N,K]_q$ MDS storage code with no collusion is given by $\frac{1-K/N}{1-(K/N)^M}$ [3].

- The capacity of a coded symmetric PIR scheme using an $[N,K]_q$ MDS storage code with no collusion is $1-K/N$ if the amount of shared common randomness among distributed nodes is at least $\frac{K}{N-K}$ times the file size, otherwise it is 0 [45].

- The capacity of a coded PIR scheme using an $[N,K]_q$ MDS storage code with $T$-collusion was conjectured to be $\frac{1-(K+T-1)/N}{1-((K+T-1)/N)^M}$ [12] and then proven in the full support-rank case [20].

- The capacity of a coded strongly linear $(T,U,B)$-robust PIR scheme using an $[N,K]_q$ MDS storage code is $1-\frac{K+T+2B+U-1}{N}$ [20].

- The capacity of a coded symmetric PIR scheme using an $[N,K]_q$ MDS storage code with $T$-collusion is given by $1-\frac{K+T-1}{N}$ if the servers share common randomness with an amount at least $\frac{K+T-1}{N-K-T+1}$ times the file size, otherwise it is 0 [44].

- The capacity of a coded $(T,U,B)$-robust $X$-secure PIR scheme using an $[N,K+X]_q$ MDS storage code is *lower bounded* by $\frac{1-(K+X+T+U+2B-1)/N}{1-((K+X+T+U+2B-1)/N)^M}$ [22].

In the QPIR setting, assuming preexisting entanglement among servers, we observe the following capacity results, which significantly enhance the capacity compared to classical PIR.

- The capacity of a symmetric QPIR scheme for a replicated storage system with no collusion is 1 [34]. In other words, there is no price for privacy, contrasting the classical case where the capacity is $1-\frac{1}{N}$.

- The capacity of a symmetric QPIR scheme for a replicated storage system protecting against any collusion of $T$ servers is $\min\left\{1,\frac{2(N-T)}{N}\right\}$, assuming preexisting entanglement among servers [35].

- The capacity of a coded symmetric QPIR scheme using an $[N,K]_q$ MDS storage code with $T$-collusion is $\min\left\{1,\frac{2(N-K-T+1)}{N}\right\}$, assuming preexisting entanglement among servers [PII].

- The capacity of a coded (symmetric and non-symmetric) QPIR scheme using an $[N,K+X]_q$ MDS storage code with $T$-collusion and $X$-security is *lower bounded* by $\min\left\{1,\frac{2(N-K-T-X+1)}{N}\right\}$, assuming preexisting entanglement among servers [PIII].

### 4.1.4 Notation

In the following sections, we will frequently deal with $M\beta\times 2N$ matrices, where sub-blocks of $\beta$ rows and the pair of columns $n$ and $N+n$ semantically belong together. We therefore index such a matrix $\mathbf{Y}$ by two pairs of indices $(i,b)$, $i\in[M]$, $b\in[\beta]$ and $(p,n)$, $p\in[2]$, $n\in[N]$, where $Y_{p,n}^{i,b}$ denotes the symbol in row $(i-1)\beta+b$ and column $(p-1)N+n$, *i.e.*, the symbol in the $b$-th row of the $i$-th sub-block of rows and the $n$-th column of the $p$-th sub-block of columns. Omitting an index implies that we take all positions, *i.e.*, $\mathbf{Y}^i$ denotes the $i$-th sub-block of $\beta$ columns, $\mathbf{Y}^{i,b}$ the row $(i-1)\beta+b$, $\mathbf{Y}_p$ the $p$-th sub-block of $N$ columns, and $\mathbf{Y}_{p,n}$ the column $(p-1)N+n$. Furthermore, $\mathbf{Y}_{\cdot,n}$ denotes the submatrix obtained by the columns $n$ and $N+n$ of $\mathbf{Y}$, while $\mathbf{Y}_{p,\mathcal{I}}$ denotes the $p$-th sub-block of columns indexed by $\mathcal{I}$. For the reader's convenience, we sometimes imply the separation of the sub-blocks of columns by a horizontal bar in the following. In Figure 4.1 we provide the general picture we consider in this thesis for an $X$-secure DSS storing $M$ files, each consisting of $2\beta K$ symbols.

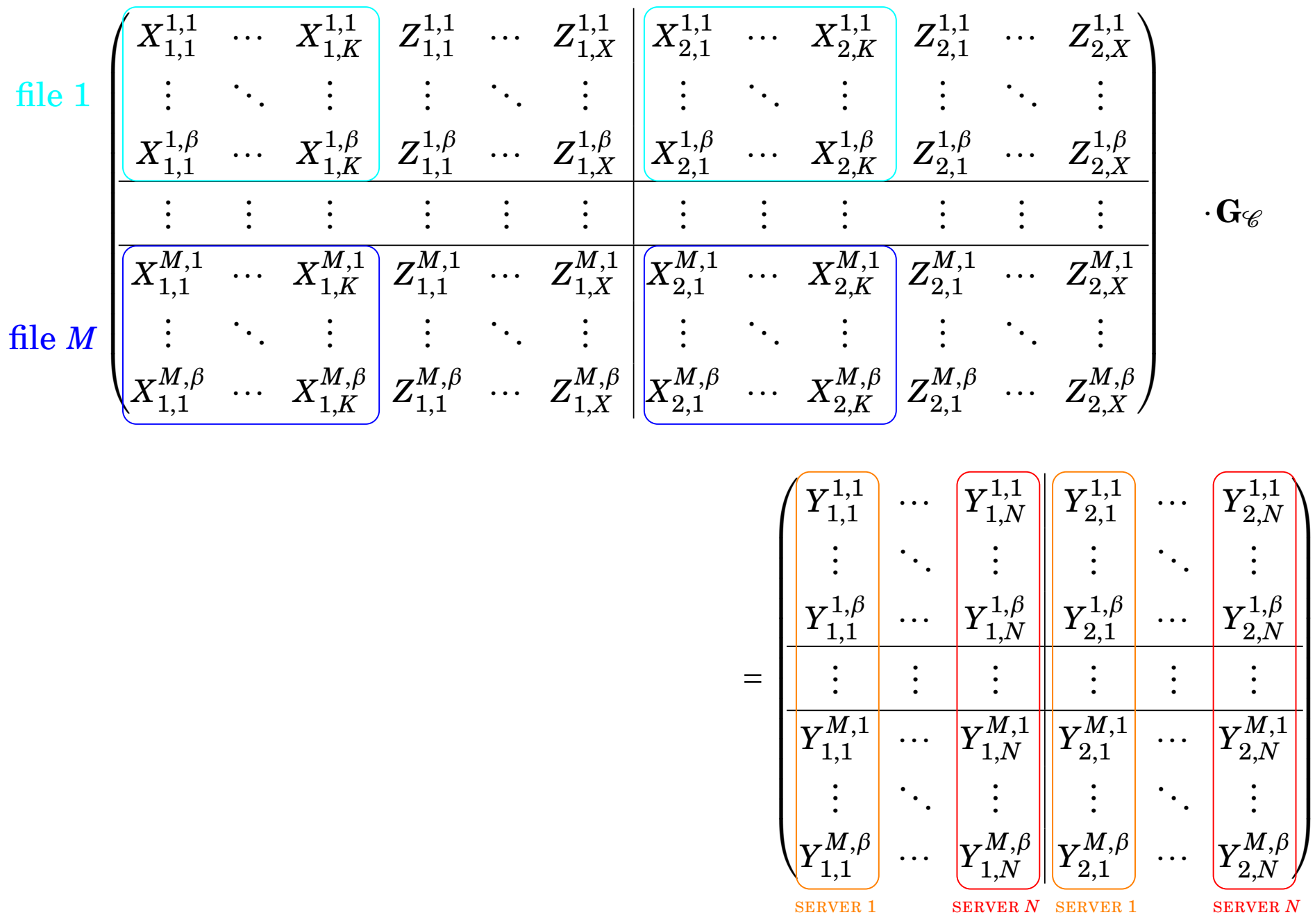

**Figure 4.1.** Illustration of an $X$-secure DSS storing $M$ files, each consisting of $2\beta K$ symbols. The matrix $\mathbf{Z}$ consists of symbols drawn uniformly at random from $\mathbb{F}_q$ to provide $X$-security. The matrix $\mathbf{G}_{\mathscr{C}}$ is a generator matrix of a $[2N, 2(K+X)]$ code $\mathscr{C}$.

We denote by $\mathbf{e}_\gamma^\lambda$ the standard basis column vector of length $\lambda$ in $\mathbb{F}_q^\lambda$ with a 1 in position $\gamma \in [\lambda]$. Given $a \in [\alpha], b \in [\beta]$, it will help our notation to call *coordinate* $(a,b)$ the position $\beta(a-1)+b$ in a vector of length $\alpha\beta$. For instance, $\mathbf{e}_{(2,1)}^{2\cdot 3} = \mathbf{e}_4^6 = (0,0,0,1,0,0)$.

Furthermore, we provide a summary of symbols in Table 4.1. The notation in this thesis may differ from the publications in the interest of coherence.

### 4.2 Qubit-based coded QPIR protocols

In Publication I we construct QPIR schemes with coded storage that protect against $T$-collusion, assuming that the servers possess only two-dimensional quantum systems, *i.e.*, qubits. The idea is that we adapt the classical PIR protocol proposed in [12] to the QPIR protocol proposed in [33].

#### 4.2.1 MDS-coded QPIR protocol with $T$-collusion

Let $N$ be the number of servers, $L = \min\{l \in \mathbb{N} : 4^l \geq N\}$ and $\mathbb{F}_{4^L}$ be the finite field with $4^L$ elements. We present an $[N,K]_{4^L}$ MDS-coded QPIR scheme that protects against $(N-K)$-collusion. We consider the underlying field $\mathbb{F}_{4^L}$ in this scheme for two reasons: firstly, MDS codes can be constructed for any code length $N$ and dimension $K$ when the size of the finite field is at least as large as the code length $N$ (cf. [29, Chapter 10]); secondly, the encoded files need to be represented as sequences of elements in $\mathbb{F}_2^2$ using the bijection defined in Equation (2.1) to enable transmission, as the Weyl gate (cf. Section 3.3.1) requires a symbol from $\mathbb{F}_2^2$ as input when applied to qubits.

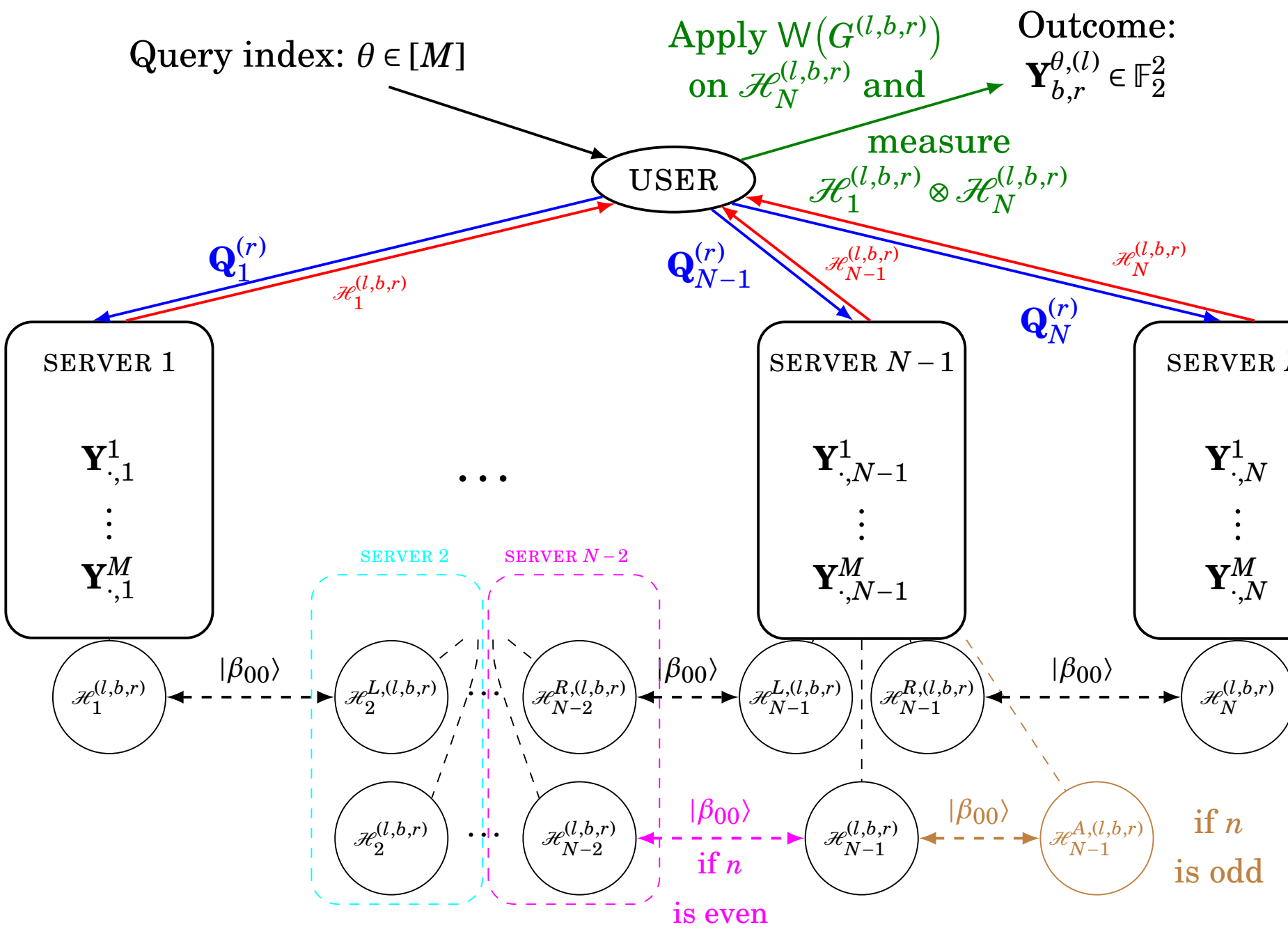

**Figure 4.2.** QPIR protocol with an $[N,K]_{4^L}$ MDS code. The queries are depicted in blue, the responses in red, and the user computations in green. For each piece $\mathbf{Y}_{\cdot,r}^{\theta}$, each server sends $L\beta$ qubits. The maximally entangled state $|\beta_{00}\rangle$ in the figure denotes $|\beta_{00}\rangle^{(l,b,r)}$.

Let $\mathscr{C}$ be such $[N,K]_{4^L}$ MDS code. Then we can represent its generator matrix as $\mathbf{G}_{\mathscr{C}} \in \mathbb{F}_{2^L}^{2K\times 2N}$. As the scheme does not provide $X$-security, the storage follows the structure depicted in Figure 4.1 with $\mathbf{G}_{\mathscr{C}}$ as the genera-tor matrix of the storage code and without the random part $\mathbf{Z}$. Since the storage code is MDS, we need any $K$ symbols of the codeword in order to retrieve the file [29]. Without loss of generality, we consider the protocol where we retrieve the symbols stored on the first $K$ servers, *i.e.*, we retrieve the symbols $\mathbf{Y}^{\theta}_{:,n}$ from server $n \in [K]$ to obtain the desired file $\mathbf{X}^{\theta}$. Thus, we need $\rho = K$ queries for each server to request the symbols stored on the first $K$ servers.

At the beginning of each round $r \in [K]$ of the protocol the servers prepare pairs of maximally entangled qubits as shown in Figure 4.2. The user generates an independent and uniformly random matrix $\mathbf{Z}^{(r)} \in \left(\mathbb{F}_{4^L}\right)^{M\times N-K}$, encodes them as codewords of the code $\mathscr{C}^{\perp}$ and adds a 1 in position $\theta$ to the query directed to server $r$. In other words, the user builds the query matrix $\mathbf{Q}^{(r)} \in \left(\mathbb{F}_{4^L}\right)^{M\times N}$ by performing a multiplication of the random matrix by the generator matrix $\mathbf{G}_{\mathscr{C}^{\perp}}$ of the dual code $\mathscr{C}^{\perp}$, *i.e.*, $\mathbf{Q}^{(r)} = \mathbf{Z}^{(r)}\mathbf{G}_{\mathscr{C}^{\perp}} + \mathbf{E}^{(r)}_{(\theta)}$, where $\mathbf{E}^{(r)}_{(\theta)}$ is an $M \times N$ matrix whose $(\theta, r)^{th}$ entry is 1 and all the other entries are 0. Then, the user sends the query vector $\mathbf{Q}^{(r)}_n$ to server $n \in [N]$. Each server computes the $\beta$ response symbols $\left\langle \mathbf{Y}^{:,b}_{:,n}, \mathbf{Q}^{(r)}_n \right\rangle \in \mathbb{F}_{4^L}$ for $b \in [\beta]$ and represents each of them to $L$ elements $\mathbf{A}^{l,b,r}_n \in \mathbb{F}_2^2$ through the bijection defined in Equation (2.1). After applying quantum operations dependent on $\mathbf{A}^{l,b,r}_n$, each server $n \in [N]$ sends a total of $L\beta$ qubits as a response during each round $r \in [K]$. The user performs further quantum operations on the received qubits and, after the final measurement, obtains $KL\beta$ outcomes $\mathbf{Y}^{\theta,(l)}_{b,r} \in \mathbb{F}_2^2$, from which they can reconstruct the desired file $\mathbf{X}^{\theta}$. For more details, we refer the reader to [PI, Section IV-A].

The protocol described above is correct, *i.e.*, the user retrieves the desired file from the responses (cf. [PI, Lemma 1]), is symmetric and protects against $(N-K)$-collusion (cf. [PI, Lemma 2]), and by [PI, Theorem 1] has rate

$$R^{\mathrm{Q}} = \begin{cases} \frac{2}{K+T}, & \text{if } K+T \text{ is even,} \\ \frac{2}{K+T+1}, & \text{if } K+T \text{ is odd.} \end{cases} \tag{4.2}$$

Let us show an example of a $[4,2]_4$ GRS-coded QPIR scheme protecting against 2-collusion.

**Example 11.** *Let us consider $N = 4$ servers. Then $L = 1$, and the underlying field is $\mathbb{F}_4 = \{0, 1, \alpha, \alpha^2\}$ where $\alpha$ is a primitive element that satisfies*

$$\alpha^2 + \alpha + 1 = 0. \tag{4.3}$$

*We want to encode the storage with a $[4,2]_4$ RS code $\mathscr{C}$ with generator matrix*

$$\mathbf{G}_{\mathscr{C}} = \begin{pmatrix} 1 & 0 & \alpha^2 & \alpha \\ 0 & 1 & \alpha & \alpha^2 \end{pmatrix}.$$

*Suppose also $\beta = 1$, so that the matrix of files is given by $\mathbf{X} \in \mathbb{F}_4^{M\times K}$. Hence,*

*each file is encoded into*

$$\mathbf{Y}^i = \mathbf{X}^i \mathbf{G}_{\mathscr{C}} = \begin{pmatrix} X_1^i & X_2^i & \alpha^2 X_1^i + \alpha X_2^i & \alpha X_1^i + \alpha^2 X_2^i \end{pmatrix} \in \mathbb{F}_4^N.$$

*Since* $\mathbb{F}_4 \cong \mathbb{F}_2^2$, *we can equivalently represent the files as* $\mathbf{X}' = \left(\mathbf{X}'_1 \,\middle|\, \mathbf{X}'_2\right) \in \mathbb{F}_2^{M\times 2K}$ *and the storage as* $\mathbf{Y}' = \left(\mathbf{X}'_1\mathbf{G}_{\mathscr{C}} \,\middle|\, \mathbf{X}'_2\mathbf{G}_{\mathscr{C}}\right) \in \mathbb{F}_2^{M\times 2N}$, *following the notation in Figure 4.1 with the storage code being* $\mathscr{C} \times \mathscr{C}$.

*Since* $K = 2$, *this protocol requires 2 rounds, and since* $L = \beta = 1$, *each server needs to send only 1 qubit per round to the user. Hence, during round* $r \in [K]$ *servers 1 and 4 possess one qubit* $\mathscr{H}_1^{(r)}$ *and* $\mathscr{H}_4^{(r)}$, *respectively, while servers 2 and 3 possess 3 qubits* $\mathscr{H}_2^{\mathscr{L},(r)}$, $\mathscr{H}_2^{\mathscr{R},(r)}$, $\mathscr{H}_2^{(r)}$ *and* $\mathscr{H}_3^{\mathscr{L},(r)}$, $\mathscr{H}_3^{\mathscr{R},(r)}$, $\mathscr{H}_3^{(r)}$, *respectively. The pairs* $\left(\mathscr{H}_1^{(r)}, \mathscr{H}_2^{\mathscr{L},(r)}\right)$, $\left(\mathscr{H}_2^{\mathscr{R},(r)}, \mathscr{H}_3^{\mathscr{L},(r)}\right)$, $\left(\mathscr{H}_3^{\mathscr{R},(r)}, \mathscr{H}_4^{(r)}\right)$ *and* $\left(\mathscr{H}_2^{(r)}, \mathscr{H}_3^{(r)}\right)$ *are prepared in the maximally entangled state* $|\beta_{00}\rangle$. *The setup is represented in Figure 4.3.*

*During round* $r$ *the user wants to retrieve the symbol* $X_r^\theta$ *of* $\mathbf{X}^\theta$, *which is stored as clear text in the server* $r$ *in this example. Since we want to protect against 2-collusion, the user generates a random matrix* $\mathbf{Z}^{(r)} \in \mathbb{F}_4^{M\times N-K}$ *and encodes it with the dual code of* $\mathscr{C}$, *which itself is self-dual. Thus, the query matrix is given by*

$$\mathbf{Q}^{(r)} = \mathbf{Z}^{(r)}\mathbf{G}_{\mathscr{C}} + \mathbf{E}_{(\theta)}^{(r)}.$$

*The user sends each query vector* $\mathbf{Q}_n$ *to server* $n \in [N]$.

*Each server proceeds to compute* $A_n^{(r)} = \left\langle \mathbf{Y}_n, \mathbf{Q}_n^{(r)} \right\rangle \in \mathbb{F}_4$. *Servers 1 and 4*

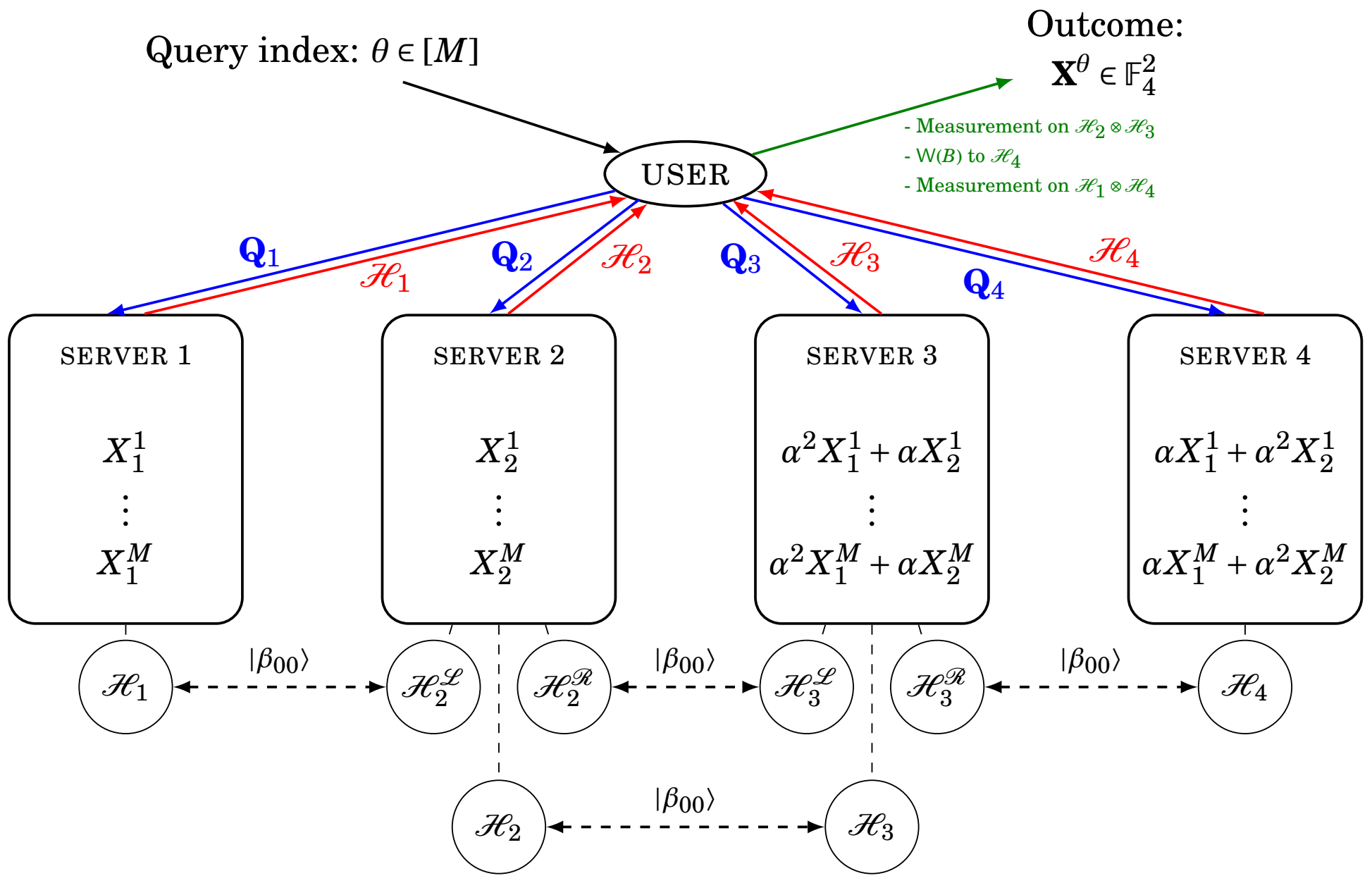

**Figure 4.3.** QPIR protocol with a $[4,2]_4$ RS code, as described in Example 11. The queries are depicted in blue, the responses in red, and the user computations in green. Here, we dropped the superscript $(r)$ for a clearer visualization.

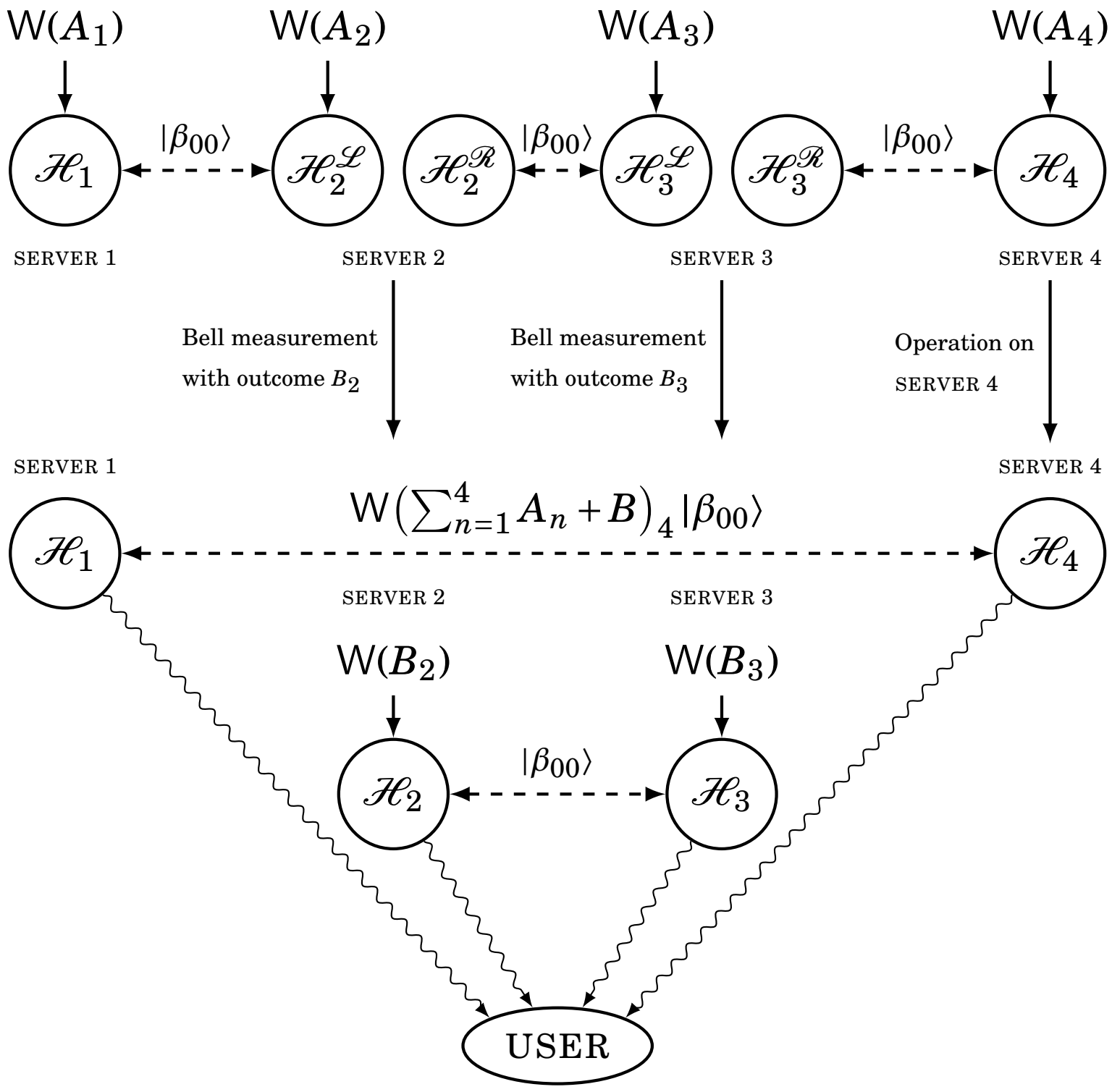

**Figure 4.4.** Servers' operations and download step for the QPIR protocol with a $[4,2]_4$ RS code, as described in Example 11. These operations are performed for every round $r \in [2]$. Here, we dropped the superscript $(r)$ for a clearer visualization. One might notice that the Bell measurements applied by servers 2 and 3 are part of a quantum teleportation protocol with an operation (cf. Section 3.5.2) where $\mathcal{H}_1$ is the reference system and the state over $\mathcal{H}_2^{\mathcal{L}}$ is first teleported to $\mathcal{H}_3^{\mathcal{L}}$ and then to $\mathcal{H}_4$. The two bits needed to obtain the final teleported state are given by $B = B_2 + B_3 \in \mathbb{F}_2^2$.

*apply Weyl gates*[1] $\mathsf{W}(A_1^{(r)})$ *and* $\mathsf{W}(A_4^{(r)})$ *to their qubit, respectively, and send them to the user. Similarly, servers 2 and 3 apply* $\mathsf{W}(A_2^{(r)})$ *and* $\mathsf{W}(A_3^{(r)})$ *to their qubit* $\mathcal{H}_2^{\mathcal{L},(r)}$ *and* $\mathcal{H}_3^{\mathcal{L},(r)}$*, respectively. Furthermore, servers 2 and 3 perform a Bell measurement (cf. Example 7) on the pairs of qubits* $\left(\mathcal{H}_2^{\mathcal{L},(r)}, \mathcal{H}_2^{\mathcal{R},(r)}\right)$, $\left(\mathcal{H}_3^{\mathcal{L},(r)}, \mathcal{H}_3^{\mathcal{R},(r)}\right)$*, from which they obtain outcomes* $B_2^{(r)}, B_3^{(r)} \in \mathbb{F}_2^2$*. Finally, servers 2 and 3 use the two-sum transmission protocol (cf. Section 3.5.3) with input qubits* $\mathcal{H}_2^{(r)} \otimes \mathcal{H}_3^{(r)}$ *to send* $B^{(r)} = B_2^{(r)} + B_3^{(r)}$ *to the user. The servers' operations and qubits' transmission are depicted in Figure 4.4.*

*The user obtains* $B^{(r)} \in \mathbb{F}_2^2$ *as output of the Bell measurement performed on* $\mathcal{H}_2^{(r)} \otimes \mathcal{H}_3^{(r)}$*, applies* $\mathsf{W}(B^{(r)})$ *to* $\mathcal{H}_4^{(r)}$ *and performs a Bell measurement on* $\mathcal{H}_1^{(r)} \otimes \mathcal{H}_4^{(r)}$*, obtaining the desired* $X_r^{\theta}$ *as output.*

*For more details, we refer the reader to [PI, Section V].*

**Remark.** *Notice that in each round of this scheme, we always retrieve only 2 bits of information. This is a consequence of the fact that the output of the*

[1] With a slight abuse of notation, we use here an element of $\mathbb{F}_4$ as input to a Weyl gate over a qubit, since any element in $\mathbb{F}_4$ can be represented as a pair of elements in $\mathbb{F}_2$.

*final measurement performed by the user is made on a pair of qubits, and by the Holevo bound [18] the user cannot obtain more than 2 bits, as 2 is the binary logarithm of the dimension of the corresponding Hilbert space. For this reason, we encode the queries with the dual of the storage code: since we are able to transmit only 2 bits at a time, we use two instances of a star-product PIR protocol with $\mathscr{C}$ as storage code and $\mathscr{C}^\perp$ as query code [12], which can transmit only 1 symbol per round as* $\dim\left(\mathscr{C}\star\mathscr{C}^\perp\right)^\perp = \dim\left(\mathscr{R}_{4^L}(N)\right) = 1$ *(cf. Section 2.2.5). For more details, we refer the reader to [PI, Remark 3].*

### 4.2.2 LRC-coded QPIR with $T$ collusion

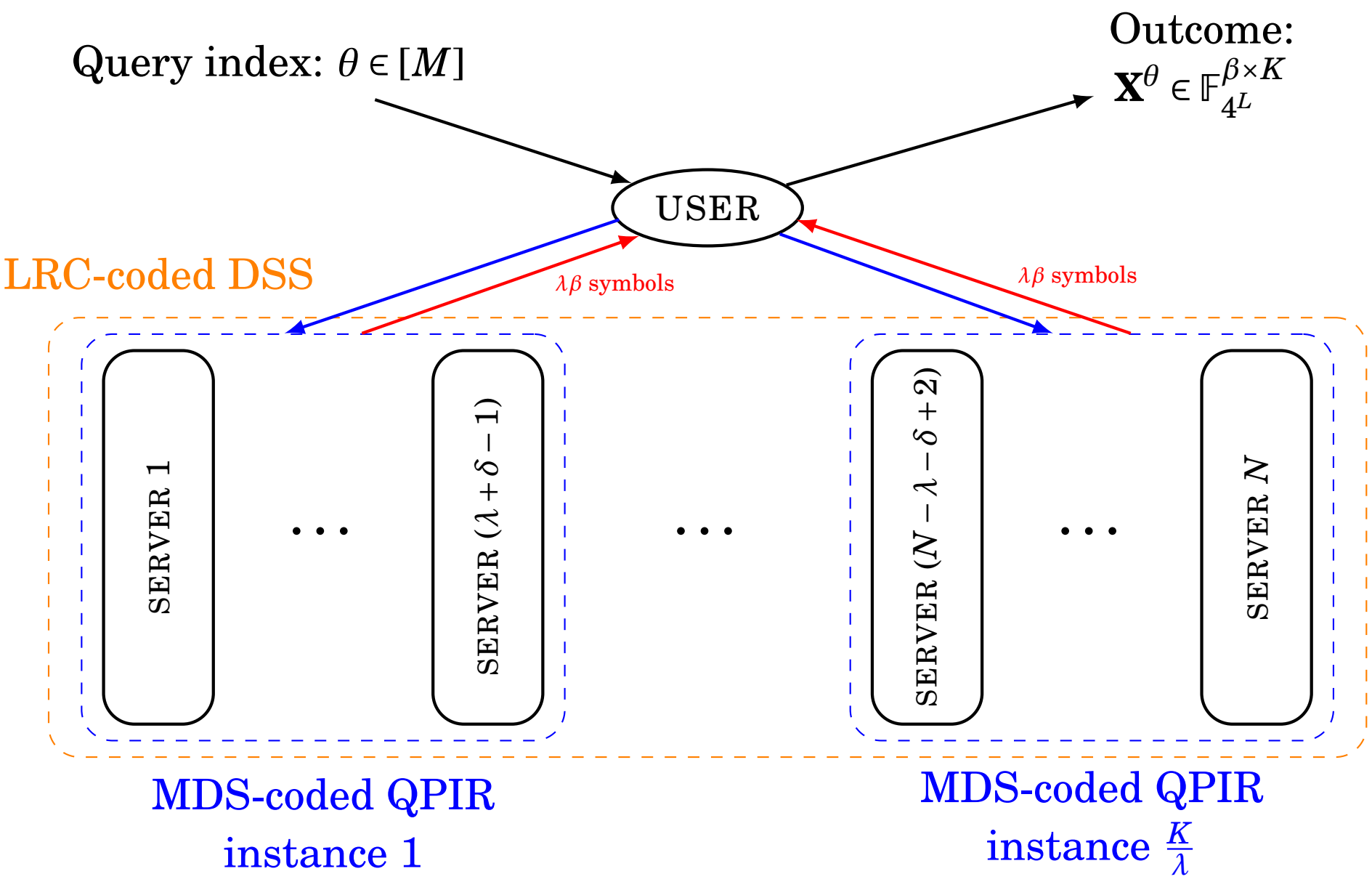

**Figure 4.5.** QPIR protocol with an optimal LRC-coded DSS. The red arrows represent the number of symbols that the user obtains from each instance of an MDS-coded QPIR protocol rather than the download step in this figure.

Let $N$ be the number of servers, $L = \min\left\{l \in \mathbb{N} : 4^l \geq N\right\}$, $K$ the dimension of the files, and $\lambda$ an integer such that $\lambda \mid K$. Let $\mathscr{C} = \mathrm{LRC}^{4^L}_{N,K}(\lambda,\delta)$ be an optimal Locally Repairable Code (cf. Section 2.3.4), *i.e.*, an LRC such that its local codes are $[\lambda+\delta-1,\lambda]_{4^L}$ MDS codes. Considering a DSS encoded with such an LRC (cf. Figure 4.5), the LRC-coded QPIR protocol proposed in [PI, Section VI] utilizes the iterative application of the scheme described in Section 4.2.1 to particular subsets of servers corresponding to the positions of the local codes, since each one of them is an MDS code. As the local codes have dimension $\lambda$, their dual codes have dimension $\lambda+\delta-1-\lambda = \delta-1$, so each instance of the QPIR protocol can protect against $(\delta-1)$-collusion. Furthermore, since $\lambda$ divides $K$, the $\lambda$ positions from each local code form a local information set, and the union of these information sets from $\frac{K}{\lambda}$ local codes form an information set for the LRC. Thus, we need to perform $\frac{K}{\lambda}$ instances of an MDS-coded QPIR to retrieve $\lambda$ symbols from each local code, from which we can reconstruct the original file with dimension $K$. It follows that the rate of an LRC-coded QPIR protecting against $(\delta-1)$-collusion is given by

$$R^{\mathrm{Q}} = \begin{cases} \frac{2}{\lambda+\delta-1}, & \text{if } \lambda+\delta-1 \text{ is even,} \\ \frac{2}{\lambda+\delta}, & \text{if } \lambda+\delta-1 \text{ is odd.} \end{cases} \tag{4.4}$$

It is important to note that with LRC-coded storage, the QPIR rate is no longer directly influenced by the code length $N$. This leads to an improved retrieval rate compared to the rate for MDS-coded storage, as the locality parameter $\lambda$ is typically much smaller than the code dimension $K$. However, there is a trade-off as the scheme becomes more susceptible to collusion and the total number of tolerated server failures decreases, due to the typically small value of $\delta$. Nevertheless, the scheme can still protect against collusion from more than $T=\delta-1$ servers, as long as no more than $T$ servers collude within a local group. Specifically, for such collusion patterns, the scheme can resist collusion of up to $\frac{KT}{\lambda}=\frac{K(\delta-1)}{\lambda}$ servers.

## 4.3 Stabilizer-based coded QPIR protocols

In Publication II we construct a capacity-achieving QPIR scheme with coded storage that protects against $T$-collusion, assuming that the servers possess $q$-dimensional quantum systems, *i.e.*, qudits. The idea is that we adapt the classical PIR protocol proposed in [12] to the QPIR protocol proposed in [35]. Furthermore, we derive the capacity of QPIR with $[N,K]_q$ MDS-coded storage and $T$-collusion.

### 4.3.1 GRS-coded QPIR protocol with $T$-collusion

Let $N$ be the number of servers and $q$ be a power of 2. Let $\mathscr{C}'=\mathrm{GRS}_K^q(\mathscr{A},\mathbf{v})$ be a GRS code (cf. Section 2.3.1), where $\mathscr{A}$ is the set of evaluation points and $\mathbf{v}$ is a vector of non-zero elements from $\mathbb{F}_q$. For a given integer $c$, which will be defined in the next paragraph, the parameter $\beta$ is fixed to $\beta=\mathrm{lcm}(c,K)/K$. Then, following the notation of Figure 4.1, the matrix of files is given by $\mathbf{X}\in\mathbb{F}_q^{M\beta\times 2K}$ and the storage code is $\mathscr{C}=\mathscr{C}'\times\mathscr{C}'$, as the scheme does not provide $X$-security.

Let $T$ be the collusion parameter with $\frac{N}{2}\le K+T-1<N$. By [PII, Lemma 5.3] there exists an $[N,T]$ GRS code $\mathscr{D}'$ such that $\mathscr{S}'=\mathscr{C}'\star\mathscr{D}'$ is a weakly self-dual $[N,K+T-1]$ GRS code, which defines $c=d_{\mathscr{S}'}-1=N-K-T+1$. Thus, we use as query code the cartesian product $\mathscr{D}=\mathscr{D}'\times\mathscr{D}'$ with a generator matrix given by $\mathbf{G}_{\mathscr{D}}=\mathrm{diag}(\mathbf{G}_{\mathscr{D}'},\mathbf{G}_{\mathscr{D}'})\in\mathbb{F}_q^{2T\times 2N}$.

Defining $\mathscr{S}=\mathscr{C}\star\mathscr{D}$, by the properties of the star product we have that $\mathscr{S}=\mathscr{C}\star\mathscr{D}=\mathscr{S}'\times\mathscr{S}'$, and it follows that $\mathscr{S}$ is weakly self-dual. Thus, its parity-check matrix $\mathbf{H}_{\mathscr{S}}$ is self-orthogonal (cf. Definition 22) and defines a self-orthogonal subspace $\mathscr{V} \subseteq \mathbb{F}_q^{2N}$. Furthermore, a generator matrix $\mathbf{G}_{\mathscr{S}}$ defines a basis for $\mathscr{V}^{\perp_{\mathbb{S}}}$, since $\mathbf{H}_{\mathscr{S}}\mathbf{J}^\top\mathbf{G}_{\mathscr{S}}^\top = \mathbf{0}$ (cf. [PII, Lemma 5.4]).

Let $\mathscr{H}_1,\dots,\mathscr{H}_N$ be $q$-dimensional quantum systems and define the mixed state $\sigma_{\text{mix}} = q^{N-2(K+T-1)} \cdot \mathbf{I}_{q^{2(K+T-1)-N}}$. By [PII, Proposition 2.2.(b)], the composite quantum system $\mathscr{H} = \mathscr{H}_1 \otimes \cdots \otimes \mathscr{H}_N$ can be decomposed as $\mathscr{H} = \mathscr{W} \otimes \mathbb{C}^{q^{2(K+T-1)-N}}$, where $\mathscr{W} = \operatorname{span}\left\{|\bar{\mathbf{s}}\rangle : \bar{\mathbf{s}} \in \mathbb{F}_q^{2N}/\mathscr{V}^{\perp_{\mathbb{S}}}\right\}$.

The protocol is composed of $\rho = \operatorname{lcm}(c,K)/c$ rounds of queries and responses to and from the servers, after which the user is able to retrieve the desired file $\mathbf{X}^\theta$. At the beginning of each round $r \in [\rho]$, the state of $\mathscr{H}$ is initialized as $|\bar{\mathbf{0}}\rangle\langle\bar{\mathbf{0}}| \otimes \sigma_{\text{mix}}$ and distributed such that server $n \in [N]$ obtains $\mathscr{H}_n$, as depicted in Figure 4.6. Then, the user generates the query $\mathbf{Q}^{(r)} \in \mathbb{F}_q^{M\beta\times 2N}$ by multiplying an uniformly random matrix $\mathbf{Z}^{(r)} \in \mathbb{F}_q^{M\beta\times 2T}$ with $\mathbf{G}_{\mathscr{D}}$ and adding an error matrix $\mathbf{E}_{(\theta)}\mathbf{M}^{(r)}$ to target specific symbols of the desired files. The matrix $\mathbf{E}_{(\theta)}$ is the same in every round and is used to target the index of the file, while $\mathbf{M}^{(r)}$ changes in each round and is used to target specific symbols. After the user sends the pair of columns $\mathbf{Q}_{1,n}^{(r)}, \mathbf{Q}_{2,n}^{(r)}$ to server $n \in [N]$, each server computes $A_{p,n}^{(r)} = \left\langle \mathbf{Y}_{p,n}, \mathbf{Q}_{p,n}^{(r)} \right\rangle \in \mathbb{F}_q$, applies $\mathbf{W}(\mathbf{A}_{\cdot,n}^{(r)})$ to its qudits $\mathscr{H}_n$, and sends it back to the user. In each round, the user can retrieve $2c/\beta = 2K/\rho$ symbols from each of the $\beta$ rows of $\mathbf{Y}^\theta$. To achieve this, the user measures the received qudits over a basis that depends on the chosen $\mathbf{M}^{(r)}$ and obtains $2K/\rho$ symbols in each round. Clearly, after $\rho$ rounds the user retrieves $2K$ symbols, and carefully designing the error matrix and the basis for the measurement in each round it is possible to obtain the matrix of symbols $\left(\mathbf{Y}_{1,[K]}^\theta \middle| \mathbf{Y}_{2,[K]}^\theta\right) \in \mathbb{F}_q^{\beta\times 2K}$. Since each $\mathbf{Y}_{p,[K]}^{\theta,b}$ is a vector of $K$ symbols of a codeword in $\mathscr{C}'$, the user is able to reconstruct the desired file $\mathbf{X}^{\theta}$. For more details, we refer the reader to [PII, Section V.B].

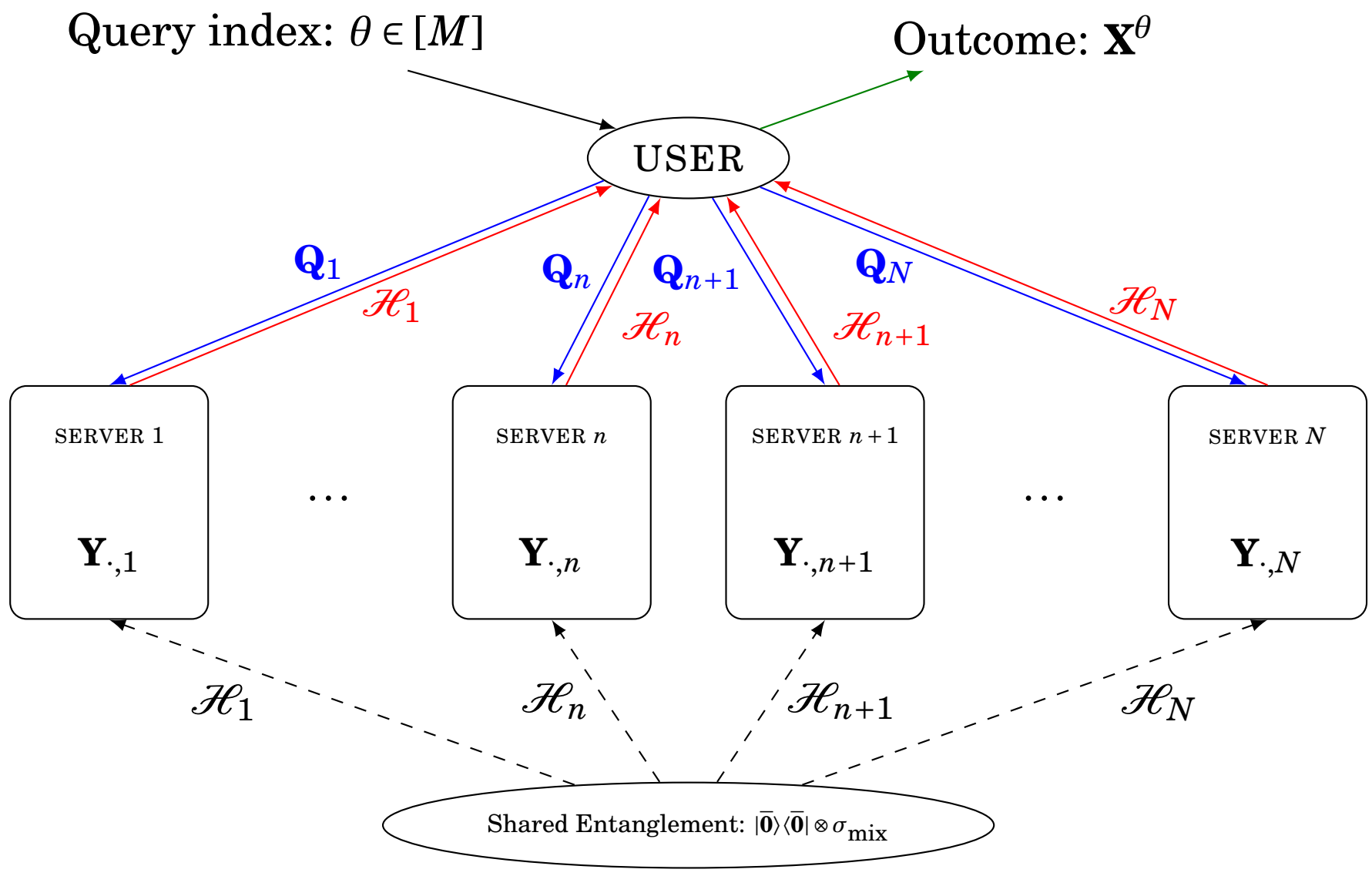

**Figure 4.6.** QPIR protocol with an $[N,K]_q$ GRS code protecting against $T$-collusion. The queries are depicted in blue, the responses in red, and the user computations in green.

The protocol described above is correct (cf. [PII, Lemma 5.5]), is symmetric and protects against $T$-collusion (cf. [PII, Lemma 5.6]), and by [PII, Theorem 5.1] has rate

$$R^{\mathrm{Q}} = \frac{2(N-K-T+1)}{N}. \tag{4.5}$$

**Remark.** *If the collusion parameter* $T$ *is such that* $1 \le K+T-1 < N/2$*, the presented scheme for* $T = N/2-K+1$ *for even* $N$ *has rate 1. Since the rate cannot be greater than 1, it is capacity-achieving. If* $N$ *is odd, we just consider* $N-1$ *servers and* $T = (N+1)/2-K$ *in order to achieve rate 1.*

Let us show an example of a $[6,3]_7$ GRS coded QPIR scheme protecting against 2-collusion.

**Example 12.** *Let us consider* $N = 6$ *servers and* $\mathscr{A} = \left\{3^{n-1} \mod 7 : n \in [N]\right\}$. *Let* $\mathscr{C}' = \mathrm{GRS}_3^7\left(\mathscr{A}, \mathbf{1}^N\right)$ *be the RS code over* $\mathbb{F}_q = \mathbb{F}_7$ *with dimension* $K = 3$ *and evaluation points given by* $\mathscr{A}$*. As the evaluation points are given by the set of powers of a primitive element of* $\mathbb{F}_7$*, this code is also referred to as a Primitive Reed–Solomon (PRS) code [29, Chapter 10.2]. The matrix of files* $\mathbf{X} \in \mathbb{F}_7^{M\beta\times 2K}$ *is encoded with* $\mathscr{C} = \mathscr{C}' \times \mathscr{C}'$, i.e., *it is multiplied by* $\mathbf{G}_{\mathscr{C}} = \mathrm{diag}(\mathbf{G}_{\mathscr{C}'}, \mathbf{G}_{\mathscr{C}'})$ *to obtain the encoded storage* $\mathbf{Y} = \mathbf{X}\mathbf{G}_{\mathscr{C}} \in \mathbb{F}_7^{M\beta\times 2N}$*, where*

$$\mathbf{G}_{\mathscr{C}'} = \begin{pmatrix} 1 & 1 & 1 & 1 & 1 & 1 \\ 1 & 3 & 2 & 6 & 4 & 5 \\ 1 & 2 & 4 & 1 & 2 & 4 \end{pmatrix}.$$

*Server* $n \in [N]$ *stores* $\mathbf{Y}_{\cdot,n}$*, following the notation of Figure 4.1.*

*Let* $\mathscr{D}' = \mathrm{GRS}_2^7\left(\mathscr{A}, \mathbf{1}^N\right)$ *be the PRS code over* $\mathbb{F}_7$ *with dimension* $T = 2$ *and evaluation points given by* $\mathscr{A}$*, which has a generator matrix* $\mathbf{G}_{\mathscr{D}'}$ *equal to the first two rows of* $\mathbf{G}_{\mathscr{C}'}$*. We choose the query code as* $\mathscr{D} = \mathscr{D}' \times \mathscr{D}'$ *which has generator matrix* $\mathbf{G}_{\mathscr{D}} = \mathrm{diag}(\mathbf{G}_{\mathscr{D}'}, \mathbf{G}_{\mathscr{D}'}) \in \mathbb{F}_7^{2T\times 2N}$*. Since* $\mathscr{S}' = \mathscr{C}' \star \mathscr{D}'$ *is a* $\mathrm{GRS}_4^7\left(\mathscr{A}, \mathbf{1}^N\right)$*, it has distance* $d_{\mathscr{S}'} = 3$*. Thus, the user can download at most* $2c = 2(N-K-T+1) = 4$ *blocks of information per round from each server. Since both* $\mathscr{C}$ *and* $\mathscr{D}$ *are cartesian products of PRS codes, also* $\mathscr{S}$ *is the cartesian product of two PRS codes generated by* $\mathscr{S}' = \mathscr{C}' \star \mathscr{D}'$*. Let*

$$\mathbf{G}_{\mathscr{S}'} = \begin{pmatrix} 1 & 3 & 2 & 6 & 4 & 5 \\ 1 & 2 & 4 & 1 & 2 & 4 \\ 1 & 1 & 1 & 1 & 1 & 1 \\ 1 & 6 & 1 & 6 & 1 & 6 \end{pmatrix} \in \mathbb{F}_7^{(K+T-1)\times N}$$

*be the generator matrix of the star product code* $\mathscr{S}'$ *(it is obtained by rearranging the rows of a* $4\times 6$ *Vandermonde matrix with evaluation points given by* $\mathscr{A}$*), and let* $\mathbf{H}_{\mathscr{S}'} \in \mathbb{F}_7^{(N-K-T+1)\times N}$ *be the submatrix with the first two rows of* $\mathbf{G}_{\mathscr{S}'}$ *and* $\mathbf{H}'_{\mathscr{S}'} \in \mathbb{F}_7^{(2(K+T-1)-N)\times N}$ *be the submatrix with the remaining*

$\mathbf{Y}^{\theta}$

| | | | | | | | | | | | |
|---|---|---|---|---|---|---|---|---|---|---|---|
| $Y^{\theta,1}_{1,1}$ | $Y^{\theta,1}_{1,2}$ | $Y^{\theta,1}_{1,3}$ | $Y^{\theta,1}_{1,4}$ | $Y^{\theta,1}_{1,5}$ | $Y^{\theta,1}_{1,6}$ | $Y^{\theta,1}_{2,1}$ | $Y^{\theta,1}_{2,2}$ | $Y^{\theta,1}_{2,3}$ | $Y^{\theta,1}_{2,4}$ | $Y^{\theta,1}_{2,5}$ | $Y^{\theta,1}_{2,6}$ |
| $Y^{\theta,2}_{1,1}$ | $Y^{\theta,2}_{1,2}$ | $Y^{\theta,2}_{1,3}$ | $Y^{\theta,2}_{1,4}$ | $Y^{\theta,2}_{1,5}$ | $Y^{\theta,2}_{1,6}$ | $Y^{\theta,2}_{2,1}$ | $Y^{\theta,2}_{2,2}$ | $Y^{\theta,2}_{2,3}$ | $Y^{\theta,2}_{2,4}$ | $Y^{\theta,2}_{2,5}$ | $Y^{\theta,2}_{2,6}$ |

**Figure 4.7.** Symbols downloaded in each round of the QPIR protocol described in Example 12. The symbols colored in red, in blue, and in green are retrieved during round 1, round 2, and round 3, respectively.

*two rows of* $\mathbf{G}_{\mathscr{S}'}$*. It is easy to check that* $\mathbf{H}_{\mathscr{S}'}$ *is a parity-check matrix of* $\mathscr{S}'$*, which makes* $\mathscr{S}'$ *a weakly self-dual PRS code. Then the generator matrix of* $\mathscr{S}$ *is given by*

$$\mathbf{G}_{\mathscr{S}} = \begin{pmatrix} \operatorname{diag}(\mathbf{H}_{\mathscr{S}'}, \mathbf{H}_{\mathscr{S}'}) \\ \operatorname{diag}\left(\mathbf{H}'_{\mathscr{S}'}, \mathbf{H}'_{\mathscr{S}'}\right) \end{pmatrix} \in \mathbb{F}_7^{2(K+T-1)\times 2N},$$

*and it is easy to see that its first* $2(N-K-T+1)$ *rows form a self-orthogonal matrix.*

*Since* $c = 2$*, we set the number of blocks per file to* $\beta = 2$ *and the number of rounds of the protocol to* $\rho = 3$*. For the first round, consider* $\mathbf{N}^{(1)} = (\mathbf{I}_2 \; \mathbf{0}_{2\times 4})$ *and* $\mathbf{M}^{(1)} = \operatorname{diag}\left(\mathbf{N}^{(1)}, \mathbf{N}^{(1)}\right) \in \mathbb{F}_7^{2c\times 12}$*. Then* $\mathbf{M}^{(1)}$ *is such that the row vectors of the matrix* $(\mathbf{G}_{\mathscr{S}} \; \mathbf{M}^{(1)})$ *form a basis for* $\mathbb{F}_7^{12}$*. To see that this is in fact a basis observe that* $\langle \mathbf{N}^{(1)} \rangle_{\mathsf{row}}$*, by definition, contains vectors of weight at most* $c$*, while* $\langle \mathbf{G}_{\mathscr{S}'} \rangle_{\mathsf{row}}$ *contains vectors of weight at least* $c+1$*. It follows that the spans of* $\mathbf{N}^{(1)}$ *and* $\mathbf{G}_{\mathscr{S}'}$ *intersect trivially, which implies that their ranks add up.*

*First, the quantum systems are prepared and distributed to the servers according to the first step of the scheme.*

*The user samples uniformly at random* $\mathbf{Z}^{(1)} \in \mathbb{F}_7^{2M\times 2T}$*. Let*

$$\mathbf{E}_{(\theta)} = \begin{pmatrix} \mathbf{e}^{2M}_{(\theta,1)} & \mathbf{e}^{2M}_{(\theta,2)} & \mathbf{e}^{2M}_{(\theta,1)} & \mathbf{e}^{2M}_{(\theta,2)} \end{pmatrix} \in \mathbb{F}_7^{2M\times 2c}.$$

*Notice that the row in coordinate* $(i,b)$ *of the product* $\mathbf{E}_{(\theta)} \cdot \mathbf{M}^{(1)}$ *is*

$$\delta_{i,\theta}\left(\delta_{b,1}\left(\mathbf{e}^{12}_{(1,1)} + \mathbf{e}^{12}_{(2,1)}\right) + \delta_{b,2}\left(\mathbf{e}^{12}_{(1,2)} + \mathbf{e}^{12}_{(2,2)}\right)\right)^{\top}.$$

*Then, with this choice, the user will retrieve the first block (with* $\delta_{b,1}$*) of the symbols stored on server 1 (with* $\mathbf{e}^{12}_{(p,1)}$*) and the second block (with* $\delta_{b,2}$*) of the symbols stored on server 2 (with* $\mathbf{e}^{12}_{(p,2)}$*) with the desired position* $\theta$ *(with* $\delta_{i,\theta}$*), as shown in Figure 4.7. The user generates the queries as* $\mathbf{Q}^{(1)} = \mathbf{Z}^{(1)}\mathbf{G}_{\mathscr{D}} + \mathbf{E}_{(\theta)}\mathbf{M}^{(1)}$*. For each pair of servers, the corresponding joint distribution of queries is the uniform distribution over* $\mathbb{F}_7^{2M\times 2T}$*.*

*The servers compute the responses* $A^{(1)}_{p,n} = \left\langle \mathbf{Y}_{p,n}, \mathbf{Q}^{(1)}_{p,n} \right\rangle \in \mathbb{F}_7$*,* $p \in [2], n \in [6]$*. Server* $n$ *applies* $\mathsf{W}(A^{(1)}_{1,n}, A^{(1)}_{2,n})$ *to its quantum system and sends it to the user.*

*As the response vector is*

$$\mathbf{A}^{(1)} \in \mathcal{V}^{\perp_{\mathbb{S}}} + \begin{pmatrix} Y^{\theta,1}_{1,1} & Y^{\theta,2}_{1,2} & Y^{\theta,1}_{2,1} & Y^{\theta,2}_{2,2} \end{pmatrix} \cdot \mathbf{M}^{(1)},$$

*the user obtains* $\left(Y_{1,1}^{\theta,1}, Y_{1,2}^{\theta,2}, Y_{2,1}^{\theta,1}, Y_{2,2}^{\theta,2}\right) \in \mathbb{F}_7^4$ *as output without error.*

*In rounds 2 and 3 the user generates different matrices* $\mathbf{M}^{(2)}, \mathbf{M}^{(3)}$ *to target the symbols according to Figure 4.7. Finally, after 3 rounds the user recovers the symbols* $\left(\mathbf{Y}_{1,[K]}^{\theta} \middle| \mathbf{Y}_{2,[K]}^{\theta}\right) \in \mathbb{F}_7^{2\times 6}$. *From these symbols, the user can easily recover the desired file* $\mathbf{X}^{\theta}$ *by solving a system of linear equations. The user downloaded a total of 18* 7*-dimensional quantum systems, giving a download cost* $D = \log_2(7^{18}) = 18\log_2(7)$, *and gathered 12 symbols of* $\mathbb{F}_7$, *as the file size in bits is* $F = 2K\beta\log_2(7) = 12\log_2(7)$. *The rate is thus given by* $R^{\mathrm{Q}} = \frac{12\log_2(7)}{18\log_2(7)} = \frac{2}{3}$.

### 4.3.2 Capacity of coded QPIR

In Publication II we propose two converses for the capacity of coded QPIR. As stated before, the QPIR protocols proposed in this thesis are an adaptation of the classical scheme in [12] to already-known uncoded QPIR protocols, thus we consider them to be *induced* by a classical scheme. For this class of QPIR protocols, we proved in [PII, Theorem 4.1] that their $\epsilon$-error capacity is upper bounded by $\min(1, 2C_{\epsilon}[\mathscr{A}])$, where $C_{\epsilon}[\mathscr{A}]$ is the $\epsilon$-error capacity (cf. Section 4.1.1) of the classical scheme with assumptions $\mathscr{A}$ that induces the QPIR scheme. For example, let $\mathscr{A}$ be the following set of assumptions:

- symmetry (*i.e.*, server privacy in addition to user privacy),
- $[N,K]$ MDS coded storage with $M$ files,
- protection against $T$-collusion,
- strong linearity.

Then it was proven in [20] that $C_M[\mathscr{A}] = \frac{N-K-T+1}{N}$. Since the QPIR protocol proposed in the previous section has rate $\frac{2(N-K-T+1)}{N}$ for $N/2 \le K+T-1 < N$ and rate 1 for $1 \le K+T-1 < N/2$, it is clear that it achieves the capacity for a PIR setting with assumptions in $\mathscr{A}$, *i.e.*,

$$C_M^{\mathrm{Q}}[\mathscr{A}] = \begin{cases} 1 & \text{if } K+T-1 < N/2, \\ \frac{2(N-K-T+1)}{N} & \text{if } N/2 \le K+T-1 < N. \end{cases} \tag{4.6}$$

The other converse result for the capacity of coded QPIR is proved in the case of no collusion, but it is valid for any QPIR scheme, not just for the ones induced by classical schemes. Let $\mathscr{A}$ be the following set of assumptions:

- symmetry,

- $[N,K]$ MDS coded storage with $M$ files and $N/2 \le K < N$,
- protection against no collusion,
- linearity.

In [PII, Theorem 4.3] we proved that the asymptotic $\epsilon$-error capacity for a coded QPIR scheme with no collusion is bounded from above as

$$\lim_{\epsilon\to 0}\lim_{M\to\infty} C^{\mathrm{Q}}_{M,\epsilon}[\mathscr{A}] \le \frac{2(N-K)}{N}.$$

By fixing the protocol proposed in the previous section with $T=1$, we can see that it is capacity-achieving, *i.e.*,

$$\lim_{\epsilon\to 0} C^{\mathrm{Q}}_{M,\epsilon}[\mathscr{A}] = \begin{cases} 1 & \text{if } K < N/2, \\ \frac{2(N-K)}{N} & \text{if } N/2 \le K < N. \end{cases} \tag{4.7}$$

.

## 4.4 $N$-sum box based QPIR protocols

In Publication III we propose the construction of an *N-sum box*, which represents an abstraction for linear computation over many-to-one quantum networks. Using such abstraction, we further construct an $X$-secure QPIR scheme with coded storage that protects against $T$-collusion, assuming that the servers possess $q$-dimensional qudits. The idea is that we adapt the classical PIR protocol proposed in [22] to employ the $N$-sum box construction and achieve higher rates.

### 4.4.1 Stabilizer-based $N$-sum boxes

An $N$-sum box is a black box that can be represented as the following linear transformation from $\mathbb{F}_q^{2N}$ to $\mathbb{F}_q^N$:

$$\begin{pmatrix} y_1 \\ y_2 \\ \vdots \\ y_N \end{pmatrix} = \begin{pmatrix} M_{1,1} & \cdots & M_{1,2N} \\ \vdots & \ddots & \vdots \\ M_{N,1} & \cdots & M_{N,2N} \end{pmatrix} \begin{pmatrix} x_1 \\ \vdots \\ x_N \\ x_{N+1} \\ \vdots \\ x_{2N} \end{pmatrix},$$

or equivalently $\mathbf{y} = \mathbf{M}\mathbf{x}$, where $\mathbf{y} \in \mathbb{F}_q^N$ is the output vector, $\mathbf{x} \in \mathbb{F}_q^{2N}$ is the input vector, and $\mathbf{M} \in \mathbb{F}_q^{N\times 2N}$ is the *transfer matrix*. The $N$-sum box operates with $N$ *transmitters*, where each transmitter $n \in [N]$ controls the elements $(x_n, x_{N+n})$ of the input vector. The resulting output vector $\mathbf{y}$ is measured by a separate party known as the *receiver*.

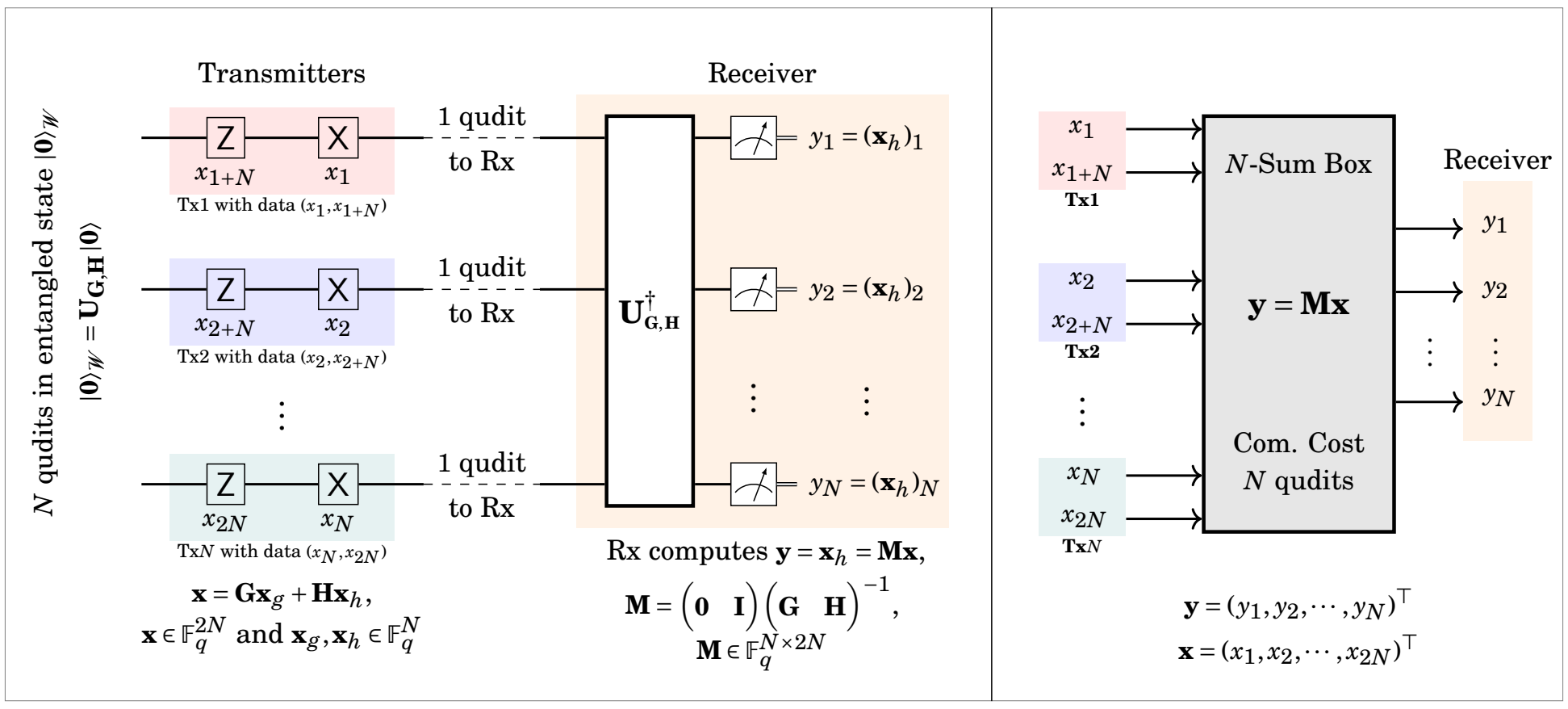

**Figure 4.8.** Quantum circuit and black-box representation for an $N$-sum box with transfer function $\mathbf{y} = \mathbf{M}\mathbf{x}$.

To initialize the $N$-sum box, the $N$ transmitters share quantum entanglement. This means that $N$ entangled $q$-dimensional qudits are prepared and distributed to the transmitters, with one qudit assigned to each transmitter. This initial entanglement is independent of the input vector $\mathbf{x}$ and any subsequent data available to the transmitters. At the beginning of the operation, the receiver does not possess any quantum resources. During the operation, each transmitter gathers data from various sources, including the receiver (*e.g.*, in the context of SPIR, each transmitter receives queries from the receiver and shared randomness to ensure server privacy), and utilizes that information to perform conditional $\mathsf{X}, \mathsf{Z}$-gate operations on their respective qudit. Subsequently, each transmitter sends its qudit to the receiver. Finally, the receiver performs a quantum measurement on the $N$ received qudits, extracting the output vector $\mathbf{y}$ from the measurement results. The quantum circuit and the black-box representation for an $N$-sum box are depicted in Figure 4.8.

The $N$-sum box represents a generalization of the two-sum transmission protocol discussed in Section 3.5.3 and is implicitly used in various stabilizer-based QPIR protocols, such as the one proposed by Song *et al.* in [35] and its extension discussed in Section 4.3.1. The black-box representation hides the details of the quantum circuit and specifies only the functionality (transfer matrix $\mathbf{M}$) and the communication cost (2 qubits), which makes it possible for non-quantum experts to design low-communication-cost coding schemes for quantum communication networks using this black box, *e.g.*, to take advantage of super-dense coding. In this thesis we consider stabilizer-based constructions for $N$-sum boxes, explore their limitations, and propose an application to QPIR. Before proceeding, let us consider an example of how to abstract the two-sum transmission protocol into a black box.

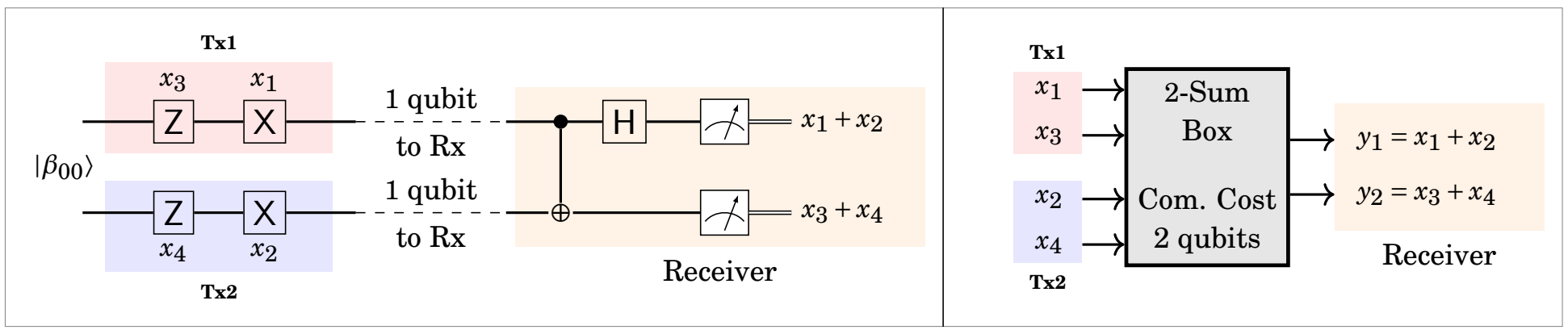

**Figure 4.9.** Quantum circuit and black-box representation for the two-sum transmission protocol with $|\beta_{00}\rangle = \frac{1}{\sqrt{2}}(|00\rangle + |11\rangle)$.

**Example 13.** *The quantum circuit and the black-box representation for the two-sum transmission protocol are shown in Figure 4.9. The two-sum protocol can be abstracted into a black box with inputs* $(x_1, x_3), (x_2, x_4)$ *controlled by Tx1 and Tx2, respectively, and output* $\mathbf{y} = \mathbf{M}\mathbf{x}$*, where* $\mathbf{M} = \left(\begin{smallmatrix} 1 & 1 & 0 & 0 \\ 0 & 0 & 1 & 1 \end{smallmatrix}\right)$ *is the transfer matrix of this* 2*-sum box and* $\mathbf{x} = (x_1, x_2, x_3, x_4)^\top$.

Next, we propose how to build an $N$-sum box construction based on the stabilizer formalism. Consider a strongly self-orthogonal matrix $\mathbf{G} \in \mathbb{F}_q^{N \times 2N}$ (cf. Definition 22), *i.e.*, a full-rank matrix such that $\mathbf{G}\mathbf{J}\mathbf{G}^\top = \mathbf{0}$, and a matrix $\mathbf{H} \in \mathbb{F}_q^{N \times 2N}$ such that the matrix $\left(\begin{smallmatrix} \mathbf{G} \\ \mathbf{H} \end{smallmatrix}\right)$ is invertible. Then in [PIII, Theorem 1] we prove that there exists a stabilizer-based construction for an $N$-sum box over $\mathbb{F}_q$ with transfer matrix

$$\mathbf{M} = \begin{pmatrix} \mathbf{0} & \mathbf{I} \end{pmatrix} \begin{pmatrix} \mathbf{G}^\top & \mathbf{H}^\top \end{pmatrix}^{-1},$$

which is the matrix comprised of the bottom $N$ rows of $\left(\mathbf{G}^\top \; \mathbf{H}^\top\right)^{-1}$. Since $\mathbf{G}$ is strongly self-orthogonal, the induced stabilizer is maximal and it serves as the foundation for constructing the $N$-sum box.

**Remark.** *A stabilizer-based construction for any feasible $N$-sum box* $\mathbf{y} = \mathbf{M}\mathbf{x}$ *achieves information-theoretic optimality as a black-box implementation in the sense that it has the smallest possible quantum download cost. In other words, since the transfer matrix* $\mathbf{M}$ *is full rank, there is no more efficient construction (in terms of download cost) for the same $N$-sum box using alternative non-stabilizer-based methods. It is impossible to deliver $N$ $q$-ary digits to the receiver with a communication cost lower than $N$ qudits, as dictated by the Holevo bound.*

### 4.4.2 Stabilizer-based $(\kappa, N)$-sum boxes

As concluded in the previous section, an $N$-sum box construction is built upon a maximal stabilizer. On the other hand, non-maximal stabilizers can also be used to build a similar abstraction, called a $(\kappa, N)$-sum box, as it outputs only $\kappa$ $q$-ary digits while discarding the remaining $N - \kappa$.

More precisely, we know from Section 3.4.1 that a self-orthogonal matrix $\mathbf{G} \in \mathbb{F}_q^{\kappa \times 2N}$ fully characterizes a non-maximal stabilizer group $\mathscr{S}(\mathcal{V})$, where $\mathcal{V} = \langle \mathbf{G} \rangle_{\mathsf{row}}$. Furthermore, as a consequence of the symplectic completion for a self-orthogonal matrix (cf. [PIII, Lemma 1]), there exists a full-rank matrix $\mathbf{G}^{\perp} \in \mathbb{F}_q^{(2N-\kappa)\times 2N}$ such that $\mathcal{V}^{\perp_{\mathbb{S}}} = \langle \mathbf{G}^{\perp} \rangle_{\mathsf{row}}$, *i.e.*, such that

1. $\mathbf{G}\mathbf{J}(\mathbf{G}^{\perp})^{\top} = \mathbf{0}$, *i.e.*, $\langle \mathbf{g}_i, \mathbf{g}'_j \rangle_{\mathbb{S}} = 0$ for $i \in [\kappa]$, $j \in [2N-\kappa]$, where $\mathbf{g}_i$ is the $i^{th}$ row of $\mathbf{G}$ and $\mathbf{g}'_j$ is the $j^{th}$ row of $\mathbf{G}^{\perp}$, and

2. $\mathbf{G} = (\mathbf{I}_\kappa \ \mathbf{0})\mathbf{G}^{\perp}$, *i.e.*, $\mathbf{G}$ is the upper $\kappa \times 2N$ submatrix of $\mathbf{G}^{\perp}$.

Consider now a pair of such matrices $\mathbf{G}, \mathbf{G}^{\perp}$, and let $\mathbf{H} \in \mathbb{F}_q^{\kappa \times 2N}$ be such that the matrix $\begin{pmatrix} \mathbf{G}^{\perp} \\ \mathbf{H} \end{pmatrix}$ is invertible. Then there exists a stabilizer-based construction for a $(\kappa, N)$-sum box over $\mathbb{F}_q$ with transfer matrix

$$\mathbf{M} = \begin{pmatrix} \mathbf{0}_{\kappa\times(2N-\kappa)} & \mathbf{I}_\kappa \end{pmatrix} \begin{pmatrix} (\mathbf{G}^{\perp})^{\top} & \mathbf{H}^{\top} \end{pmatrix}^{-1} \in \mathbb{F}_q^{\kappa \times 2N},$$

which is the matrix comprised of the bottom $\kappa$ rows of $\begin{pmatrix} (\mathbf{G}^{\perp})^{\top} & \mathbf{H}^{\top} \end{pmatrix}^{-1}$.

As an example, let us show the implementation of a $(1,2)$-sum box over 2 qubits.

**Example 14.** *Suppose we have two parties, Tx1 and Tx2, both possessing a pair of bits* $\mathbf{a} = (x_1, x_3)$, $\mathbf{b} = (x_2, x_4) \in \mathbb{F}_2^2$, *respectively. Let* $\mathcal{S} = \langle \mathbf{W}(1,1,0,0) \rangle$ *be a 1-dimensional stabilizer over 2 qubits. The symplectic matrix that completes the transpose of the generator matrix* $\mathbf{G}$ *of the stabilizer can be chosen as*

$$\mathbf{F} = \begin{pmatrix} 1 & 0 & 0 & 1 \\ 1 & 0 & 0 & 0 \\ 0 & 1 & 0 & 0 \\ 0 & 1 & 1 & 0 \end{pmatrix}.$$

*Thus, we can choose*

$$\begin{pmatrix} (\mathbf{G}^{\perp})^{\top} & \mathbf{H}^{\top} \end{pmatrix} = \begin{pmatrix} 1 & 0 & 1 & 0 \\ 1 & 0 & 0 & 0 \\ 0 & 1 & 0 & 0 \\ 0 & 1 & 0 & 1 \end{pmatrix},$$

*which has an inverse so that*

$$\begin{pmatrix} 0 & 0 & 0 & 1 \end{pmatrix} \begin{pmatrix} (\mathbf{G}^{\perp})^{\top} & \mathbf{H}^{\top} \end{pmatrix}^{-1} \begin{pmatrix} x_1 \\ x_2 \\ x_3 \\ x_4 \end{pmatrix} = x_3 + x_4.$$

*One might notice that the transpose of the first two columns of* $\mathbf{F}$ *is the same matrix as the generator matrix of a maximal stabilizer over 2 qubits*

*(cf. Example 9). Indeed, the quantum circuit for such a* $(1,2)$*-sum box involves preparing an initial state by applying the unitary* $\mathsf{U} = \mathsf{CNOT}(\mathsf{H} \otimes \mathsf{I})$ *to a pair of qubits in the mixed state* $|0\rangle\langle 0| \otimes \frac{\mathbf{I}}{2}$*, which is the same unitary we need to prepare a two-sum box from the pure state* $|\mathbf{0}\rangle$ *over two qubits. Since one of the two qubits is originally prepared in the maximally mixed state, the output of a measurement on that qubit is uniformly random and can be discarded. It follows that the output of a Bell measurement at the end of the operation of a* $(1,2)$*-sum box is given by a single bit.*

### 4.4.3 $X$-secure CSA-coded QPIR protocol with $T$-collusion

Let $N$ be the number of distributed servers that compute local answers $A_n$ for a user's query. Each answer $A_n$ is a linear combination of $L$ desired symbols $\delta_1, \delta_2, \dots, \delta_L$, along with $N-L$ interference symbols $v_1, v_2, \dots, v_{N-L}$. The linear combinations exhibit a Cauchy-Vandermonde structure, characteristic of Cross Subspace Alignment (CSA) schemes (cf. Section 2.3.2), where the desired terms align with the Cauchy terms and the interference aligns with the Vandermonde terms. Thus, let $\mathscr{F} = \{f_1, \dots, f_L\} \subseteq \mathbb{F}_q^L$ and $\mathscr{A} = \{\alpha_1, \dots, \alpha_{N-L}\} \subseteq \mathbb{F}_q^{N-L}$ be sets of distinct elements and such that $\mathscr{F} \cap \mathscr{A} = \emptyset$. The $b$-th instance of a CSA scheme can be represented as $\mathbf{A}^b = \mathbf{X}^b_{\delta,v} \mathbf{G}_{\mathrm{CSA}^q_{N-L}(\mathscr{A},\mathscr{F})}$, where $\mathbf{A}^b$ represents the servers' answers, $\mathbf{G}_{\mathrm{CSA}^q_{N-L}(\mathscr{A},\mathscr{F})}$ is the generator matrix of a CSA code of dimension $N$ with Cauchy-Vandermonde structure, and $\mathbf{X}^b_{\delta,v}$ denotes the vector of desired and interference symbols. After downloading the answers $A_n$ from each server $n$, the user can recover the desired symbols $\delta^b_1, \dots, \delta^b_L$ by inverting the matrix $\mathbf{G}_{\mathrm{CSA}^q_{N-L}(\mathscr{A},\mathscr{F})}$. Thus, each instance $b$ of the CSA scheme allows the user to retrieve $L$ desired symbols at the cost of downloading $N$ symbols, achieving a rate of $L/N$.

Such CSA scheme is used in various private information retrieval (PIR) schemes and in secure distributed batch matrix multiplication (SDBMM). To adapt a CSA scheme to the $N$-sum box construction, we need to generalize the generator matrix of a CSA code to obtain a self-orthogonal submatrix. Noticing that the interference terms align with the Vandermonde structure which induces an RS code, we can introduce column multipliers in the generator matrix to obtain a submatrix which induces a GRS code. As GRS codes can be made self-dual by choosing specific column multipliers (cf. Corollary 2), we introduce a QCSA code as the set of codewords

$$\mathrm{QCSA}^q_K(\mathscr{A},\mathscr{F},\mathbf{v}) := \left\{ \mathbf{m}\mathbf{V}(\mathscr{A},\mathscr{F})\,\mathrm{diag}(\mathbf{v}) : \mathbf{m} \in \mathbb{F}_q^{K+L} \right\} \subseteq \mathbb{F}_q^{K+L}, \tag{4.8}$$

where $\mathbf{v}$ is a vector with non-zero entries in $\mathbb{F}_q$.

Thus, to create a matrix that matches the $\mathbf{G},\mathbf{H}$ construction for an $N$-sum box, we consider two vectors of column multipliers $\mathbf{u}, \mathbf{v}$ such that

$$v_j = \frac{1}{u_j} \left( \prod_{i \in [N], i \neq j} (\alpha_j - \alpha_i) \right)^{-1} \quad \forall j \in [N]. \tag{4.9}$$

Then, letting $\mathbf{Q}_N^{\mathbf{u}} = \mathbf{G}_{\mathrm{QCSA}_{N-L}^q(\mathscr{A},\mathscr{F},\mathbf{u})}$, $\mathbf{Q}_N^{\mathbf{v}} = \mathbf{G}_{\mathrm{QCSA}_{N-L}^q(\mathscr{A},\mathscr{F},\mathbf{v})}$, we have that the submatrices $\mathbf{G}_{\mathrm{GRS}_{\lceil N/2 \rceil}^q(\mathscr{A},\mathbf{u})}, \mathbf{G}_{\mathrm{GRS}_{\lfloor N/2 \rfloor}^q(\mathscr{A},\mathbf{v})}$ obtained by extracting the first $\lceil N/2 \rceil$ and $\lfloor N/2 \rfloor$ rows of the Vandermonde parts of the two QCSA matrices satisfy the relation

$$\mathbf{G}_{\mathrm{GRS}_{\lceil N/2 \rceil}^q(\mathscr{A},\mathbf{u})} \mathbf{G}_{\mathrm{GRS}_{\lfloor N/2 \rfloor}^q(\mathscr{A},\mathbf{v})}^\top = \mathbf{0}.$$

It follows that the matrix $\operatorname{diag}\left( \mathbf{G}_{\mathrm{GRS}_{\lceil N/2 \rceil}^q(\mathscr{A},\mathbf{u})}, \mathbf{G}_{\mathrm{GRS}_{\lfloor N/2 \rfloor}^q(\mathscr{A},\mathbf{v})} \right)$ is strongly self-orthogonal, and since the two matrices $\mathbf{Q}_N^{\mathbf{u}}, \mathbf{Q}_N^{\mathbf{v}}$ are invertible, we obtain a feasible $N$-sum box with transfer matrix

$$\mathbf{M} = \left( \begin{array}{ccc|ccc} \mathbf{I}_L & \mathbf{0}_{L \times \lceil N/2 \rceil} & \mathbf{0} & \mathbf{0} & \mathbf{0} & \mathbf{0} \\ \mathbf{0} & \mathbf{0} & \mathbf{I}_{\lfloor N/2 \rfloor - L} & \mathbf{0} & \mathbf{0} & \mathbf{0} \\ \mathbf{0} & \mathbf{0} & \mathbf{0} & \mathbf{I}_L & \mathbf{0}_{L \times \lfloor N/2 \rfloor} & \mathbf{0} \\ \mathbf{0} & \mathbf{0} & \mathbf{0} & \mathbf{0} & \mathbf{0} & \mathbf{I}_{\lceil N/2 \rceil - L} \end{array} \right) \begin{pmatrix} \left(\mathbf{Q}_N^{\mathbf{u}}\right)^\top & \mathbf{0} \\ \mathbf{0} & \left(\mathbf{Q}_N^{\mathbf{v}}\right)^\top \end{pmatrix}^{-1}. \tag{4.10}$$

For more details, we refer the reader to [PIII, Theorem 6].

Now, we know there exists a CSA scheme for an $X$-secure MDS-coded PIR protocol that protects against $T$-collusion [22]. By combining two instances of such a CSA scheme with parameters matching the ones described in the above construction, we can create an $X$-secure MDS-coded QPIR protocol that protects against $T$-collusion and achieves a rate of

$$R^{\mathrm{Q}} = \min\left\{ 1, 2\left( 1 - \left( \frac{X+T+K-1}{N} \right) \right) \right\}, \tag{4.11}$$

where $K$ is the dimension of the uncoded files (cf. [PIII, Corollary 1]).

In a similar manner as above, we can create an $(2L, N)$-sum box to output only the desired symbols of the two instances of the scheme (cf. [PIII, Section VI.B]). In this way, we can create an $X$-secure MDS-coded *symmetric* QPIR protocol that protects against $T$-collusion and achieves a rate of

$$R^{\mathrm{Q}} = \min\left\{ 1, 2\left( 1 - \left( \frac{X+T+K-1}{N} \right) \right) \right\}. \tag{4.12}$$

**Remark.** *It is worth noting that in this scenario, there is no requirement for supplying extra information to the servers in order to establish symmetry for the protocol. In fact, the only necessary modification is to change the initial entangled state distributed among the servers, eliminating the need for additional shared randomness that would require additional computations and storage. This is in sharp contrast to classical PIR.*

### 4.5 BRM-coded QPIR protocols

In Publication IV we propose the best practices to create a $(T,U,B)$-robust BRM-coded (*i.e.*, with a Binary Reed–Muller storage code) PIR protocol along with an example, following the work on BRM-coded PIR by Freij-Hollanti *et al.* [11]. We extend now that work using the $N$-sum box abstraction to propose an example of a BRM-coded QPIR scheme that protects against $T$-collusion, assuming that the servers possess qubits.

#### 4.5.1 $(T,U,B)$-robust BRM-coded PIR protocol

Let $N = 2^{\mathsf{m}}$ be the number of servers for some integer $\mathsf{m}$ (which is lower bounded by the parameters of the scheme, as we discuss later), and let $\mathscr{C} = \text{BRM}(\mathsf{r},\mathsf{m})$ be the storage code. Let $K = \sum_{i=0}^{\mathsf{r}} \binom{\mathsf{m}}{i}$ be the dimension of $\mathscr{C}$, and choose $\beta$ and $\rho$ optimally by setting $\beta = \text{lcm}(c,K)/K$ and $\rho = \text{lcm}(c,K)/c$ for some $c$ we define later. The parameters $\beta,\rho,c$ can be thought of as the number of rows per file, the number of iterations in the scheme, and the number of symbols that can be retrieved in each round, similarly to the protocol in Section 4.3.1. Then, we encode the matrix of files $\mathbf{X} \in \mathbb{F}_2^{M\beta\times K}$ by multiplying it with the generator matrix $\mathbf{G}_{\mathscr{C}}$ and distribute the encoded matrix $\mathbf{Y} \in \mathbb{F}_2^{M\beta\times N}$ to the servers.

Let $\mathscr{D} = \text{BRM}(\mathsf{r}',\mathsf{m})$ be the query code. In order to protect against $T$-collusion, we need to set $\mathsf{r}' \geq \log(T+1)-1$, since the minimum distance of $\mathscr{D}^{\perp}$ must be greater than or equal to $T+1$. Furthermore, we need a symbol retriever vector $\mathbf{e}$ such that $\mathscr{C}\star\mathbf{e}$ is contained in some suitable RM code, and such that $\mathscr{C}\star\mathscr{D}+\mathscr{C}\star\mathbf{e}$ has minimum distance greater or equal than $U+2B+1$ to enable error-correcting capabilities. Assuming that $\mathscr{C}\star\mathscr{D}+\mathscr{C}\star\mathbf{e} \subseteq \text{BRM}(\mathsf{r}+\mathsf{r}_e,\mathsf{m})$ for some $\mathsf{r}_e$, we have that $\mathsf{r}_e \leq \mathsf{m}-\mathsf{r}-\log(U+2B+1)$ to ensure the minimum distance requirement. To ensure that $\mathscr{C}\star\mathbf{e}$ does not vanish when projecting to $(\mathscr{C}\star\mathscr{D})^{\perp}$ to remove the randomness from the response, we need $\mathsf{m} \geq \mathsf{r}+\log((T+1)(U+2B+1))$. In order to minimize the number of servers required, we can establish a precise equality between $\mathsf{m}$ and $\mathsf{r}_e$ by taking the ceiling of the logarithm. Specifically, we set $\mathsf{m} = \mathsf{r}+\lceil\log((T+1)(U+2B+1))\rceil$ and $\mathsf{r}_e = \mathsf{m}-\mathsf{r}-\lceil\log(U+2B+1)\rceil$, which fixes $c = \sum_{i=\mathsf{r}+\mathsf{r}'+1}^{\mathsf{r}+\mathsf{r}_e} \binom{\mathsf{m}}{i}$.

As usual, the user generates the queries by adding an error matrix $\mathbf{E}_{(\theta)}^{(r)}$ to the $M\beta\times N$ matrix of random codewords $\mathbf{Z}^{(r)}\mathbf{G}_{\mathscr{D}}$. In order to have a clear description of the error matrix, we represent the codewords as Reed–Muller polynomials. As stated in [PIV, Remark 1], the existence of such polynomials and their explicit formula in the general case remains an open problem. Nevertheless, we can offer some general guidelines to consider when selecting these polynomials, which will become clear in the discussion of Example 15

1. It appears that obtaining linear combinations of desired symbols in the

response is inevitable. However, this is not an issue if the queries are designed in a way that allows these linear dependencies to be solved later.

2. It is crucial to download the coefficients of higher degree terms first. Downloading lower-degree terms would involve pushing higher-order terms beyond the error-correction capability. Consequently, if the coefficients of the higher degree terms are unknown, they cannot be subtracted, rendering error correction infeasible.

3. It is expected that symbols will be downloaded from only a subset of the stripes in a single round, as sending polynomials to additional stripes would reduce the number of explicitly downloaded symbols and increase the number of downloaded linear combinations.

After the user sends the queries to the servers, the $n^{th}$ server computes $\left\langle \mathbf{Y}_n, \mathbf{Q}_n^{(r)} \right\rangle$ during round $r \in [\rho]$ as usual, and sends the response back to the user. To decode the responses, the user initiates the decoding process by subtracting the known terms from the response. The resulting vector is then error-corrected and projected onto the code $(\mathscr{C} \star \mathscr{D})^{\perp}$ by left-multiplying it with its generator matrix $\mathbf{G}_{(\mathscr{C}\star\mathscr{D})^{\perp}}$. In order to recover the desired symbols, it is necessary for an information set of the matrix containing these symbols to be included within an information set of $(\mathscr{C} \star \mathscr{D})^{\perp}$. This allows us to simply consider an invertible submatrix of $\mathbf{G}_{(\mathscr{C}\star\mathscr{D})^{\perp}}$ formed by selecting the columns corresponding to the information set. By multiplying this submatrix with its inverse, we can successfully recover the desired symbols. Due to the conditions $\mathsf{r} + \mathsf{r}_e \leq \mathsf{m} - \mathsf{r} - \mathsf{r}' - 1$ and the recursive properties of Reed-Muller codes, it is always possible to identify an information set that satisfies the requirements outlined above.

Thus, if the query polynomial exists, [PIV, Theorem 1] states that this $(T,U,B)$-robust BRM-coded PIR protocol has rate

$$R = \frac{\sum_{i=\mathsf{r}+\mathsf{r}'+1}^{\mathsf{r}+\mathsf{r}_e} \binom{\mathsf{m}}{i}}{2^{\mathsf{m}}}$$

for given parameters $\mathsf{r}, \mathsf{r}', \mathsf{r}_e, \mathsf{m}$ depending on the PIR parameters $N,K,T,U,B$. Let us show an example of a $(1,1,1)$-robust BRM-coded PIR scheme

**Example 15.** *Consider the case where* $\mathsf{m} = 4$, $T = U = B = 1$ *and* $\mathscr{C} = \mathrm{BRM}(1,4)$*, so that* $\mathsf{r} = 1$, $\mathsf{r}' = 0$, $K = 5$ *and* $N = 16$*. Furthermore, the query code is* $\mathscr{D} = \mathrm{BRM}(0,4)$ *and we find that* $\mathsf{r}_e = 1$, $c = 6$, $\beta = 6$ *and* $\rho = 5$*. Hence, each row* $\mathbf{X}^{i,b} = \left(X_1^{i,b}, \ldots, X_5^{i,b}\right)$ *of the file matrix* $\mathbf{X} \in \mathbb{F}_2^{M\beta\times K}$ *is encoded by the polynomial* $p^{i,b}(z) = X_1^{i,b} + X_2^{i,b} z_1 + X_3^{i,b} z_2 + X_4^{i,b} z_3 + X_5^{i,b} z_4$ *and the* $n^{th}$ *server has the evaluation* $p^{i,b}(P_n)$ *for some* $P_n \in \mathscr{P} = \mathbb{F}_2^4$*. Notice that, since this is classical PIR, there is only one instance of the database as an* $M\beta \times K$ *matrix, so we drop the first subscript that is used to refer to the instance.*

*There are* $c = 6$ *monomials whose coefficients can be downloaded in each round, namely* $\mathscr{I} = \{z_1z_2, z_1z_3, z_1z_4, z_2z_3, z_2z_4, z_3z_4\}$. *Considering that the polynomials* $p^{i,b}$ *have degree one, we notice that we need degree-one monomial terms in the error matrix* $\mathbf{E}^{(1),\theta}_{(\theta)} \in \mathbb{F}_2^{\beta\times N}$ *containing the evaluations of the error polynomials* $e^{(1),b}(z)$ *so that* $e^{(1),b}(z)p^{\theta,b}(z)$ *is one of the polynomials defined in* $\mathscr{I}$. *To avoid downloading the same symbol multiple times from a single row, we divide the monomial terms into three stripes. The same argument is valid for rounds 2, 3, and 4. Thus, we define error matrices for the first 4 rounds as follows:*

$$\mathbf{E}^{(1),\theta}_{(\theta)} = \begin{pmatrix} \mathrm{eval}_{\mathscr{P}}(z_1) \\ \mathrm{eval}_{\mathscr{P}}(z_2) \\ \mathrm{eval}_{\mathscr{P}}(z_3) \\ \mathbf{0} \\ \mathbf{0} \\ \mathbf{0} \end{pmatrix}, \quad \mathbf{E}^{(2),\theta}_{(\theta)} = \begin{pmatrix} \mathrm{eval}_{\mathscr{P}}(z_2) \\ \mathrm{eval}_{\mathscr{P}}(z_3) \\ \mathrm{eval}_{\mathscr{P}}(z_4) \\ \mathbf{0} \\ \mathbf{0} \\ \mathbf{0} \end{pmatrix}$$

$$\mathbf{E}^{(3),\theta}_{(\theta)} = \begin{pmatrix} \mathbf{0} \\ \mathbf{0} \\ \mathbf{0} \\ \mathrm{eval}_{\mathscr{P}}(z_1) \\ \mathrm{eval}_{\mathscr{P}}(z_2) \\ \mathrm{eval}_{\mathscr{P}}(z_3) \end{pmatrix}, \quad \mathbf{E}^{(4),\theta}_{(\theta)} = \begin{pmatrix} \mathbf{0} \\ \mathbf{0} \\ \mathbf{0} \\ \mathrm{eval}_{\mathscr{P}}(z_2) \\ \mathrm{eval}_{\mathscr{P}}(z_3) \\ \mathrm{eval}_{\mathscr{P}}(z_4) \end{pmatrix},$$

*and the rest of each overall matrix* $\mathbf{E}^{(r)}_{(\theta)}$ *is zero.*

*During round 1, the response polynomial has degree-two monomials with coefficients* $X_5^{\theta,1}$,$X_5^{\theta,2}$, $X_5^{\theta,3}$, $\left(X_3^{\theta,1}+X_2^{\theta,2}\right)$, $\left(X_4^{\theta,1}+X_2^{\theta,3}\right)$ *and* $\left(X_4^{\theta,2}+X_3^{\theta,3}\right)$, *while during round 2 it has degree-two monomials with coefficients* $X_2^{\theta,1}$, $X_2^{\theta,2}$, $X_2^{\theta,3}$, $\left(X_4^{\theta,1}+X_3^{\theta,2}\right)$, $\left(X_5^{\theta,1}+X_3^{\theta,3}\right)$ *and* $\left(X_5^{\theta,2}+X_4^{\theta,3}\right)$. *After error-correcting and projecting to* $(\mathscr{C}\star\mathscr{D})^{\perp}$, *the user retrieves these coefficients, and it is easy to see that he can recover all the bits of a part of the degree-one terms of* $p^{\theta,b}$ *corresponding to a part of the file by combining the linear combinations (cf. Figure 4.10). During rounds 3 and 4 he can recover the remaining degree-one terms of* $p^{\theta,b}$.

*Finally, in the last round, the user can download the coefficients of the degree-zero terms by sending each of the degree-two query polynomials to different rows. This means that the response will have degree-three terms which cannot be error-corrected, but since we know those terms from previous rounds, the user can subtract the higher degree terms before error*

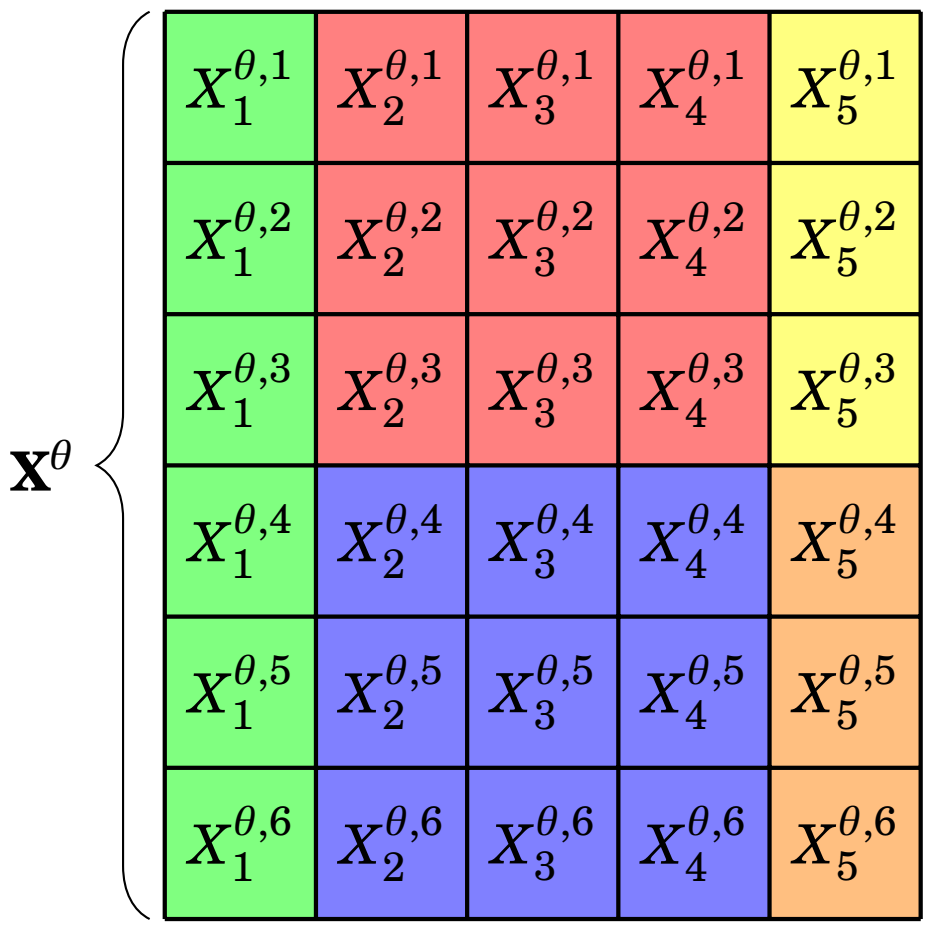

**Figure 4.10.** Bits downloaded in each round of the PIR protocol described in Example 15. The bits colored in yellow, in red, in orange, in blue, and in green are retrieved during round 1, round 2, round 3, round 4, and round 5, respectively.

*correction. The error matrix* $\mathbf{E}_{(\theta)}^{(5)}$ *for round 5 is thus defined with*

$$\mathbf{E}_{(\theta)}^{(5),\theta} = \begin{pmatrix} \mathrm{eval}_{\mathscr{P}}(z_1z_2) \\ \mathrm{eval}_{\mathscr{P}}(z_1z_3) \\ \mathrm{eval}_{\mathscr{P}}(z_1z_4) \\ \mathrm{eval}_{\mathscr{P}}(z_2z_3) \\ \mathrm{eval}_{\mathscr{P}}(z_2z_4) \\ \mathrm{eval}_{\mathscr{P}}(z_3z_4) \end{pmatrix}.$$

*Notice that the choice of which degree two monomial is sent to which stripe is irrelevant. We conclude that the user retrieves all the desired file's bits after 5 rounds, as shown in Figure 4.10*

**Remark.** *The rate achieved by the protocol in Example 15 is* $R = \frac{3}{8}$*, while the rate achieved by the protocol with the same robustness properties and GRS coding is* $R = \frac{4}{8}$ *[40]. However, the GRS scheme would require a field size of at least* $q \geq 16$ *whereas the BRM scheme works over the binary field. Further comparison demonstrates that the BRM scheme works better with more errors and less collusion. More precisely, an advantage of BRM-coded PIR with* $T$*-collusion is that it resists against collusion of more than* $T$ *servers with high probability, while GRS-coded PIR with* $T$*-collusion does not resist against collusion of* $T+1$ *servers.*

### 4.5.2 BRM-coded QPIR protocol with $T$-collusion

In this section, we propose an example of BRM-coded QPIR protocol with $T$-collusion which is obtained by adapting [11, Example 5] to the $N$-sum box construction. The example can be easily generalized by combining the general BRM-coded PIR protocol described in [11] with the $N$-sum box construction. Such a generalization follows directly from the fact that BRM codes are weakly self-dual by construction, from which we can easily define self-orthogonal matrices to build $N$-sum boxes.

First, let us show how to generate an $N$-sum box from BRM codes. The main requirement is to find a strongly self-orthogonal matrix, which can be found using the recursive structure of BRM codes as follows.

**Proposition 1.** *Let* $\mathbf{v}_0^{\mathsf{m}},\dots,\mathbf{v}_{\mathsf{m}}^{\mathsf{m}}$ *be vectors defined following the recursive structure*

$$\mathbf{v}_i^{\mathsf{m}} = \begin{cases} \mathbf{1}^{2^{\mathsf{m}}}, & \text{if } i=0,\ \mathsf{m}\ge 0,\\ \left(\mathbf{v}_i^{\mathsf{m}-1},\mathbf{v}_i^{\mathsf{m}-1}\right), & \text{if } i\in\{1,\dots,\mathsf{m}-1\},\ \mathsf{m}>1,\\ \left(\mathbf{1}^{2^{\mathsf{m}-1}},\mathbf{0}^{2^{\mathsf{m}-1}}\right), & \text{if } i=\mathsf{m},\ \mathsf{m}\ge 1.\end{cases}$$

*The generator matrix of* BRM(r,m) *can also be described through the star product of vectors* $\mathbf{v}_0^{\mathsf{m}},\dots,\mathbf{v}_{\mathsf{m}}^{\mathsf{m}}$ *[1] as follows:*

$$\mathbf{G}_{\mathrm{BRM}(\mathsf{r},\mathsf{m})} = \begin{pmatrix} \mathbf{v}_0^{\mathsf{m}}\\ \mathbf{v}_{\mathsf{m}}^{\mathsf{m}}\\ \vdots\\ \mathbf{v}_1^{\mathsf{m}}\\ \mathbf{v}_{\mathsf{m}}^{\mathsf{m}}\star\mathbf{v}_{\mathsf{m}-1}^{\mathsf{m}}\\ \vdots\\ \mathbf{v}_2^{\mathsf{m}}\star\mathbf{v}_1^{\mathsf{m}}\\ \vdots\\ \mathbf{v}_{\mathsf{m}}^{\mathsf{m}}\star\dots\star\mathbf{v}_{\mathsf{m}-\mathsf{r}+1}^{\mathsf{m}}\\ \vdots\\ \mathbf{v}_{\mathsf{r}}^{\mathsf{m}}\star\dots\star\mathbf{v}_1^{\mathsf{m}} \end{pmatrix}. \tag{4.13}$$

*Using such a definition, it is easy to prove that the top half of the matrix* $\mathbf{G}_{\mathrm{BRM}(\mathsf{m},\mathsf{m})}\in\mathbb{F}_2^{2^{\mathsf{m}}\times 2^{\mathsf{m}}}$ *is strongly self-orthogonal.*

The $N$-sum box construction follows easily from the proposition as follows.

**Theorem 7.** *Let* $\mathbf{Q}_{\mathsf{m}}\in\mathbb{F}_2^{2^{\mathsf{m}}\times 2^{\mathsf{m}}}$ *be the transpose of generator matrix of a* BRM(m,m) *code,* i.e.*,* $\mathbf{Q}_{\mathsf{m}}=\mathbf{G}^{\top}_{\mathrm{BRM}(\mathsf{m},\mathsf{m})}$. *Then there exists a feasible $N$-sum box* $\mathbf{y}=\mathbf{Mx}$ *in* $\mathbb{F}_2$ *where* $N=2^{\mathsf{m}}$ *and the transfer matrix is given by*

$$\mathbf{M}=\left(\begin{array}{cc|cc} \mathbf{0}_{N/2} & \mathbf{I}_{N/2} & \mathbf{0}_{N/2} & \mathbf{0}_{N/2}\\ \mathbf{0}_{N/2} & \mathbf{0}_{N/2} & \mathbf{0}_{N/2} & \mathbf{I}_{N/2}\end{array}\right)\begin{pmatrix}\mathbf{Q}_{\mathsf{m}} & \mathbf{0}\\ \mathbf{0} & \mathbf{Q}_{\mathsf{m}}\end{pmatrix}^{-1}. \tag{4.14}$$

**Example 16.** *Let* $\mathsf{r}=1,\ \mathsf{r}'=1,\ \mathsf{m}=4,\ N=2^{\mathsf{m}}=16$*, and let* $\mathscr{C}=\mathscr{D}=\mathrm{BRM}(1,4)$

*be the storage and query codes. Consider*

$$\mathbf{G}_{\mathrm{BRM}(4,4)} = \begin{pmatrix} 1&1&1&1&1&1&1&1&1&1&1&1&1&1&1&1 \\ 1&1&1&1&1&1&1&1&0&0&0&0&0&0&0&0 \\ 1&1&1&1&0&0&0&0&1&1&1&1&0&0&0&0 \\ 1&1&0&0&1&1&0&0&1&1&0&0&1&1&0&0 \\ 1&0&1&0&1&0&1&0&1&0&1&0&1&0&1&0 \\ 1&1&1&1&0&0&0&0&0&0&0&0&0&0&0&0 \\ 1&1&0&0&1&1&0&0&0&0&0&0&0&0&0&0 \\ 1&1&0&0&0&0&0&0&1&1&0&0&0&0&0&0 \\ 1&0&1&0&1&0&1&0&0&0&0&0&0&0&0&0 \\ 1&0&1&0&0&0&0&0&1&0&1&0&0&0&0&0 \\ 1&0&0&0&1&0&0&0&1&0&0&0&1&0&0&0 \\ 1&1&0&0&0&0&0&0&0&0&0&0&0&0&0&0 \\ 1&0&1&0&0&0&0&0&0&0&0&0&0&0&0&0 \\ 1&0&0&0&1&0&0&0&0&0&0&0&0&0&0&0 \\ 1&0&0&0&0&0&0&0&1&0&0&0&0&0&0&0 \\ 1&0&0&0&0&0&0&0&0&0&0&0&0&0&0&0 \end{pmatrix} \in \mathbb{F}_2^{N\times N}.$$

*Then* $\mathbf{G}_{\mathscr{C}}$ *and* $\mathbf{G}_{\mathscr{D}}$ *are obtained by extracting the first* $K = \sum_{i=0}^{\mathsf{r}} \binom{\mathsf{m}}{i} = 5$. *Furthermore,* $\mathscr{C} \star \mathscr{D} = \mathrm{BRM}(2,4)$, *whose dual is* $(\mathscr{C} \star \mathscr{D})^{\perp} = \mathrm{BRM}(1,4)$ *which has dimension 5. This means that the classical PIR rate that can be achieved is* $\frac{5}{16}$.

*Let us first rewrite [11, Example 5] to match the requirements for the N-sum box. Let* $\mathbf{X} \in \mathbb{F}_2^{M\times K}$ *be the uncoded storage with M messages* $\mathbf{X}^i \in \mathbb{F}_2^K$ *and let* $\mathbf{Z} \in \mathbb{F}_2^{M\times K}$ *be the randomness for the query generation. The coded storage is thus given by* $\mathbf{Y} = \mathbf{X}\mathbf{G}_{\mathscr{C}}$, *and column* $\mathbf{Y}_n$ *is stored on server* $n \in [N]$. *The query* $\mathbf{Q}$ *should be the sum of the coded randomness plus some error matrix that targets the desired file,* i.e., $\mathbf{Q} = \mathbf{Z}\mathbf{G}_{\mathscr{D}} + \mathbf{E}_{(\theta)}$, *where each row of* $\mathbf{E}_{(\theta)}$ *should be a codeword of the error code* $\mathscr{E}$.

*For* $i \in [M]$, *notice that*

$$\mathbf{Y}^i \star \mathbf{Q}^i = \underbrace{(\mathbf{X}^i\mathbf{G}_{\mathscr{C}}) \star (\mathbf{Z}^i\mathbf{G}_{\mathscr{D}})}_{\in \mathscr{C}\star\mathscr{D}} + \underbrace{(\mathbf{X}^i\mathbf{G}_{\mathscr{C}}) \star \mathbf{E}^i_{(\theta)}}_{\mathscr{C}\star\mathscr{E}}. \tag{4.15}$$

*Notice that the first part is a codeword of the star product* $\mathscr{C} \star \mathscr{D}$, *whose generator matrix is obtained by extracting the first* $\sum_{i=0}^{\mathsf{r}+\mathsf{r}'} \binom{\mathsf{m}}{i} = 11$ *rows of* $\mathbf{G}_{\mathrm{BRM}(4,4)}$. *Summing over* $i \in [M]$, *the first part of the sum in Equation* (4.15) *is a codeword in* $\mathscr{C} \star \mathscr{D}$.

*Let us choose* $\mathscr{E}$ *to be the code generated by the last* $\dim(\mathscr{C} \star \mathscr{D})^{\perp} = 5$ *rows of* $\mathbf{G}_{\mathrm{BRM}(4,4)}$, *which we denote* $\mathbf{G}_{\mathscr{E}}$, *and choose* $\mathbf{E}_{(\theta)}$ *such that*

$$\mathbf{E}^i_{(\theta)} = \begin{cases} \mathbf{0} & \text{if } i \neq \theta, \\ \mathbf{1}^K\mathbf{G}_{\mathscr{E}} & \text{if } i = \theta, \end{cases}$$

*where* $\theta \in [M]$ *is the index of the requested file. Then, following the notation for the row vectors of Equation* (4.13), *for* $i = \theta$ *we have that*

$$(\mathbf{X}^\theta\mathbf{G}_{\mathscr{C}}) \star \mathbf{E}^\theta_{(\theta)} = \mathbf{X}^\theta \begin{pmatrix} 1&1&1&1&1 \\ 1&1&1&0&0 \\ 1&1&0&1&0 \\ 1&0&1&1&0 \\ 0&1&1&1&0 \end{pmatrix} \mathbf{G}_{\mathscr{E}} = \mathbf{X}^\theta\mathbf{T}\mathbf{G}_{\mathscr{E}}.$$

*We conclude that, thanks to the star-product characterization of the generator matrix of a Reed–Muller code, the code $\mathscr{C} \star \mathscr{E}$ is actually $\mathscr{E}$ itself. Thus, we have the relation*

$$\mathbf{G}_{\mathrm{BRM}(4,4)} = \begin{pmatrix} \mathbf{G}_{\mathscr{C}\star\mathscr{D}} \\ \mathbf{G}_{\mathscr{E}} \end{pmatrix}.$$

*At this point, we can sum the elements in Equation* (4.15) *as*

$$\sum_{i=1}^{M} \mathbf{Y}^i \star \mathbf{Q}^i = \Big(\underbrace{* \;\cdots\; *}_{\in \mathbb{F}_2^{11}} \;\; \underbrace{\mathbf{X}^\theta \mathbf{T}}_{\in \mathbb{F}_2^5}\Big) \begin{pmatrix} \mathbf{G}_{\mathscr{C}\star\mathscr{D}} \\ \mathbf{G}_{\mathscr{E}} \end{pmatrix}.$$

*Using two instances of this scheme into an $N$-sum box with storages $\mathbf{X}^1, \mathbf{X}^2$, using Equation* (4.14) *with $\mathbf{Q} = \mathbf{G}^\top_{\mathrm{BRM}(4,4)}$, we have that*

$$\begin{aligned}
\mathbf{A} &= \mathbf{M} \begin{pmatrix} \mathbf{G}^\top_{\mathscr{C}\star\mathscr{D}} & \mathbf{G}^\top_{\mathscr{E}} & \mathbf{0} & \mathbf{0} \\ \mathbf{0} & \mathbf{0} & \mathbf{G}^\top_{\mathscr{C}\star\mathscr{D}} & \mathbf{G}^\top_{\mathscr{E}} \end{pmatrix} \begin{pmatrix} * \\ \mathbf{T}^\top \left(\mathbf{X}^{1,\theta}\right)^\top \\ * \\ \mathbf{T}^\top \left(\mathbf{X}^{2,\theta}\right)^\top \end{pmatrix} \\
&= \left(\begin{array}{cc|cc} \mathbf{0}_{N/2} & \mathbf{I}_{N/2} & \mathbf{0}_{N/2} & \mathbf{0}_{N/2} \\ \mathbf{0}_{N/2} & \mathbf{0}_{N/2} & \mathbf{0}_{N/2} & \mathbf{I}_{N/2} \end{array}\right) \begin{pmatrix} * \\ \mathbf{T}^\top \left(\mathbf{X}^{1,\theta}\right)^\top \\ * \\ \mathbf{T}^\top \left(\mathbf{X}^{2,\theta}\right)^\top \end{pmatrix} \\
&= \Big(\underbrace{* \;\cdots\; *}_{\in \mathbb{F}_2^3} \;\; \underbrace{\mathbf{X}^{1,\theta}\mathbf{T}}_{\in \mathbb{F}_2^5} \;\; \underbrace{* \;\cdots\; *}_{\in \mathbb{F}_2^3} \;\; \underbrace{\mathbf{X}^{2,\theta}\mathbf{T}}_{\in \mathbb{F}_2^5}\Big)^\top,
\end{aligned}$$

*where $*$ denotes redundant interference. Finally, we can retrieve the desired file $\left(\mathbf{X}^{1,\theta}, \mathbf{X}^{2,\theta}\right)$ by multiplying the answers with the inverse of* $\mathbf{T}$.

# 5. Conclusion

In the context of classical Private Information Retrieval (PIR), a user wants to access a file from a database or a Distributed Storage System (DSS) while maintaining the confidentiality of the file's identity, thus ensuring that the servers holding the data remain unaware of the specific file being retrieved. In this thesis, we proposed a novel example of a $(T,U,B)$-robust PIR protocol with a Binary Reed–Muller (BRM) storage code, *i.e.*, a PIR protocol retrieving a file from a BRM-coded DSS and protecting against $T$-collusion, $U$ unresponsive servers and $B$ adversarial servers.

Quantum Private Information Retrieval (QPIR) extends this classical setup by introducing a framework where users can privately retrieve classical files by receiving quantum information from the servers. This approach leverages the unique properties of quantum systems to achieve enhanced privacy and security in the retrieval process. The initial QPIR setting considered in this thesis has been introduced by Song *et al.*, who have addressed the scenarios of replicated servers with different collusion patterns. Their work has contributed valuable insights into developing QPIR protocols that effectively preserve privacy and enable users to retrieve information securely in both non-colluding and colluding server scenarios.

In this thesis, the QPIR setting was extended to account for coded servers, *i.e.*, servers storing files that are encoded and distributed according to a linear code. Similarly to the case with replicated storage considered by Song *et al.*, the proposed QPIR protocols achieve rates that are better than those known or conjectured in the classical counterparts using the underlying superdense-coding gain. In Chapter 4 we presented the following protocols.

- The first protocol works for any $[N,K]_{4^L}$ Maximum Distance Separable (MDS) code, assuming $(N-K)$-collusion among the servers possessing only qubits. We then used such a protocol as a subroutine for a QPIR protocol with a Locally Repairable Code (LRC) as storage code, where each local code is assumed to be MDS.

- The second protocol is a capacity-achieving strongly linear QPIR protocol with a Generalized Reed–Solomon (GRS) storage code that protects against $T$-collusion, assuming that the servers possess qudits. This protocol leverages the properties of the stabilizer formalism combined with the star-product PIR protocols to achieve a better rate than the classical counterpart [12].

- The third protocol is based on the combination of a PIR protocol with a Cross Subspace Alignment (CSA) storage code proposed in [22] and the $N$-sum box, a quantum protocol based on the stabilizer formalism to improve the communication rates of classical many-to-one networks employing quantum communication. This is an $X$-secure CSA-coded QPIR protocol that protects against $T$-collusion, assuming the servers possess qudits.

- The fourth protocol is an example of a BRM-coded QPIR protocol that protects against $T$-collusion with bounded $T$, assuming that the servers possess qubits, as BRM codes are binary. This example was built by combining a classical BRM-coded PIR protocol [11] and the $N$-sum box.

## 5.1 Future directions and open problems

The first direction involves determining the general capacity of coded QPIR, with or without symmetry, and possibly considering further assumptions, such as $X$-security or $(T,U,B)$-robustness. This problem remains unsolved in its full generality even in the classical setting. Although significant progress has been made in the case of full support-rank schemes [20] that essentially cover the general capacity-achieving schemes available in the literature, a complete resolution necessitates addressing intricate linear dependencies arising from the combination of collusion and coded storage. It is worth noting that the capacities established in the recent works are dependent on the number of files $M$, and it is plausible that they may exceed the asymptotic QPIR capacity derived in this study for a small number of files.

Another possibility is to explore non-stabilizer based QPIR schemes. The majority of existing multi-server QPIR schemes relies on constructions based on the stabilizer formalism. Thus, the search for non-stabilizer QPIR schemes becomes crucial as a preliminary step towards establishing the achievability aspect of a general non-asymptotic capacity theorem.

Furthermore, we can study the trade-off between the amount of entanglement and the capacity in QPIR protocols. However, even in the case of only two servers, determining the capacity under restricted entanglement poses a significant challenge.

These open problems present promising directions for future research, addressing fundamental questions in the capacity and achievability of QPIR schemes. Resolving these challenges would significantly advance the understanding and practical implementation of secure and efficient QPIR protocols. As the era of extensive data collection continues to unfold, individuals are becoming increasingly conscious of privacy concerns, so it is expected that the advancement of related services and product development will accelerate in the future. Anticipating further progress in quantum computation and the emergence of robust results regarding the capacity of PIR for practical applications, the problem addressed by this thesis can only gain even greater significance in the future.

# Publication I

Matteo Allaix, Lukas Holzbaur, Tefjol Pllaha, and Camilla Hollanti. Quantum Private Information Retrieval from Coded and Colluding Servers. *IEEE Journal on Selected Areas in Information Theory* , 1, no. 2: 599–610, August 2020.

# Publication II

Matteo Allaix, Seunghoan Song, Lukas Holzbaur, Tefjol Pllaha, Masahito Hayashi, and Camilla Hollanti. On the Capacity of Quantum Private Information Retrieval from MDS-Coded and Colluding Servers. *IEEE Journal on Selected Areas in Communications*, 40, no. 3: 885–898, January 2022.

# Publication III

Matteo Allaix, Yuxiang Lu, Yuhang Yao, Tefjol Pllaha, Camilla Hollanti, and Syed Jafar. $N$-Sum Box: An Abstraction for Linear Computation over Many-to-one Quantum Networks. Submitted to *IEEE Transactions on Information Theory*, June 2023.

# Publication IV

Perttu Saarela, Matteo Allaix, Ragnar Freij-Hollanti, and Camilla Hollanti. Private Information Retrieval from Colluding and Byzantine Servers with Binary Reed–Muller Codes. In *2022 IEEE International Symposium on Information Theory (ISIT)*, Espoo, Finland, pp. 2839–2844, June 2022.